\documentclass[aps,prd,twocolumn,nofootinbib,floatfix]{revtex4} 

\usepackage{graphics}
\usepackage{graphicx}

\usepackage{latexsym}
\usepackage[cp866]{inputenc}
\usepackage[english]{babel}

\input epsf

\begin{document}

\title{Strong isospin symmetry breaking in
light scalar meson production}
\author{N.~N. Achasov\,\footnote{E-mail: achasov@math.nsc.ru}
and G.~N. Shestakov\,\footnote{E-mail: shestako@math.nsc.ru}
}\affiliation{Laboratory of Theoretical Physics, Sobolev Institute
for Mathematics, Siberian Branch of the Russian Academy of Sciences,
prosp. Akademika Koptyuga 4, 630090 Novosibirsk, Russian Federation}

\begin{abstract}
Isospin symmetry breaking is discussed as a tool for studying the
nature and production mechanisms of light scalar mesons. We are
concerned with isospin breaking effects with an amplitude
$\sim\sqrt{m_d-m_u}$ (instead of the usual $\sim m_d-m_u$), where
$m_u$ and $m_d$ are the $u$ and $d$ quark masses, whose magnitude
and phase vary with energy in a resonance-like way characteristic of
the $K\bar K$ threshold region. We consider a variety of reactions
that can experimentally reveal (or have revealed) the mixing of
$a^0_0(980)$ and $f_0(980) $ resonances that breaks the isotopic
invariance due to the mass difference between $K^+$ and $K^0$
mesons. Experimental results on the search for $a^0_0(980)-f_0(980)$
mixing in $f_1(1285)\to f_0(980)\pi^0\to\pi^+\pi^-\pi^0$ and
$\eta(14 05)\to f_0(980)\pi^0\to\pi^+\pi^-\pi^0$ decays suggest a
broader perspective on the isotopic symmetry breaking effects due to
the $K^+$ and $K^0$ mass difference. It has become clear that not
only the $a^0_0(980)-f_0(980)$ mixing but also any mechanism
producing $K\bar K$ pairs with a definite isospin in an S wave gives
rise to such effects, thus suggesting a new tool for studying the
nature and production mechanisms of light scalars. Of particular
interest is the case of a large isotopic symmetry breaking in the
$\eta(1405)$\,$ \to$\,$f_0(980) \pi^0$\,$\to$\,$\pi^+\pi^-\pi^0$
decay due to the occurrence of anomalous Landau thresholds
(logarithmic triangle singularities), i.e., due to the $\eta(1405)
\to(K^*\bar K+\bar K^*K)\to(K^+K^-+K^0\bar K^0)\pi^0\to f_0(980)
\pi^0\to\pi^+\pi^- \pi^0$ transition (where it is of fundamental
importance that the $K^*$ meson has a finite width).\\

\end{abstract}

\maketitle

\begin{description}
\item{\bf 1.} {\bf Introduction. Nature of scalar mesons}
\item{\bf 2.} {\bf\boldmath The $a^0_0(980)-f_0(980)$ mixing as
a threshold phenomenon. Amplitude and phase
of the $a_0^0(980)$\,$\to $\,$(K^+K^-+K^0\bar
K^0)$\,$\to$\,$f_0(980)$ transition. First search proposals}\\
2.1. Peripheral reactions $\pi^\pm
N\to(a^0_0(980),$ $f_0(980))(N,\Delta)\to\eta\pi^0(N,\Delta)$;\\
2.2. Reactions $(K^-,\bar K^0)\,N\to(f_0(980),a^0_0 (980))$
$(\Lambda,\Sigma,\Sigma(1385))\to(\pi^+\pi^-/\eta\pi^0)\,
(\Lambda,\Sigma,\Sigma(1385))$;\\ 2.3. Reactions of $\bar pn$
annihilation at rest $\bar
pn\to(\pi^-,\rho^-)f_0(980)\to(\pi^-,\rho^-)\eta\pi^0 $;\\ 2.4.
Decay $f_1(1285)\to a^0_0(980)\pi^0\to3\pi$.
\item{\bf 3.}
{\bf\boldmath The $a^0_0(980)-f_0(980)$ mixing in polarization
phenomena. Reaction $\pi^-p\uparrow$\,$\to$\,$\eta\pi^0n$}
\item{\bf 4.} {\bf\boldmath Detection of $a^0_0(980)-f_0(980)$ mixing}\\
4.1. VES experiment in Protvino: the reaction
$\pi^-N\to\pi^-f_1(1285)
N\to\pi^-f_0(980)\pi^0N\to\pi^-\pi^+\pi^-\pi^0N$;\\
4.2. BESIII experiment in Beijing: reactions $J/\psi$\,$\to$\,$\phi
f_0(980)$\,$\to$\,$\phi a_0(980)$\,$\to$\,$\phi\eta\pi^0\,\,\,$and
$\chi_{c1}$\,$\to$\,$ a_0(980)\pi^0$\,$\to$\,$f_0(980)\pi^0
$\,$\to$\,$\pi^+\pi^-\pi^0$;\\ 4.3. Data analysis
\item{\bf 5.} {\bf\boldmath Strong violation of isotopic invariance
according to BESIII data for the reactions\\ $J/\psi$\,$\to$\,$\phi
f_1(1285)$\,$\to$\,$\phi f_0(980)\pi^0$\,$\to$\,$\phi\,3\pi$ and \\
$J/\psi$\,$\to$\,$\gamma\eta(1405)$\,$\to$\,$\gamma
f_0(980)\pi^0$\,$\to
$\,$\gamma\,3\pi$;}\\
5.1. Mechanisms of $f_1(1285)$\,$\to$\,$f_0(980)\pi^0$\,
$\to$\,$3\pi$ decay;\\ 5.2. Consistency condition;\\ 5.3. $K\bar
K$-loop mechanism of isotopic invariance violation in the decay
$\eta(1405 )$\,$\to$\,$f_0(980)\pi^0$\,$\to$\,$3\pi$ and the role of
anomalous Landau thresholds
\item{\bf 6.} {\bf\boldmath Manifestation of $a^0_0(980)-f_0(980)$
mixing in decays of charmed mesons}\\ 6.1. Decay $D^+_s\to\eta\pi^0\pi^+$;\\
6.2. Decays $D^0\to K^0_S\pi^+\pi^-$ and $D^0\to K^0_S\eta\pi^0$
\item{\bf 7.} {\bf\boldmath Bottomonium decay $\Upsilon(10860)\to\Upsilon(1S)f_0(980
)\to\Upsilon(1S)\eta\pi^0$}
\item{\bf 8.} {\bf\boldmath Reactions violating isotopic invariance
in the central region}\\ 8.1. Reactions $pp\to
p(f_1(1285)/f_1(1420)) p\to p(\pi^+\pi^-\pi^0)p$;\\ 8.2. Reaction
$pp\to p(a^0_0(980))p\to p(\eta\pi^0)p$
\item{\bf 9. Conclusion}
\item{\bf References}
\end{description}

{\it \hspace*{1.5cm} In memory of Dmitrii Vasil'evich Shirkov} 

\vspace{0.3cm} \noindent{\large \bf 1. Introduction. Nature of
scalar mesons}\vspace{0.2cm}

\noindent In April 2016, one of the authors of this review (NNA)
delivered a plenary report ``37 Years with Light Scalar Mesons. The
Lessons Learned'' at the International conference on the Physics of
Fundamental Interaction devoted to the 60th anniversary the Joint
Institute for Nuclear Research in Dubna \cite{A2016}. The main topic
of the report was a threshold phenomenon discovered in experiments
not long before the conference, which is now known as the mixing of
$a^0_0(980)$ and $f_0(980)$ resonances. It was theoretically
predicted in \cite{ADS79} as early as 1979 (see also \cite{ADS81,
ADS84}). A special report on the problem related to strong violation
of isotopic invariance in the production of light scalar mesons
\cite{AS16} was delivered at the 14th International Workshop on
Tau-Lepton Physics held in Beijing in September 2016. An extended
version of that report is the basis of this review.

The problem of light scalar mesons with masses $\lesssim$\,1 GeV,
$\sigma(600)$, $\kappa(800)$, $a_0(980)$, and $f_0(980)$, has
remained one of the most intriguing topics of hadron spectroscopy
for several decades. The physics of light scalar mesons has been
repeatedly discussed in {\it Physics--Uspekhi} \cite{ADS84,AS91,
Ac98,AS11}. An impressive array of data on light scalars has been
collected to date \cite{PDG16}. The nontrivial nature of these
states is virtually uncontested. In particular, there is much
evidence of their four-quark $q^2\bar q^2$ structure, and it is a
matter for lively discussions. The number of publications devoted to
light scalar mesons is truly immense. Some understanding of how
theoretical and experimental explorations related to light scalar
mesons have been developing can be gained, for example, from studies
and reviews \cite{A2016,ADS79,ADS81,AS16,ADS84,AS91,Ac98,AS11,
PDG16,CT02,AT04,Ma04,Far08,Ja05,KZ07,tH08,Wie13,DP14,Wol16,Pel16,
Ja77,AI89,Fl72,Gay76,Fl76,MOS77,ADS80a,ADS80b,ADS84b,KK17}.

The search for light s and k mesons commenced as early as the 1960s,
and at the same time preliminary information about these mesons
appeared in Particle Data Group (PDG) reviews. The linear
$\sigma$-model \cite{GL60,Ge64,Le67} was used as a theoretical tool
to search for scalar mesons; this model takes spontaneous breaking
of chiral symmetry into account and contains pseudoscalar mesons as
Goldstone bosons. It has remarkably proved to be an effective
low-energy realization of quantum chromodynamics (QCD). Narrow
scalar resonances, the isovector $a_0(980)$ and the isoscalar
$f_0(980)$, were discovered in the late 1960s and early 1970s.

In regard to $\sigma$ and $\kappa$ mesons, protracted unsuccessful
attempts to confirm their existence in a conclusive way resulted in
general disappointment, and information about those particles
disappeared from the PDG reviews. The main argument against their
existence was that the phase shifts in both $\pi\pi$ and $K\pi$
scattering fail to pass through 90$^\circ$ at the assumed resonance
masses. Nevertheless, experimental and theoretical studies of
reactions in which the $\sigma$ and $\kappa$ states could manifest
themselves continued.

The situation changed completely when the $S$-wave amplitude of the
$\pi\pi$ scattering with the isospin $I$\,=\,0 in the linear
$\sigma$-model was shown to contain a negative background phase
\cite{AS94a} that conceals the $\sigma$ meson, as a result of which
the $\pi\pi$ scattering phase fails to pass through 90$^\circ$ at
the assumed resonance mass. This observation clearly showed that the
(background) screening of the lightest broad scalar mesons is a
phenomenon inherent in chiral dynamics. This idea has been embraced
to initiate a new wave of theoretical and experimental studies of
$\sigma$ and $\kappa$ mesons. As a result, beginning in 1996, the
light $\sigma$ resonance, and beginning in 2004, the light $\kappa$
resonance reappeared in particle-physics reviews \cite{PDG96,PDG04}.
In study \cite{AS07}, which focused on the lightest scalar in the
$SU(2)_L\times SU(2)_R$ linear $\sigma$-model, the existence of
chiral screening of the $\sigma$ resonance was demonstrated in a
joint description of low-energy data on the
$\pi\pi$\,$\to$\,$\pi\pi$ and $\gamma\gamma $\,$ \to$\,$\pi^0\pi^0$
reactions. It was discovered that the $\sigma(600)\to\gamma\gamma$
decay is a four-quark transition \cite{AS07,AS11}.

Hadron scalar channels in the range $\lesssim$\,1 GeV became a
stumbling block for QCD, because neither the perturbation theory nor
the sum rules for individual resonances are applicable in these
channels. However, the nature of $\sigma(600)$, $\kappa(800)$,
$a_0(980)$, and $f_0(980)$ light scalar mesons is the central point
for understanding the chiral symmetry mechanism that emerges as a
result of confinement, and hence for understanding confinement
itself. Jaffe noted in 1977 that the quantum MIT bag model, which
incorporates confinement in a phenomenological way, contains a nonet
of light four-quark scalar states \cite{Ja77}. Jaffe suggested that
$a_0(980)$ and $f_0(980)$ could be members of a nonet with the
following symbolic quark structures:
$a^+_0(980)$\,=\,$u\bar{d}s\bar{s}$, $a^0_0(980)$\,=\,$(u\bar{u}
$\,-\,$d\bar{d})s\bar{s}/\sqrt{2}$, $a^-_0(980)$\,=\,$d\bar{u}s
\bar{s}$ and $f_0(980)$\,=\,$ (u\bar{u}$\,+\,$d\bar{d})s\bar{s}
/\sqrt{2}$. From that time on, the $a^0_0(980)$ and $f_0(980)$
resonances have been the light-quark spectroscopy favorites.

Light scalar mesons have been explored in reactions of virtually all
conceivable types using accelerators currently under operation.
First, strong reactions in which pairs of pseudoscalar mesons
$\pi^+\pi^-$, $K\bar K$, $\pi\eta$, $K\pi$, etc. are produced in
$\pi N$, $KN$, and $N\bar N $ collisions have been studied (see,
e.g., review \cite{ADS84}). It was shown in the late 1980s that
studying the radiation decays $\phi(1020)$\,$\to$\,$\gamma a_0(980)
$\,$\to$\,$\gamma\pi^0\eta$ and $\phi(1020)$\,$\to$\,$\gamma
f_0(980) $\,$\to$\,$\gamma\pi\pi$ can shed light on the problem of
$a_0(980)$ and $f_0(980)$ mesons \cite{AI89}. This issue was under
consideration from different perspectives for a decade, until the
first experimental results appeared in 1988
\cite{BGP92,CIK93,LN94,AGS97,AGShev97,AG97, AG98}.

The decays $\phi(1020)$\,$\to$\,$\gamma a_0(980)
$\,$\to$\,$\gamma\pi^0\eta$ and $\phi(1020)$\,$\to$\,$\gamma
f_0(980) $\,$\to$\,$\gamma\pi\pi$ have been studied to date not only
theoretically but also experimentally with $e^+e^-$ colliders using
the Spherical Neutral Detector (SND) \cite{Ach98a,Ach98b,Ach00a,
Ach00b} and the CMD-2 (Cryogenic Magnetic Detector) \cite{Akh99a,
Akh99b} at the Novosibirsk- based Budker Institute of Nuclear
Physics, Siberian Branch of the Russian Academy of Sciences, and the
KLOE detector (K LOng Experiment) at the Frascati $\phi$-factory
(Italy) \cite{Al02a,Al02b,Am06,Am07a, Amb09a,Ambr09b,Bo08}.

The data from these experiments motivated theoretical studies that
yielded the first arguments in favor of the four-quark nature of the
$f_0(980)$ and $a_0(980)$ states \cite{A2016,Ac98,AI89,AGShev97,
AK07b,AK08,Ac02,Ac03a,Ac04,AGS97,AG98,AG97,AG01,AG01a,AK03,AK06,
AK12b}. It was shown in \cite{AGShev97,AK07b,AK08}, for example,
that light scalar mesons in $\phi(1020)$-meson radiative decays,
$\phi(1020)\to\gamma f_0(980)$ and $\phi(1020)\to\gamma a^0_0(980)$,
are generated at small distances, implying that the light scalar
mesons are compact states rather than loosely bound molecules. It
was also shown in \cite{Ac02,Ac03a,Ac04} that radiative decays of
the $\phi(1020)$ meson into light scalar mesons are four-quark
transitions and, in the approach based on a large number of colors
$N_c$, correspond to the four-quark nature of light scalars.

Experimental studies of light scalar mesons in photon--photon
collisions, more specifically in the reactions $\gamma\gamma$\,$
\to$\,$\pi^+\pi^-$, $\gamma\gamma$\,$\to$\,$\pi^0\pi^0$ and
$\gamma\gamma$\,$\to$\,$\pi^0\eta$, which commenced as early as the
1980s, are still in progress. It was predicted in 1982 \cite{ADS82a,
ADS82b} that if the $a_0(980)$ and $f_0(980)$ mesons have a
four-quark structure, the rates of their production in
photon--photon collisions must be suppressed by an order of
magnitude compared to the case of a two-quark structure. An estimate
$\Gamma_{a^0_0\to\gamma\gamma}\approx\Gamma_{f_0\to\gamma\gamma}
\approx0.27$ keV was obtained in a four-quark model \cite{ADS82a,
ADS82b}, which was supported by experiment \cite{An86,Ma90,Bo90,
PDG16}. The situation with $a_0(980)$ that emerged after the Crystal
Ball collaboration \cite{An86} had for the first time measured the
cross section of the reaction $\gamma\gamma\to a^0_0(980)\to\pi^0
\eta$ was analyzed in detail in \cite{AS88}. A dynamic model was
proposed in that study for the amplitude of a two-photon decay of
the $a_0(980)$ resonance (including the $K^+K^-$-loop mechanism of
the transition $a_0(980)\to K^+K^-\to\gamma\gamma$), which agrees
well with the data and demonstrates that the production of
$a_0(980)$ in photon--photon collisions is suppressed.

Studies of light scalar mesons in photon--photon collisions at the
B-factory in Japan \cite{Mo07a,Mo07b,Ue08,Ue09,PMUW08}, which opened
the era of super-precise statistics, not only have confirmed
predictions made 35 years earlier but also enabled reaching the
conclusion \cite{AS05,AS08,AS11} that the production mechanisms are
four-quark transitions, thus signaling the four-quark nature of
light scalar mesons.

It was shown in \cite{AK12,AK14} that data on semileptonic decays of
heavy quarkonia are evidence of the four-quark nature of isoscalar
$\sigma$ [or $f_0(500)$] and $f_0(980)$ mesons and provide a unique
opportunity to explore the nature of light isovector scalar mesons
in the nearest future.

The $f_0(980)$, $a_0(980)$, $\sigma(600)$, and $\kappa(800)$
resonances are now actively being studied in the numerous decays of
heavy quarkonia, $J/\psi$, $D_s$, $D$, $B_s$, and $B$ by the LHCb
(Large Hadron Collider beauty experiment) collaboration, BaBar,
FOCUS, CLEO, Belle, BES III (Beijing Spectrometer III), etc. (see,
e.g., reviews \cite{PDG16,As09,Nogu15,Reis16,Lois16}). We note that
the studies relating to the four-quark structure of light scalar
mesons have paved the way for the `bold' search for similar objects
in the families of mesons that contain heavy c and b quarks. The
majority of theoretical and experimental teams all over the world
that explore elementary particle physics are now studying different
four-quark states. One might expect that the Belle II detector on
the SuperKEK B accelerator (Japan) will provide new data on
production in photon--photon collisions of tensor four-quark states,
including the tensor exotic four-quark state (E) with the isospin
$I=2$ and mass in the vicinity of the rr threshold \cite{AS91,
ADS82a,ADS82b,ADS84c,ADS85}. Along with the predicted suppression of
the two-photon widths of $a_0(980)$ and $f_0(980)$ resonances, a
spectacular manifestation of that exotic state in the reactions
$\gamma\gamma\to\rho^0 \rho^0$ and $\gamma\gamma\to\rho^+\rho^-$ was
predicted in \cite{ADS82a,ADS82b}. It was confirmed later in the
experiments of three groups, JADE (Japan, Deutschland, England)
\cite{Ko84}, ARGUS (A Russian--German--United States--Swedish
collaboration) \cite{Al89}, and CELLO \cite{Be89}, which were
performed at the PETRA accelerator (an abbreviation of
Positron--Elektron Tandem Ring Anlage, the German name of the
facility) at the DESY (Deutsches Elecktronen Synchrotron) particle
physics research center (see, in relation to this, \cite{AS91,
ADS84c,ADS85}). Without a doubt, to conclusively prove the existence
of a state with the quantum numbers $I^G( J^{PC})=2^+(2^{++})$ and
mass $\approx1.5$ GeV, it has to be discovered in charged $\rho^\pm
\rho^\pm$ \cite{AS91,AS92} and $\rho^\pm \rho^0$ \cite{AS91,AS99}
channels. Because the $E^\pm$ coupling to the $\gamma\rho^\pm$
channel is strong, we believe that photoproduction reactions $\gamma
N\to E^\pm N\to\rho^\pm \rho^0N$ and $\gamma N\to E^\pm\Delta\to
\rho^\pm \rho^0\Delta$ (which occur owing to $\rho^\pm$ exchange) at
the Jefferson Laboratory (USA) are the most promising ones for its
search.

Each upgrade of accelerators and detectors triggers a new wave in
studying light scalar mesons. The mechanisms of their production and
decay and the shapes of corresponding mass distributions are studied
at a new level of accuracy. The immense wealth of data with large
statistics obtainable using modern-day accelerators will undoubtedly
enable researchers to substantially improve the accuracy of
scalar-resonance parameters and fruitfully continue the already
commenced studies of fine threshold phenomena related to the
violation of isotopic invariance in the region of $a_0(980)$ and
$f_0(980)$ resonances in the $\pi^+\pi^-$ and $\eta\pi^0$ mass
spectra.


\vspace{0.4cm} \noindent{\large \bf 2. {\boldmath The $a^0_0(980)-
f_0(980)$ mixing as a threshold phenomenon. Amplitude and phase of
the $a_0^0(980)$\,$\to $\,$(K^+K^-+K^0\bar K^0)$\,$\to$\,$ f_0(980)$
transition. First search proposals}}\vspace{0.2cm}

\noindent Experimental studies of effects related to the mixing of
particles with different isotopic spins and close masses were always
the focus of attention. The characteristic `mass' (amplitude)
squared of transitions such as $\rho^0-\omega$ and $\pi^0-\eta$ is
approximately \footnote{This means that the order of magnitude of
the amplitudes of those transitions is determined by the differences
between the masses squared of the particles in isotopic meson
multiplets. For example, the amplitude of the $\pi^0-\eta$
transition is $|\Pi_{\pi^0\eta} |\approx m^2_\pi(m_d-m_u)/[
\sqrt{3}(m_d+m_u)] \approx(m^2_{K^0}-m^2_{K^+}+ m^2_{\pi^+}
-m^2_{\pi^0})/\sqrt{3}$ \cite{Io01}.} 0.003 GeV$^2$ or
$\approx(m_d-m_u) \times 1$ GeV. We here discuss the effects of
isotopic symmetry breaking, the amplitude of which is
$\sim\sqrt{m_d-m_u}$ (rather than $\sim(m_d-m_u)$, as usually
occurs), and begin with the amplitude of transitions between
$a_0^0(980)$ and $f_0(980)$ resonances.

The $a_0^0(980)$ and $f_0(980)$ resonances are located at the $K\bar
K$ channel threshold and therefore experience its strong influence
(see, e.g., \cite{Fl76,MOS77,ADS80a,ADS80b}). This essentially
distinguishes the $a^0_0(980)-f_0(980)$ mixing from the well-studied
$\rho^0-\omega$ mixing. The point is that due to the vicinity of
$a^0_0(980)$ and $f_0(980)$ resonances to the $K\bar K$ threshold,
the $a^0_0(980)-f_0(980)$ transition is significantly determined by
contributions from the $K^+K^-$ and $K^0\bar K^0$ intermediate
states: $a^0_0(980)\to(K^+K^-+K^0\bar K^0)\to f_0(980)$ (Fig. 1). In
the region between the $K^+K^-$ and $K^0 \bar K^0$ thresholds, whose
width is 8 MeV, the amplitude of the $a^0_0(980)-f_0(980)$
transition $\Pi_{a^0_0f_0}(m)$ (here and below, $m$ is the invariant
virtual mass of the scalar resonance) is $\sim \sqrt{m_d-m_u}$
rather than $\sim(m_d-m_u)$, as could be expected from general
considerations. Outside this region, $\Pi_{a^0_0 f_0}(m)$ rapidly
decreases and tends to $\sim(m_d-m_u)$. Indeed, the sum of the
diagrams shown in Fig. 1 converges and, given the isotopic symmetry
of the coupling constants $g_{a^0_0K^+K^-}=-g_{a^0_0K^0\bar K^0}$
and $g_{f_0K^+K^-}=g_{f_0K^0 \bar K^0}$, yields the following
contribution to the amplitude of the $a^0_0(980)-f_0 (980)$
transition:
\begin{eqnarray}\label{Eq1} \hspace*{-20pt} &&
\Pi_{a^0_0f_0}(m)=\frac{g_{a^0_0K^+K^-}g_{f_0K^+K^-
}}{16\pi}\Biggl[\,i\,\Bigl(\rho_{K^+K^-}(m)\nonumber\\
\hspace*{-20pt} && -\rho_{K^0\bar K^0}(m)\Bigr)- \frac{
\rho_{K^+K^-}(m)}{\pi}\ln\frac{1+\rho_{K^+K^-}(m)}{1-\rho_{K^+K^-
}(m)}\nonumber\\ \hspace*{-20pt} && +\frac{\rho_{K^0 \bar K^0}
(m)}{\pi}\ln\frac{1+\rho_{K^0\bar K^0}(m)}{1-\rho_{K^0\bar
K^0}(m)}\,\,\Biggr],\end{eqnarray} where $\rho_{K\bar
K}(m)=\sqrt{1-4m_K^2/m^2}$ for $0\leq m\leq2m_K$ and $\rho_{K\bar
K}(m)$ must be replaced with $i|\rho_{K\bar K}(m)|$ for $0\leq
m\leq2m_K$. The resonance-like behavior of the absolute value and
phase of amplitude (1) is visually displayed in Fig. 2. We represent
$\Pi_{a^0_0f_0}(m)$ near the $K\bar K$ threshold as a series in
$\rho_{K^+K^-}(m)$ and $\rho_{K^0\bar K^0}(m)$:
\begin{figure}
\hspace*{0.32cm}\includegraphics[width=19pc]{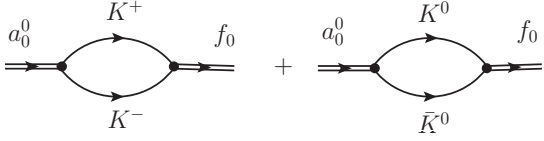}
\caption{\label{Fig1} $K\bar K$-loop mechanism of the $a^0_0(980)-
f_0(980)$ mixing.}\end{figure}
\begin{figure}
\hspace*{-0.15cm}
\includegraphics[width=21.2pc]{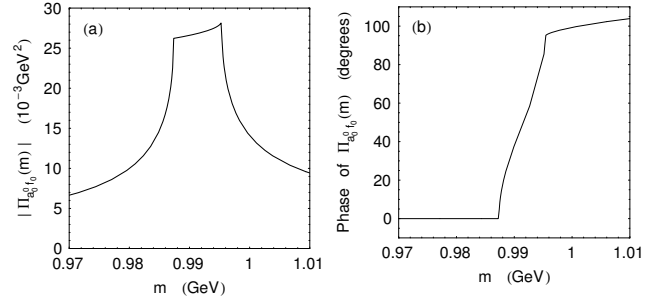}
\caption{\label{Fig2} (a) Absolute value of the
$a^0_0(980)-f_0(980)$ mixing amplitude: an example. (b) The phase of
the $a^0_0(980)-f_0(980)$ mixing changes by $90^\circ$ between the
$K^+K^-$ and $K^0 \bar K^0$ thresholds, where the absolute value of
the amplitude is practically constant and has the maximum value.
}\end{figure}
\begin{eqnarray}\label{Eq2}\hspace*{-15pt}
&& \Pi_{a^0_0f_0}(m)=\frac{g_{a^0_0K^+K^-}g_{f_0K^+K^-}}{16\pi}
\Biggl[\,i\,\Bigl(\rho_{K^+K^-}(m)\nonumber\\ \hspace*{-15pt} &&
-\rho_{K^0\bar K^0}(m)\Bigr)-\frac{2}{\pi}
\Bigl(\rho^2_{K^+K^-}(m)-\rho^2_{ K^0\bar K^0}(m)\Bigr)\nonumber\\
\hspace*{-15pt} && +\,O\Bigl(\rho^4_{K^+K^-}(m)-\rho^4_{K^0\bar
K^0}(m)\Bigr)+...\Biggr]\nonumber\\
\hspace*{-15pt} && =\frac{g_{a^0_0K^+K^-}g_{f_0K^+K^-
}}{16\pi}\Biggl[\,i\,\Bigl(\rho_{K^+K^-}(m)-\rho_{K^0\bar
K^0}(m)\Bigr)\nonumber\\
\hspace*{-15pt} && +\,\frac{8\,(m^2_{K^0}-m^2_{K^+})}{\pi\,m^2}\nonumber\\
\hspace*{-15pt} && +\,O\Bigl(\rho^4_{K^+K^-}(m)-\rho^4_{K^0\bar
K^0}(m)\Bigr)+...\Biggr].
\end{eqnarray}
In the region between the $K^+K^-$ and $K^0 \bar K^0$ thresholds,
the absolute value of the first term in (2) is
\begin{eqnarray}\label{Eq3}
\hspace*{-25pt} && \qquad |\Pi_{a^0_0f_0}(m)|\approx\frac{|g_{
a^0_0K^+K^-}g_{f_0K^+K^-}|}{16\pi}\sqrt{\frac{2(m_{K^0}-
m_{K^+})}{m_{K^0}}}\nonumber\\ \hspace*{-25pt} &&
\qquad\approx0.127\,\frac{|g_{a_0K^+K^-}g_{f_0K^+K^-}|}{
16\pi}\simeq0.03\mbox{ GeV}^2\nonumber\\
\hspace*{-25pt} &&\qquad\approx m_K\sqrt{m_{K^0}^2-m_{K^+}^2}\approx
m_K^20.127\nonumber\\ \hspace*{-25pt} &&\qquad\approx
m_K^{3/2}\sqrt{2(m_d-m_u)}.\end{eqnarray} In deriving this estimate
for $|\Pi_{a^0_0f_0}(m\approx2m_K)|$ and plotting the
$|\Pi_{a^0_0f_0}(m)|$ function in Fig. 2, we used the values of the
$g_{a^0_0K^+K^-}$ and $g_{f_0K^+K^-}$ constants that were determined
in \cite{AKS16} from the analysis of the first data of the BES III
collaboration \cite{Ab1} on the $a^0_0(980)-f_0(980)$ mixing (see
details in Section 4.3).

As was noted above, for the $\rho^0-\omega$ and $\pi^0-\eta$
transitions, $|\Pi_{\rho^0\omega}|\approx|\Pi_{\pi^0\eta}|\approx
0.003\mbox{ GeV}^2\approx (m_d-m_u)\times1$ GeV.

We emphasize that the first term in the right-hand side of Eqn (2)
is theoretically determined in an unambiguous way using the
$a^0_0(980)$ and $f_0(980)$ coupling constants with the $K\bar K$
channel, as follows from the unitarity condition. Numerous data
analyses indicate the four-quark nature of $f_0(980)$
and$a^0_0(980)$ resonances and their strong coupling with the
superallowed decay into the $K\bar K$ channel \cite{A2016,Ac98,
Ja77}. The contribution from other intermediate states is, generally
speaking, of the same order as the second term in (2), i.e., an
order of magnitude smaller. Therefore, the $a^0_0(980)-f_0(980)$
transition is virtually completely driven by the first term in the
right-hand side of Eqn (2). As was noted above, the amplitude
$\Pi_{a^0_0 f_0}(m)$ rapidly decreases outside the region of the
$K\bar K$ thresholds, and we can naturally expect the violation of
isotopic invariance to be relatively small here, of the order of
$(m_{K^0}-m_{K^+})/m_{K^0}\approx1/125$. Such contributions, which
are incalculable in practice, can be ignored in the first
approximation.

Because the amplitude $\Pi_{a^0_0f_0}(m)$ is not small between the
$K^+K^-$ and $K^0 \bar K^0$ thresholds, all orders of the
$a^0_0(980)-f_0(980)$ mixing must be taken into account in the
propagator $G_{a^0_0f_0}(m)$ that describes the $a^0_0(980)\to
f_0(980)$ and reverse transitions \cite{ADS79,ADS81},
\begin{eqnarray}\label{Eq4}
G_{a^0_0f_0}(m)=\frac{\Pi_{a^0_0f_0}(m)}{D_{a^0_0}(m)
D_{f_0}(m)-\Pi^2_{a^0_0f_0}(m)},
\end{eqnarray} where $D_r(s)$ is the inverse propagator of the
unmixed resonance $r$ with mass $m_r$ $(r=a^0_0(980),f_0 (980))$,
\begin{eqnarray}\label{Eq5}
D_r(m)=m^2_r-m^2+\sum_{ab}[\mbox{Re}\Pi^{ab}_r(m_r)-\Pi^{ab}_r(m)],
\end{eqnarray} $ab=(\eta\pi^0,\,K^+K^-,\,K^0\bar K^0,\,\eta'\pi^0)$
for $r=a^0_0(980)$ and
$ab=(\pi^+\pi^-,\,\pi^0\pi^0,\,K^+K^-,\,K^0\bar K^0,\,\eta\eta)$ for
$r=f_0(980)$, $\Pi^{ab}_r(m)$ is the diagonal matrix element of the
polarization operator of a resonance r, corresponding to the
contribution of the $ab$ intermediate state.\,\footnote{The
expressions for $\Pi^{ab}_{r}(m)$ in different m regions are related
to each other by analytic continuation \cite{ADS80a,AKS16}. If
$m>m_a+m_b$, then
\begin{eqnarray}\label{FN1}\Pi^{ab}_{r}(m)=\frac{g^2_{r
ab}}{16\pi} \left[\frac{m_{ab}^{(+)}m_{ab}^{(-)}}{\pi
m^2}\ln\frac{m_b}{m_a}\right.\qquad\qquad\ \ \nonumber\\
\left.+\rho_{ab}(m)\left(i-\frac{1}{\pi}\,\ln\frac{\sqrt{m^2-m_{ab}^{(-)
\,2}}+\sqrt{m^2-m_{ab}^{(+)\,2}}}{\sqrt{m^2-m_{ab}^{(-)\,2}}-\sqrt{m^2
-m_{ab}^{(+)\,2}}}\right)\right],\end{eqnarray} where $g_{rab}$ is
the constant of $r$ coupling to $ab$, $\rho_{ab}(m)$\,=\,$
(m^2-m_{ab}^{(+)\,2})^{1/2}(m^2-m_{ab}^{(-)\,2})^{1/2}/m^2$,
$m_{ab}^{(\pm)}$\,=\,$m_a\pm m_b$, and $m_a\geq m_b$;
\begin{eqnarray}\label{FN1a}
\mbox{Im} \,\Pi^{ab}_r(m)=m\Gamma_{r\to ab}(m)=\frac{g^2_{r
ab}}{16\pi}\rho_{ab}(m). \end{eqnarray} If
$m_{ab}^{(-)}<m<m_{ab}^{(+)}$, then
\begin{eqnarray}\label{FN2}\Pi^{ab}_{r}(m)=\frac{g^2_{r
ab}}{16\pi} \left[\frac{m_{ab}^{(+)}m_{ab}^{(-)}}{\pi
m^2}\ln\frac{m_b}{m_a}\right.\qquad\nonumber\\ \left.
-\rho_{ab}(m)\left(1-\frac{2}{\pi}\arctan\frac{\sqrt{
m_{ab}^{(+)\,2}-m^2}}{\sqrt{m^2-m_{ab}^{(-)\,2}}}\right)\right],
\end{eqnarray}
where  $\rho_{ab}(m)$\,=\,$(m_{ab}^{(+)\,2}-m^2)^{1/2}
(m^2-m_{ab}^{(-)\,2})^{1/2}/m^2$.\\ If $m\leq m_{ab}^{(-)}$, then
\begin{eqnarray}\label{FN3}\Pi^{ab}_{r}(m)=\frac{g^2_{r
ab}}{16\pi} \left[\frac{m_{ab}^{(+)}m_{ab}^{(-)}}{\pi
m^2}\ln\frac{m_b}{m_a}\right.\qquad\quad \nonumber\\ \left.
+\rho_{ab}(m)\frac{1}{\pi}\,\ln\frac{
\sqrt{m_{ab}^{(+)\,2}-m^2}+\sqrt{m_{ab}^{(-)\,2}-m^2}}
{\sqrt{m_{ab}^{(+)\,2}-m^2}-\sqrt{m_{ab}^{(-)\,2}-m^2}}\right],
\end{eqnarray}
where $\rho_{ab}(m)$\,=\,$(m_{ab}^{(+)\,2}-m^2)^{1/2}(
m_{ab}^{(-)\,2}-m^2)^{1/2}/m^2$.}

The propagators of $a^0_0(980)$ and $f_0(980)$ resonances,
$1/D_{a^0_0}(m)$ and $1/D_{f_0}(m)$, built with consideration for
the finite width of the particles, satisfy the
K\"{a}ll\'{e}n--Lehmann representation in a broad range of the
coupling constant $g_{rab}$ and therefore ensure that the decay
probabilities $\sum_{ab}BR(r\to ab)$ sum up to unity ($BR$ is the
branching ratio) \cite{AKi04}.

We now present estimates for the probabilities of isospin-violating
decays $f_0(980)\to\eta\pi^0$ and $a^0_0(980)\to\pi^+\pi^-$ that are
due to the $a^0_0(980)-f_0(980)$ mixing. Using the constants of the
coupling of $a^0_0(980)$ and $f_0(980)$ resonances to $\pi\pi$,
$K\bar K$, $\eta\pi$, etc. channels \cite{AKS16}, we obtain
\begin{eqnarray}\label{Eq7}
BR(f_0(980)\to K\bar K\to a^0_0(980)\to\eta\pi^0)\nonumber\\
\hspace*{-8pt}=\int\left|G_{a^0_0f_0}(m)\right|^2\,\frac{2m^2\Gamma_{a^0_0
\to\eta\pi^0}(m)}{\pi}\,dm\approx 0.3\%\,, \end{eqnarray}
\begin{eqnarray}\label{Eq8}
BR(a^0_0(980)\to K\bar K\to f_0(980)\to\pi^+\pi^-)\nonumber\\
\hspace*{-8pt}=\int\left|G_{a^0_0f_0}(m)\right|^2\,
\frac{2m^2\Gamma_{f_0\to\pi^+\pi^-}(m)}{\pi}\,dm\approx 0.14\%\,.
\end{eqnarray}
Figure 3 shows the mass spectra that correspond to integrands in
Eqns (10) and (11).
\begin{figure} 
\includegraphics[width=6.5cm]{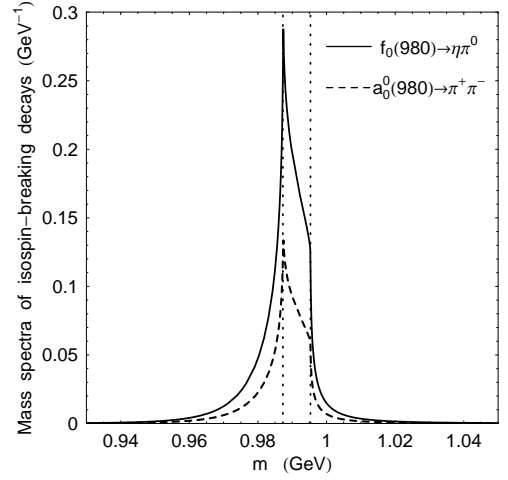}
\caption{\label{Fig3} Mass spectra in isospin-violating decays
$f_0(980)\to\eta\pi^0$ and $a_0(980)\to\pi^+\pi^-$ driven by the
$a^0_0(980)-f_0(980)$ mixing. The solid and dashed curves are
basically similar to each other. The dotted vertical lines show
positions of the $K^+K^-$ and $K^0\bar K^0$ thresholds.}
\end{figure}
A characteristic feature of those decays is that the $\eta\pi^0$ and
$\pi^+\pi^-$ mass spectra are dominated by narrow resonance
structures in the vicinity of the $K\bar K$ thresholds.

It is also of interest to estimate the cross section of the
$G$-parity-forbidden $S$-wave process $\pi^+\pi^-\to\eta\pi^0$
(which is not directly observable) that can occur owing to the
$a^0_0(980)-f_0(980)$ mixing mechanism:
\begin{eqnarray}\label{Eq9}
\sigma(\pi^+\pi^-\to\eta\pi^0)=\frac{16\pi
}{\rho^2_{\pi^+\pi^-}(m)}\qquad\nonumber\\
\times\Gamma_{f_0\to\pi^+\pi^-}(m)\Gamma_{a^0_0
\to\eta\pi^0}(m)\left|G_{a^0_0f_0}(m)\right|^2.
\end{eqnarray}
The resonance-like behavior of $\sigma(\pi^+\pi^-\to\eta\pi^0)$ as a
function of energy is virtually undistinguishable from that of the
$\eta\pi^0$ mass spectrum (see the solid curve in Fig. 3). The
estimated average value of $\sigma(\pi^+\pi^-\to\eta\pi^0)$ in the
range between the $K^+K^-$ and $K^0\bar K^0$ thresholds is
$\approx0.39$ mb. This value is too large for a $G$-parity-
forbidden process. The quoted value is about 8\% of the unitary
limit for the cross section of the allowed $S$-wave process
$\pi^+\pi^-\to\pi^0\pi^0$, which is $32\pi/[3m \rho_{\pi^+\pi^-}
(m)]^2\approx4.8$ mb at $m\approx2m_K$.

If $a^0_0(980)$ and $f_0(980)$ resonances can be produced
concurrently in a reaction (for example, in $\pi^-p$ collisions),
their contributions mutually interfere in $\eta\pi^0$ and
$\pi^+\pi^-$ mass spectra owing to the $a^0_0(980)-f_0(980)$ mixing.
The corresponding interference effects are determined by the
$a^0_0(980)-f_0(980)$ mixing amplitude and relative values of the
initial $a^0_0(980)$ and $f_0(980)$ production amplitudes. The
possibility of detecting the $a^0_0(980)-f_0(980)$ mixing is then
significantly dependent on the choice of the particular reaction.

If selection rules allow production in a reaction of the
$a^0_0(980)$ or $f_0(980)$ resonance alone, that reaction provides,
in principle, an opportunity to directly measure
$BR(a^0_0(980)\to\pi^+ \pi^-)$ or $BR(f_0(980)\to\eta\pi^0)$ that
are due to the $a^0_0(980)-f_0(980)$ mixing.

We discuss various examples of reactions of both types in this
review. \vspace*{0.3cm}

\noindent{\bf 2.1. \boldmath Peripheral reactions $\pi^\pm
N\to(a^0_0(980),$ $f_0(980))(N,\Delta)\to\eta\pi^0(N,\Delta)$}

\noindent We begin with high-energy reactions
\begin{equation}\label{Eq10} \pi^\pm
N\to(a^0_0(980),f_0(980))N\to\eta\pi^0N,
\end{equation}where one-pion exchange in the t channel becomes possible
owing to the $a^0_0(980)-f_0(980)$ mixing, $\pi^\pm N\to
f_0(980)N\to\eta\pi^0N$ \cite{ADS79,ADS81,AS97}. The one-pion
exchange amplitudes are known to be large. Therefore, the effect of
the $G$-parity-forbidden amplitude $\pi^\pm N\to
f_0(980)N\to\eta\pi^0N$ is enhanced on the background of the allowed
process $\pi^\pm N\to a^0_0(980)N\to\eta\pi^0N$ at $|t|\lesssim0.1$
GeV$^2$ (where $t$ is the momentum squared transferred from the
$\pi^\pm$ system to the $\eta\pi^0$ system). The differential cross
section of reaction (13) can be represented as
\begin{eqnarray}\label{Eq11}
\frac{d\sigma}{dtdm}=\left|\pm M^{a^0_0}_{+-}(b_1)+M^{f_0}_{+-}(\pi)
\right|^2+\left|\pm M^{a^0_0}_{++}(\rho_2)\right|^2.\end{eqnarray}
where $M^r_{+-}$ and $M^r_{++}$ are the $s$-channel helicity
amplitudes with and without the nucleon helicity flip, which
describe production of the resonance $r$ and its subsequent decay
into $\eta\pi^0$. The symbols in parentheses indicate the type of
Regge exchange for the given helicity amplitude (for this, we use
symbols of the lightest particles lying on the Regge trajectories
with the corresponding quantum numbers); the $\pm$ sign means that
the amplitudes of $a^0_0(980)$ production have different signs in
the reactions $\pi^-p\to\eta\pi^0n$ and $\pi^+n \to\eta\pi^0p$. We
recall that the $\rho_2$ state corresponds in today's notations to
the quantum numbers $I^G(J^{PC})=1^+(2^{--})$; however, this state
has not been identified yet as a peak in corresponding multiparticle
mass spectra \cite{PDG16}.

Regge amplitudes with $\rho_2$-exchange quantum numbers in the t
channel (earlier referred to as the $Z$ exchange \cite{Ir76}) are
nevertheless needed to describe data on differential cross sections
of the reactions $\pi^-p\to a^0_0(980)n$, $\pi^-p\to\omega n$, $\pi
N\to a_2(1320)N$, etc. These amplitudes can also be due to Regge
cuts. The properties of the Regge amplitude $M^{a^0_0}_{++}
(\rho_2)$ for the reaction $\pi^-p\to a^0_0(980)n$ have been
considered in detail in \cite{AS97}.

The reaction $\pi^-p\to\eta\pi^0n$ has been studied at the
Brookhaven National Laboratory at the laboratory-system momentum of
the incident pion $P^{\pi^-}_{lab}=18.3$ GeV$/c$ \cite{Dz95,Te98},
at the Protvino-based Institute of High Energy Physics at 32 and 38
GeV$/c$ \cite{Sad98,Al99}, and at CERN at 100 GeV$/c$ \cite{Al99}.
The contribution to the cross section of the $\pi^-p\to\eta\pi^0n$
process from the reggeized one-pion exchange (OPE) has the form
\cite{AS97}
\begin{eqnarray}\label{Eq12}
\frac{d\sigma^{\mbox{\scriptsize{(OPE)}}}}{dtdm}=\frac{1}{\pi
s^2}\frac{g^2_{\pi NN}}{4\pi}\left[\frac{-te^{\Lambda_\pi(s)
(t-m^2_\pi)}}{(t-m^2_\pi)^2}\right]\nonumber\\
\times m^3\rho_{\pi^+ \pi^-}(m)\sigma(\pi^+\pi^-\to\eta\pi^0),
\qquad \end{eqnarray} where $s$ is the energy squared in the
$\pi^-p$ center-of-mass system, $g^2_{\pi NN}/(4\pi)\approx 14.6$,
$\Lambda_\pi(s)/2$ is the slope of the Regge $\pi$-pole residue, and
the $\sigma(\pi^+\pi^-\to\eta\pi^0)$ cross section is given by Eqn
(12) [see also the discussion that follows Eqn (12)]. According to a
very conservative estimate quoted in \cite{AS04a,AS04b}, at
$P^{\pi^-}_{lab} =18.3$ GeV$/c$, the value of
$d\sigma^{\mbox{\scriptsize{(OPE)}}} /dt$ at the maximum located at
$t\approx-0.015$ GeV$^2$ is $\approx 140$ nb/GeV$^2$, and the total
cross section of the one-pion exchange is
$\sigma^{\mbox{\scriptsize{(OPE)}}}\approx11$ nb. These values
respectively make about 15\% of $[d\sigma/dt(\pi^-p\to a^0_0
(980)n\to\eta \pi^0n)]|_{t\approx0}\approx940$ nb/GeV$^2$ and about
5.5\% of the total cross section $\sigma(\pi^-p\to
a^0_0(980)n\to\eta \pi^0n)\approx 200$ nb at 18.3 GeV$/c$ (see
details in \cite{AS97,AS04a,AS04b}). The BNL data on the reaction
$\pi^-p\to a^0_0 (980)n\to\eta\pi^0n$ were normalized in \cite{AS97}
(see also \cite{AS04a,AS04b}, where the agreement of that
normalization with the estimated cross section presented in
\cite{Sad98} was noted).

Thus, $G$-parity-violating one-pion exchange can play an important
role in forming the forward peak in the $d\sigma/dt(\pi^-p\to
a^0_0(980)n\to\eta\pi^0n)$ cross section and influence the shape of
the $\eta\pi^0$ mass spectrum in the region of the $a^0_0(980)$
resonance. Unfortunately, due to insufficient resolution in the
invariant mass of the $\eta\pi^0$ system and variable $t$ and
limited statistics (errors in cross sections), it is not possible to
separate the signal from the $a^0_0(980)-f_0(980)$ mixing in the
data of the BNL \cite{Dz95, Te98} and IHEP \cite{Sad98,Al99}
experiments. More precise measurements of the $\pi^-p\to
a^0_0(980)n\to\eta\pi^0n$ reaction remain a promising and
interesting option in what regards detection of the
$a^0_0(980)-f_0(980)$ mixing.

We note that the interference of the contributions from the $\pi$
and $b_1$ exchanges that occurs in the amplitude $M_{+-}$ [see (14)]
was discussed in detail in \cite{ADS81} (see also subsequent studies
\cite{AS97, AS04a,AS04b}). Study \cite{ADS81} also contained
estimates of possible manifestations of $a^0_0(980)-f_0(980)$ mixing
in the reactions $\pi^\pm N\to(a^0_0
(980),f_0(980))\Delta\to\eta\pi^0\Delta$, where the interference of
the $\pi$ and $b_1$ exchanges at small $t$ can be more significant
than in $\pi^\pm N\to(a^0_0(980),f_0(980)) N\to\eta
\pi^0N$.\vspace*{0.3cm}

\noindent{\bf 2.2. \boldmath Reactions $(K^-,\bar K^0)\,N\to(f_0
(980),a^0_0 (980))$ $(\Lambda,\Sigma,\Sigma(1385))\to(\pi^+
\pi^-/\eta\pi^0)\, (\Lambda,\Sigma,\Sigma(1385))$}

\noindent A somewhat different picture of $a^0_0(980)-f_0(980)$
interference can be expected at high energies in the reactions
\begin{eqnarray}\label{Eq13} (K^-,\bar K^0)N\to(f_0
(980),a^0_0 (980))(\Lambda,\Sigma,\Sigma(1385))\nonumber \\
\to\left\{\begin{array}{ll}(\pi^+\pi^-)(\Lambda,\Sigma,\Sigma
(1385)),\quad\\ (\eta\pi^0)(\Lambda,\Sigma,\Sigma(1385)).
\end{array}\right.\quad\end{eqnarray}
The allowed $t$-channel exchanges are in this case the same as for
the production of $a^0_0(980)$ and $f_0(980)$ resonances. If the
Regge $K$-pole exchange dominates (exchanges with natural parity are
forbidden), the interference phenomena in the $\pi^+\pi^-$ and
$\eta\pi^0$ mass spectra are calculated in an unambiguous way
\cite{ADS81}:
\begin{eqnarray}\label{Eq14}
\frac{dN_{\pi^+\pi^-}}{dm}=C\,g^2_{f_0K^+K^-}\frac{2m^2\Gamma_{f_0\to
\pi^+\pi^-}(m)}{\pi|D_{f_0}(m)|^2} \nonumber \\
\times\left|1\pm\frac{g_{a^0_0K^+K^-}}{g_{f_0K^+K^-}}
G_{a^0_0f_0}(m)D_{f_0}(m)\right|^2,
\end{eqnarray}
\begin{eqnarray}\label{Eq15}
\frac{dN_{\eta\pi^0}}{dm}=C\,g^2_{a^0_0K^+K^-}\frac{2m^2\Gamma_{a^0_0\to
\eta\pi^0}(m)}{\pi|D_{a^0_0}(m)|^2} \nonumber \\
\times\left|1\pm\frac{g_{f_0K^+K^-}}{g_{a^0_0K^+K^-}}
G_{a^0_0f_0}(m)D_{a^0_0}(m)\right|^2,
\end{eqnarray} The $+$ and $-$ signs in Eqns (17) and (18) correspond to
reactions (16) with $K^-$ and $\bar K^0$ mesons. It is generally
assumed that the Regge residues are proportional to the coupling
constants of the scalar resonances and $K^+K^-$. As a result of the
interference, the resonance peaks in the reactions caused by $K^-$
and $\bar K^0$ mesons are separated by 10 to 15 MeV and have
different shapes \cite{ADS81}.\vspace*{0.3cm}

\noindent{\bf 2.3. \boldmath Reactions of $\bar pn$ annihilation at
rest $\bar pn\to(\pi^-,\rho^-)f_0(980)\to(\pi^-,\rho^-)\eta\pi^0$}

\noindent In the reactions of $\bar pn$ annihilation at rest,
\begin{eqnarray}\label{Eq16}
\bar pn\to\pi^-f_0(980)\to\pi^-\eta\pi^0,\\ \label{Eq17} \bar
pn\to\rho^-f_0(980)\to\rho^-\eta\pi^0,\,
\end{eqnarray}
which occur owing to the $a^0_0(980)-f_0(980)$ mixing, a narrow peak
is supposed to be observed in the $\eta\pi^0$ mass spectrum at
$m\approx2m_K$, similar to that shown in Fig. 3. The $\bar pn$
annihilation at rest occurs in reaction (19) in the state with the
$\pi^-$-meson quantum numbers, and in reaction (20) with the
$\rho^-$-meson quantum numbers. Therefore, production of the
$\eta\pi^0$ system via the $a^0_0(980)$ resonance is forbidden in
these reactions due to $G$-parity. Thus, reactions (19) and (20)
provide an opportunity to directly measure the value of
$BR(f_0(980)\to\eta\pi^0)$ \cite{ADS81}. \vspace*{0.3cm}

\noindent{\bf 2.4. \boldmath Decay $f_1(1285)\to
a^0_0(980)\pi^0\to3\pi$}

\noindent As was noted in \cite{ADS79}, the $a^0_0(980)-f_0(980)$
mixing enables the $f_1(1285)$ meson with the quantum numbers
$I^G(J^{PC} )=0^+(1^{++})$ to decay into $3\pi$:
\begin{eqnarray}\label{Eq18}
f_1(1285)\to a^0_0(980)\pi^0\to f_0(980)\pi^0\to3\pi.
\end{eqnarray}
The decay $f_1(1285)\to a_0(980)\pi\to\eta\pi\pi$ is one of the main
decay channels for the $f_1(1285)$ meson \cite{PDG16}. According to
the estimate given in \cite{ADS81},
\begin{eqnarray}\label{Eq19}
\frac{\Gamma_{f_1(1285)\to a^0_0\pi^0\to\pi^+\pi^-\pi^0}}{\Gamma_{
f_1(1285)\to a^\pm_0(980)\pi^\mp\to\eta\pi^+\pi^-}}\approx1-3\%
\end{eqnarray}
depending on the parameters of $f_0(980)$ and $a^0_0(980)$
resonances, which were varied in a rather broad range. This result
initiated studies of the isospin-violating decay $f_1(1285)\to
f_0(980)\pi^0\to\pi^+ \pi^-\pi^0$. Resumed later (after more than 25
years), these studies yielded a rather interesting result. The
modern-day situation with the $f_1(1285)\to3\pi$ decay is described
in Sections 4 and 5.

In concluding this section, we note that the interest in the
phenomenon of $a^0_0(980)-f_0(980)$ mixing started gradually
increasing beginning in 1995. For example, the $a^0_0(980)-f_0(980)$
mixing and the ways to observe it in experiment were repeatedly
discussed in the decade that followed \cite{AS04a,AS04b,AT04,AS97,
KerT,CK1,Gr,AK02, BHS,Ku,Bu1,Ko,Ha,Bu2,WYW,Dz95,Ad04}.

A conceptually new proposal to search for $a^0_0(980)-f_0(980)$
mixing was made in 2004. We proposed in \cite{AS04a,AS04b} to
conduct an experiment using the $\pi^-p\to\eta\pi^0n$ reaction on a
polarized target, which allowed uniquely identifying the
$a^0_0(980)-f_0(980)$ by a large and abrupt increase in the
azimuthal asymmetry in the $\eta\pi^0$ production cross section in
an $S$-wave with the $\eta\pi^0$ invariant mass varying in the
region of the $K\bar K$ thresholds. The behavior of this asymmetry
is primarily driven by the 90$^\circ$ change in the amplitude of the
$a^0_0(980)-f_0(980)$ transition between the $K\bar K$ thresholds.
This proposal is discussed in Section 3. \vspace*{0.3cm}

\noindent{\large \boldmath \bf 3. The $a^0_0(980)-f_0(980)$ mixing
in polarization phenomena. Reaction $\pi^-p\uparrow$\,$\to$\,$
\eta\pi^0n$} \vspace{0.2cm}

\noindent The surge in the phase of the $a^0_0(980)-f_0(980)$
transition amplitudes shown in Fig. 2b has prompted an exploration
of $a^0_0(980)-f_0(980)$ mixing in polarization phenomena
\cite{AS04a,AS04b}. If the amplitude of the process with this spin
configuration is dominated by the contribution due to the
$a^0_0(980)-f_0(980)$ mixing, the spin asymmetry of the cross
section must exhibit a surge near the $K\bar K$ thresholds. An
example of this phenomenon is the $\pi^-p
\uparrow\to(a^0_0(980),f_0(980))n\to a^0_0(980)n\to\eta \pi^0n$
reaction on a polarized target. The corresponding differential cross
section is
\begin{equation}\label{Eq20}
\frac{d\sigma}{dtdmd\psi}=\frac{1}{2\pi}\left[\frac{d\sigma}{dtdm}
+I(t,m)\,P\cos\psi\right], \end{equation} where $\psi$ is the angle
between the normal to the reaction plane determined by the momenta
of the incident $\pi^-$ meson and the final $\eta\pi^0$ system and
the direction of the polarization of target protons perpendicular to
the $\pi^-$-beam axis, $P$ is the degree of that polarization,
$I(t,m)=2\,\mbox{Im}(M_{++}M^*_{+-})$ describes the interference
contribution that yields the azimuthal asymmetry of the cross
section, and $d\sigma/dtdm=|M_{+-}|^2+|M_{++}|^2$ is the cross
section on an unpolarized target. The values of $I(t,m)$ and
$d\sigma/dtdm$ are used to determine the dimensionless normalized
spin asymmetry $A(t,m)=I(t,m)/[d^2\sigma/dt dm]$, $\,-1\leq
A(t,m)\leq1$.

\begin{figure} 
\hspace{-2.5mm}
\includegraphics[width=6.5cm]{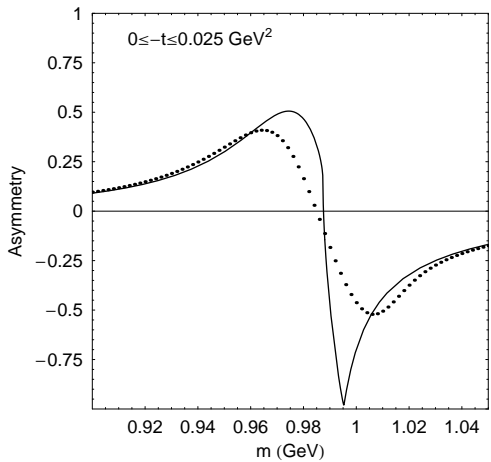}
\caption{\label{Fig4} Manifestation of $a^0_0(980)-f_0(980) $ mixing
in the reaction $\pi^-p\uparrow$\,$\to$\,$\eta\pi^0n$ on a polarized
target at $P^{\pi^-}_{lab}=18.3$ GeV$/c$ in the model of $\rho_2$
and $\pi$ Regge exchanges \cite{AS04a,AS04b}. The solid curve shows
the spin asymmetry $A(0\leq-t\leq0.025 \,\mbox{GeV}^2,m)$ as a
function of the invariant mass of the $\eta\pi^0 $ system m (the
dotted curve is the result of smoothing in $m$ using a Gauss
distribution with a dispersion of 10 MeV).}
\end{figure}


Figure 4 illustrates the surge in the asymmetry as a function of $m$
in the region of the $K\bar K$ thresholds at $0\leq-t\leq0.025
\,\mbox{GeV}^2$. It emerges as a result of interference between the
isospin-conserving amplitude $M_{++}=M^{a^0_0}_{++}(\rho_2 )$ and
the amplitude $M_{+-}=M^{f_0 }_{+-}(\pi)$ due to the $a^0_0(980)
-f_0(980)$ mixing in the model of $\rho_2$ and $\pi$ Regge exchanges
[see (14)]. This selection of the Regge mechanism model is not
fortuitous: a description of the BNL \cite{Dz95} and IHEP
\cite{Al99} data for the differential cross section
$d\sigma/dt(\pi^-p\to a^0_0(980)n\to \eta\pi^0n)$ on an unpolarized
target does not require introducing $M^{a^0_0 }_{+-}(b_1)$, an
amplitude with the quantum numbers of the $b_1$ Regge exchange [see
(14)]. Those data are perfectly well approximated in the $0\leq-t
\leq0.6\,\mbox{GeV}^2$ range using a simple exponential function
$d\sigma/dt=C\exp(\Lambda t)$, i.e., they can be described using the
amplitude $M^{a^0_0}_{++}(\rho_2)$, which does not vanish as $t\to0$
(see, e.g., the solid curve in Fig. 5), while the contribution of
the spin-flip amplitude is $M^{a^0_0 }_{+-}(b_1)\sim\sqrt{-t}$. The
value of the polarization effects in the model of $\rho_2$ and $\pi$
Regge exchanges does not change much within the intervals
$0\leq-t\leq0.05$, 0.1, and 0.2 GeV$^2$. All the details of versions
that occur when the contribution from the b1 exchange is taken into
account (a version of this kind is illustrated in Fig. 5) can be
found in \cite{AS04a,AS04b}. The main conclusion that can be drawn
from the analysis made in the papers cited above is that the spin
asymmetry emerging in any interval $0\leq-t\leq
0.025,...,0.100\,\mbox{GeV}^2$ is in any event predicted to exhibit,
due to an admixture of the $\pi$ exchange, an increase by a value
close to unity in the $m$ range from 0.965 to 1.010 GeV.

\begin{figure} 
\hspace{-2.5mm}
\includegraphics[width=6.5cm]{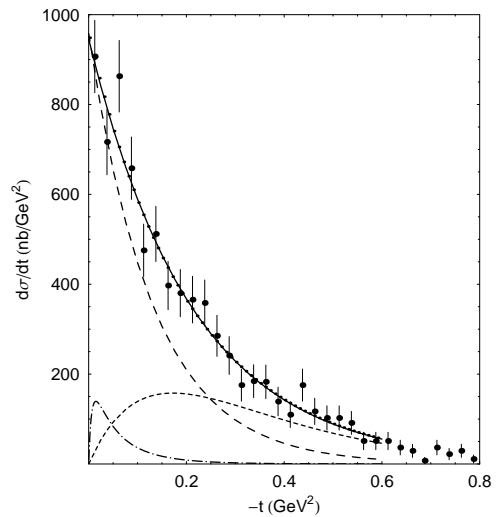}
\caption{\label{Fig5}  Dots with error bars are the normalized BNL
data [100] for the differential cross section $d\sigma/dt$ of the
reaction $\pi^-p\to a^0_0(980)n\to\eta\pi^0n$ at $P^{
\pi^-}_{lab}=18.3$ GeV$/c$ \cite{AS97}. The solid curve is the
result of data approximation in the model of the $\rho_2$ Regge
exchange. The dotted curve that virtually coincides with the solid
one is an example of data approximation in the model of $\rho_2$ and
$b_1 $ Regge exchanges, whose contributions to $d\sigma/dt$ are
respectively presented by the long-dashed and short-dashed curves.
The dot-dashed curve shows the cross section $d\sigma^{
\mbox{\scriptsize{(OPE)}}}/dt$ of the reaction $\pi^-p\to
f_0(980)n\to a^0_0(980)n\to\eta\pi^0n$ at $P^{\pi^-}_{lab}=18.3$
GeV$/c$ (see details in \cite{AS04b}).}\end{figure}


It is noteworthy that detecting the spin asymmetry surge does not
require high resolution in the invariant mass of the $\eta\pi^0$
system (see Fig. 4), which is of importance for identifying
manifestations of the $a^0_0(980)-f_0(980)$ mixing in the $\eta\pi^0
$ mass spectrum in an unpolarized experiment. Moreover, it is
expected that the energy dependence of the polarization effect is
weak. Therefore, it can be explored in a broad $P^{\pi^-}_{lab}$
range, for example, from 8 to 100 GeV$/c$.

The reactions with polarized targets are still waiting to be
explored experimentally.


\vspace*{0.3cm} \noindent{\large \boldmath \bf 4. Detection of
$a^0_0(980)-f_0(980)$ mixing} \vspace*{0.2cm}

\noindent Interest in the isospin-violating $a^0_0(980)-f_0(980)$
mixing has not abated over the last decade
\cite{AS16,AKS16,Ab1,HKP07,Nik07,Do08,Nik09,Do11,WZZ07,WZ08,Ab2,WLZZ12,
ALOWZ12,WWZZ13,Ro13,CK15,SKum1,SKum2,ADO15,Ab3,AKS15,MA15,Wa16,
ADHKM17,AM17,AS17a,AS17b,SOL17,BaD17}. Effects related to this
phenomenon have been discovered and experimentally studied using the
VES (Vertex Spectrometer) detector in Protvino \cite{Nik07,Do08,
Nik09,Do11} and the BES III detector in Beijing \cite{Ab1,Ab2,Ab3}
in the reactions
\begin{eqnarray}
& \pi^-N\to\pi^-f_1(1285)N\to\pi^-f_0(980)\pi^0N\nonumber\\
& \to\pi^-\pi^+\pi^-\pi^0N\ \ \mbox{\cite{Nik07,Do08,Nik09,Do11}},\label{Eq21}\\
& J/\psi\to\phi f_0(980)\to\phi a_0(980)\to\phi\eta\pi^0\ \
\mbox{\cite{Ab1}},\label{Eq22}\\ & \chi_{c1}\to a_0(980)\pi^0\to
f_0(980)\pi^0 \to\pi^+\pi^-\pi^0\ \ \mbox{\cite{Ab1}},\label{Eq23}\\
& J/\psi\to\gamma\eta(1405)\to \gamma f_0(980)\pi^0\to\gamma\,3\pi\
\ \mbox{\cite{Ab2}},\label{Eq24} \\ & J/\psi\to\phi f_1(1285)\to\phi
f_0(980)\pi^0\to\phi\,3\pi\ \ \mbox{\cite{Ab3}}.\label{Eq25}
\end{eqnarray}

The results of these experiments have shown quite clearly
\cite{AS16,AKS15,AKS16} that similar isospin violation effects,
which are due to the mass difference of the $K^+$ and $K^0$ mesons,
can emerge not only as a result of the $a^0_0(980)-f_0(980)$ mixing
alone but also from any mechanism of $K\bar K$-pair production with
a definite isospin in the $S$-wave:
\begin{eqnarray}\label{Eq26}
X_{I=0}\to(K^+K^-+K^0\bar K^0)\to a^0_0(980)\to\eta\pi^0,\ \ \\
\label{Eq27} X_{I=1}\to(K^+K^-+K^0\bar K^0)\to f_0(980)\to\pi^+
\pi^-.\end{eqnarray} Each mechanism of this kind generates both a
narrow resonance peak in the $\eta\pi^0$ or $\pi^+\pi^-$ mass
spectrum (see Fig. 3 as an example) and a surge in the phase
amplitude between the $K^+K^-$ and $K^0\bar K^0$ thresholds (Fig.
2b). This implies that a new tool for studying the production
mechanism and nature of light scalars is now available.

We discuss the effects discovered in reactions (24)--(28) in
Sections 4.1--4.3 and 5.

\vspace*{0.3cm} \noindent{\boldmath \bf 4.1. VES experiment in
Protvino: the reaction $\pi^-N\to\pi^-f_1(1285)N\to\pi^-f_0(980)
\pi^0N\to\pi^- \pi^+\pi^-\pi^0N$}

\noindent The first preliminary observations of the
isotopic-symmetry-violating decay $f_1(1285)\to\pi^+\pi^-\pi^0$
obtained using the VES detector at the Protvino accelerator
\cite{Nik07} were reported in 2007 at the 12th International
Conference on Hadron Spectroscopy (Hadron 2007) (see also
\cite{Do08,Nik09}). The reaction of the diffraction production of
the $\pi^-f_1(1285)$ system, $\pi^-N\to\pi^-f_1(1285)N$, was used as
a source of $f_1(1285)$ mesons. This reaction was studied in
$\pi^-$Be collisions at $P^{\pi^-}_{lab}\,$= 27, 37, and 41 GeV/$c$.
The signal from the $f_1(1285)$ resonance was sought in the
$\pi^+\pi^-\pi^0$ decay channel and in the `reference'
$\eta\pi^+\pi^-$ channel. It was found that a distinguishing feature
of the $f_1(1285)\to\pi^+ \pi^-\pi^0$ decay is the presence in the
$\pi^+\pi^-$ mass spectrum of a resonance-type structure in the
vicinity of $f_0(980)$, which is characteristic of the
isospin-violating $a^0_0(980)-f_0(980)$ mixing, $f_1(1285)\to
a^0_0(980)\pi^0\to f_0(980)\pi^0\to \pi^+\pi^-\pi^0$ (see Section
2.4).

The final results of the VES experiment reported in \cite{Do11} can
be formulated as follows. The $f_1(1285)\to\pi^+\pi^-\pi^0$ decay,
which violates isotopic symmetry, has been discovered. Its branching
ratio
\begin{eqnarray}\label{Eq28}
\frac{BR(f_1(1285)\to f_0(980)\pi^0\to\pi^+\pi^-\pi^0)}{
BR(f_1(1285)\to\eta\pi^+\pi^-)}\nonumber
\\ =(0.86\pm0.16\pm0.20)\%\qquad\qquad\
\end{eqnarray}
and the probability of the $f_1(1285)\to\pi^+\pi^-\pi^0$ decay have
been measured:
\begin{eqnarray}\label{Eq29}
BR(f_1(1285)\to f_0(980)\pi^0\to\pi^+\pi^-\pi^0)\nonumber \\
=(0.30\pm0.055\pm0.074)\%.\qquad\ \ \
\end{eqnarray}

The PDG data \cite{PDG16} on $BR(f_1(1285)\to a^0_0(980)\pi^0\to\eta
\pi^0\pi^0)$ yield
\begin{eqnarray}\label{Eq29a}
\frac{BR(f_1(1285)\to f_0(980)\pi^0\to\pi^+\pi^-\pi^0)}{
BR(f_1(1285)\to a^0_0(980)\pi^0\to\eta\pi^0\pi^0)}\nonumber
\\ =(2.5\pm0.9)\%.\qquad\qquad\qquad\ \
\end{eqnarray}

\vspace*{0.3cm} \noindent{\boldmath \bf 4.2. BESIII experiment in
Beijing: reactions $J/\psi$\,$\to$\,$\phi f_0(980)$\,$\to$\,$\phi
a_0(980)$\,$\to$\,$\phi\eta\pi^0\,$ and $\chi_{c1}$\,$\to$\,$
a_0(980)\pi^0$\,$\to$\,$f_0(980)\pi^0 $\,$\to$\,$\pi^+\pi^-\pi^0$}

\noindent A proposal was made in \cite{WZZ07} to search for
$a^0_0(980)-f_0(980)$ mixing in the reaction $J/\psi\to\phi
f_0(980)\to\phi a_0(980)\to\phi\eta\pi^0$, which as estimates show
can be measured at the Beijing-based upgraded electron+positron
collider using the BES III detector. It was later proposed to use
the same facility to search for the $a^0_0(980)-f_0( 980)$ mixing in
the reaction $\chi_{c1}\to a_0(980)\pi^0\to f_0(980)\pi^0\to\pi^+
\pi^-\pi^0$ \cite{WZ08}. Corresponding measurements were performed
shortly after that, among other initial experiments with the BESIII
detector \cite{Ab1}. The rates $\xi_{fa}$ and $\xi_{af}$ of the
$f_0(980)\to a^0_0(980)$ and $a^0_0(980)\to f_0(980)$ transitions
were determined as a result of those experiments:
\begin{eqnarray}\label{Eq30}
\xi_{fa}=\frac{BR(J/\psi\to\phi f_0(980)\to\phi a^0_0(980)\to
\phi\eta \pi^0)}{BR(J/\psi\to\phi f_0(980)\to\phi\pi\pi)}
\nonumber\\ =(0.60\pm0.20(stat.)\pm0.12(sys.)\pm0.26(para.))\%,
\quad \end{eqnarray}
\begin{eqnarray}\label{Eq31} \hspace*{-10pt} &&
\xi_{af}=\frac{BR(\chi_{c1}\to a^0_0(980)\pi^0\to f_0(980)\pi^0
\to\pi^+\pi^-\pi^0)}{BR(\chi_{c1}\to a^0_0(980)\pi^0\to\eta \pi^0
\pi^0)}\nonumber\\ \hspace*{-10pt} &&
=(0.31\pm0.16(stat.)\pm0.14(sys.)\pm0.03(para.))\%,\end{eqnarray}
where the abbreviations stat., syst., and param. mean statistical,
systematic, and parameterization-related errors. The values of the
denominators in (34) and (35) where respectively taken from
\cite{Ab1} and \cite{Ab4,PDG10}.

Enhancing the accuracy of the data in (34) and (35) obtained by the
BESIII collaboration in the first experiments for reactions (25) and
(26) is an interesting and challenging task.

\vspace*{0.3cm} \noindent{\boldmath \bf 4.3. Data analysis}

\noindent Some specific values of the couplings to the $\pi\pi$,
$K\bar K$, and $\eta\pi$ decay channels are typically used in
standard calculations related to $f_0(980)$ and $a_0(980) $
resonances. We first quote the estimates for these constants
obtained in \cite{AKS16} directly from the BES III data \cite{Ab1}
on $a^0_0(980)-f_0(980)$ mixing and then proceed to discussing the
unexpected consequences of the VES data \cite{Do11}.

Because the $a^0_0(980)-f_0(980)$ mixing is primarily determined by
the contribution of the $K\bar K$ loops, we set
\begin{eqnarray}\label{Eq32}
\xi_{fa}=\frac{BR(f_0(980)\to K\bar K\to a^0_0(980)
\to\eta\pi^0)}{BR(f_0(980)\to\pi\pi)},\\
\label{Eq33} \xi_{af}=\frac{BR(a^0_0(980)\to K\bar K\to
f_0(980)\to\pi^+\pi^-)}{BR(a^0_0(980)\to\eta\pi^0)},\end{eqnarray}
where
\begin{eqnarray}\label{Eq34}
&& BR(f_0(980)\to K\bar K\to a^0_0(980)\to\eta\pi^0)\qquad\ \nonumber\\
&& \ \ =\int\left|G_{a^0_0f_0}(m)\right|^2\,\frac{2m^2\Gamma_{a^0_0
\to\eta\pi^0}(m)}{\pi}\,dm,\end{eqnarray}
\begin{eqnarray}\label{Eq35} &&
BR(a^0_0(980)\to K\bar K\to f_0(980)\to\pi^+\pi^-)\qquad\ \nonumber\\
&& \ \ =\int\left|G_{a^0_0f_0}(m)\right|^2\,\frac{2m^2\Gamma_{f_0
\to\pi^+\pi^-}(m)}{\pi}\,dm,\end{eqnarray}
\begin{eqnarray}\label{Eq36}
BR(a^0_0(980)\to\eta\pi^0)=\int\frac{2m^2\Gamma_{a^0_0\to\eta\pi^0}
(m)}{\pi|D_{a^0_0}(m)|^2}\,dm, \end{eqnarray}
\begin{eqnarray}\label{Eq37}
BR(f_0(980)\to\pi\pi)=\int\frac{2m^2\Gamma_{f_0\to\pi\pi}
(m)}{\pi|D_{f_0}(m)|^2}\,dm, \end{eqnarray} [see also Eqns
(4)--(9)].\,\footnote{In passing from (34) and (35) to (36) and
(37), we disregard the dependence of the kinematic factors contained
in the $J/\psi\to\phi f_0(980)$ and $\chi_{c1}\to a^0_0(980)\pi^0$
vertices on the invariant virtual masses of scalar mesons, because
it is not significant in the decays of heavy $J/\psi(3097)$ and
$\chi_{c1}(3511)$ resonances. The values of the virtual masses were
taken to be those of the scalar mesons. The corresponding kinematic
factors in the equations above simply drop out under this
approximation. We also disregard the interference of the
contributions to the $\chi_{c1}\to a^0_0(980)\pi^0\to\eta\pi^0
\pi^0$ decay related to the permutation of identical $\pi^0$ mesons.
This interference is small because the relative momenta of the
$a^0_0(980)$ and $\pi^0$ mesons in the decay $\chi_{c1}(3510)\to
a^0_0(980)\pi^0$ are large.}

Substituting the central values of $\xi_{fa}$ and $\xi_{af}$ from
(34) and (35) in the left-hand sides of Eqns (36) and (37), we
obtain equations for the coupling constants of $a^0_0(980)$ and
$f_0(980)$ resonances, which must be solved numerically. Proceeding
in this way, we found the following approximate values in
\cite{AKS16}:
\begin{eqnarray}\label{Eq38}
\frac{g^2_{f_0\pi\pi}}{16\pi}\equiv\frac{3}{2}\frac{g^2_{f_0\pi^+\pi^-}}
{16\pi}=0.098\mbox{\ GeV}^2,\\ \label{Eq39} \frac{g^2_{f_0 K\bar
K}}{16\pi}\equiv2\frac{g^2_{f_0 K^+K^-}}{16\pi}=0.4\mbox{\ GeV}^2,
\\ \label{Eq40}
\frac{g^2_{a^0_0\eta\pi^0}}{16\pi}=0.2\mbox{\ GeV}^2, \quad\quad\  \\
\label{Eq41} \frac{g^2_{a^0_0 K\bar
K}}{16\pi}\equiv2\frac{g^2_{a^0_0 K^+K^-}}{16\pi}=0.5\mbox{\ GeV}^2.
\end{eqnarray}
The masses of $a^0_0(980)$ and $f_0(980)$ resonances,
$m_{a^0_0}=0.985$ GeV and $m_{f_0}=0.985$ GeV, were kept fixed in
this procedure, and for the constants of their coupling to the
$\eta\,'\pi^0$ and $\eta\eta$ channels we used the relations that
follow from the $q^2\bar q^2$ model: $g^2_{a^0_0\eta'\pi^0}=
g^2_{a^0_0\eta\pi^0}$ and $g^2_{f_0\eta\eta}=g^2_{f_0 K^+K^-}$ (see,
e.g., \cite{ADS81, AI89}). Integration in (38) and (39) included the
region from 0.9 GeV to 1.05 GeV (see Fig. 3) and, in (40) and (41),
from the respective $\eta\pi^0$ and $\pi\pi$ threshold to 1.3 GeV.

These are the values of the constants that we used in Section 2 to
obtain the absolute value of the $a_0^0(980)-f_0(980)$ transition in
Fig. 2, the mass spectra for isospin-violating decays of $f_0(980)$
and $a^0_0(980)$ resonances in Fig. 3, and estimates (3), (10), and
(11). The mass spectra of the isospin-preserving decays of
$f_0(980)$ and $a^0_0(980) $ resonances as a function of $m$
calculated for the obtained values of the constants are displayed in
Fig. 6. The curves plotted there represent the integrands in (40)
and (41) and similar expressions for the mass spectra in the decays
into $K\bar K$. There is nothing special in these spectra.

\begin{figure} 
\includegraphics[width=6.5cm]{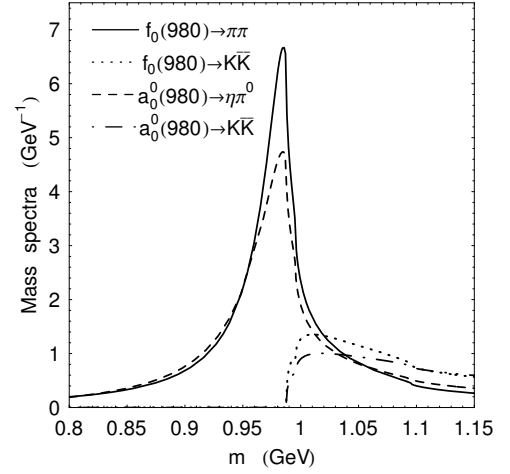}
\caption{\label{Fig6} Mass spectra in isospin-conserving decays of
$f_0(980)$ and $a^0_0(980)$ resonances.}
\end{figure}

There are many studies in which the constants of the $f_0(980)$ and
$a_0(980)$ coupling to the $\pi\pi$, $K\bar K$, and $\eta\pi$
channels have been calculated, estimated, and determined by fitting
(see, e.g., \cite{ADS79,ADS81,WZZ07,WZ08,ADS80a,AKi04,Fl76,Ja77,
MOS77,AI89,AG97,AG01,AK12b,AS04a,AS04b}, this list is far from
complete). The obtained values of the constants are dispersed in a
rather broad range. The values of the constants determined in
different ways and from various reactions often disagree by a factor
of two and sometimes even more. The values in (42)--(45) fall
somewhere in the middle of the range of those available in the
literature; in our opinion, it is natural to use them as a guide in
making estimates.

As regards the decay $f_1(1285)\to f_0(980)\pi^0\to\pi^+\pi^-
\pi^0$, the VES data \cite{Do11} [see (33)] require a `huge'
$a^0_0(980)-f_0(980)$ mixing:
\begin{eqnarray}\label{Eq42} \hspace*{-10pt} &&
\frac{BR(f_1(1285)\to a^0_0(980)\pi^0\to f_0(980)\pi^0\to
\pi^+\pi^-\pi^0) }{BR(f_1(1285)\to a^0_0(980)\pi^0\to
\eta\pi^0\pi^0)}\nonumber \\ \hspace*{-10pt} && \qquad\qquad
\qquad\quad =(2.5\pm0.9)\%\nonumber \\ \hspace*{-10pt} &&
\approx\frac{BR(a^0_0(980)\to K\bar K\to f_0(980)\to\pi^+\pi^-)}
{BR(a^0_0(980)\to\eta\pi^0)}
\end{eqnarray}
[compare (46) with (35) and (37)] and, as a consequence, the
constants of the $f_0(980)$ and $a^0_0(980)$ resonance coupling to
pseudoscalar mesons that are unsatisfactory in many aspects:
\begin{eqnarray}\label{Eq43}
\frac{g^2_{f_0\pi\pi}}{16\pi}\equiv\frac{3}{2}\frac{g^2_{f_0\pi^+\pi^-}}
{16\pi}=0.46\mbox{\ GeV}^2,\\ \label{Eq44} \frac{g^2_{f_0 K\bar
K}}{16\pi}\equiv2\frac{g^2_{f_0 K^+K^-}}{16\pi}=2.87\mbox{\ GeV}^2,
\\ \label{Eq45}
\frac{g^2_{a^0_0\eta\pi^0}}{16\pi}=0.48\mbox{\ GeV}^2, \quad\quad\  \\
\label{Eq46} \frac{g^2_{a^0_0 K\bar
K}}{16\pi}\equiv2\frac{g^2_{a^0_0 K^+K^-}}{16\pi}=4.97\mbox{\
GeV}^2.\end{eqnarray} For example, due to the very large values of
$g^2_{f_0 K\bar K}/(16\pi)$ and $g^2_{a^0_0 K\bar K}/(16\pi)$ in
(48) and (50), the width of the $a^0_0(980)$ resonance in the
$\eta\pi^0$ mass spectrum only proves to be about 15 MeV, and the
value of $\xi_{af}$ calculated using Eqn (37) is about nine times
larger than its central experimental value in (35).

The ratio
\begin{eqnarray}\label{EqEmpty1} \hspace*{-10pt} &&
\frac{BR(f_1(1285)\to a^0_0(980)\pi^0\to f_0(980)\pi^0\to
\pi^+\pi^-\pi^0) }{BR(f_1(1285)\to a^0_0(980)\pi^0\to
\eta\pi^0\pi^0)}\nonumber\end{eqnarray} calculated using the set of
constants in (42)--(45) is \cite{AKS16}\,\footnote{Because the
$f_1(1285)$ resonance, whose mass is $m_{f_1}=(1281.9\pm0.5)$ MeV
\cite{PDG16}, is located only 160 MeV higher than the nominal
$a_0(980) \pi$-production threshold, the calculations took both the
$P$-wave character of the $f_1(1285)\to a_0(980)\pi$ decays and the
interference of two amplitudes in the $f_1(1285)\to a_0(980)\pi\to
\eta\pi\pi$ channel into account. More details can be found in
Section 5.}
\begin{eqnarray}\label{Eq47} \hspace*{-10pt} &&
\frac{BR(f_1(1285)\to a^0_0(980)\pi^0\to f_0(980)\pi^0\to
\pi^+\pi^-\pi^0)}{BR(f_1(1285)\to a^0_0(980)\pi^0\to
\eta\pi^0\pi^0)}\nonumber \\ \hspace*{-10pt} && \qquad\qquad
\qquad\qquad\ \ \approx0.29\%. \end{eqnarray}

This ratio is close to the central value of $\xi_{af}$ in (35);
however, it is approximately one order of magnitude smaller than the
value required to explain the VES data \cite{Do11} [see (33)]. Thus,
we can conclude that the VES collaboration \cite{Do11} has
discovered a significant violation of isotopic symmetry in the
region of $a^0_0(980)$ and $f_0(980)$ resonances, which can hardly
be explained by the $a^0_0(980)-f_0(980)$ mixing alone.

In Section 5, following experimental indications, we consider
additional $K\bar K$-loop mechanisms of the $f_1(1285)\to
f_0(980)\pi^0\to\pi^+\pi^-\pi^0$ decay that are due, similarly to
the $a^0_0(980)-f_0(980) $ mixing, to the mass difference of $K^+$-
and $K^0$ mesons. We also use the decay $f_1(1285)\to
f_0(980)\pi^0\to\pi^+\pi^-\pi^0$ to demonstrate the general approach
to estimating the total contribution of the $K\bar K$-loop
mechanisms that violate isotopic invariance.


\vspace*{0.3cm} \noindent{\large \boldmath \bf 5. Strong violation
of isotopic invariance according to BESIII data for the reactions
$J/\psi\to\phi f_1(1285)\to\phi f_0(980)\pi^0\to\phi\,3\pi$ and
$J/\psi\to\gamma\eta(1405)\to\gamma f_0(980)\pi^0\to
\gamma\,3\pi$} \vspace*{0.2cm} 

\noindent{\boldmath \bf 5.1. Mechanisms of
$f_1(1285)$\,$\to$\,$f_0(980)\pi^0$\, $\to$\,$3\pi$ decay}

\noindent After the VES experiment \cite{Do11} had been completed,
the isospin-violating decay $f_1(1285)\to\pi^+\pi^-\pi^0$ was also
observed by the BESIII collaboration \cite{Ab3} in the reaction
$J/\psi\to\phi f_1(1285)\to\phi f_0(980)\pi^0\to\phi\,3\pi$. The
experiment yielded the branching function ratio
\begin{eqnarray}\label{Eq48}
\frac{BR(f_1(1285)\to f_0(980)\pi^0\to\pi^+\pi^-\pi^0)}{
BR(f_1(1285)\to a^0_0(980)\pi^0\to\eta\pi^0\pi^0)}\nonumber
\\ =(3.6\pm1.4)\%\qquad\quad\qquad
\end{eqnarray}
It was also indicated that a specific feature of the
$f_1(1285)\to\pi^+\pi^-\pi^0$ transition is that the $\pi^+\pi^-$
mass spectrum is dominated by a narrow resonance structure located
in the region of $K\bar K$ thresholds.

Another indication of the decay $f_1(1285)/\eta(1295)\to\pi^+\pi^-
\pi^0$ came from the BESIII experiment \cite{Ab2} combined with data
on the reaction $J/\psi\to\gamma\eta(1405)\to\gamma f_0(980)\pi^0\to
\gamma\,3\pi$. If it is considered to be due to $f_1(1285)$ alone,
we obtain
\begin{eqnarray}\label{Eq49} \frac{BR(f_1(1285)\to
f_0(980)\pi^0\to\pi^+\pi^-\pi^0)}{ BR(f_1(1285)\to
a^0_0(980)\pi^0\to\eta\pi^0\pi^0)}\nonumber
\\ =(1.3\pm0.7)\%.\qquad\quad\qquad
\end{eqnarray}
Thus, according to data from the first experiments (33), (52), and
(53), the ratio of the isospin-forbidden decay $f_1(1285)\to
f_0(980)\pi^0\to\pi^+\pi^-\pi^0$ and the isospin-allowed decay
$f_1(1285)\to a^0_0(980)\pi^0\to\eta\pi^0 \pi^0$ can be as large as
one to four percent. This value is too high for a quantity that is
apparently expected to be of the order of $10^{-4}$. The data rather
convincingly suggest the existence of mechanisms that enhance the
$f_1(1285)\to\pi^+\pi^- \pi^0$ decay. Nevertheless, they still need
to be confirmed. Below, in comparing theoretical estimates with
experimental data, for certainty we use the VES data \cite{Do11}
[see (33)] as mean values.

We now proceed to estimating the contribution of the following
possible $K\bar K$-loop mechanisms to the decay $f_1(1285)\to
f_0(980)\pi^0\to \pi^+\pi^-\pi^0$ in detail:

(1) the $a^0_0(980)-f_0(980)$ mixing, $f_1(1285)\to
a_0(980)\pi^0\to(K^+K^-+K^0\bar K^0)\pi^0\to f_0(980)\pi^0\to\pi^+
\pi^-\pi^0$;

(2) the transition $f_1(1285)\to(K^+K^-+K^0\bar K^0)\pi^0\to
f_0(980)\pi^0\to\pi^+ \pi^-\pi^0$, which is due to the pointlike
$f_1(1285)\to K\bar K\pi^0$ decay;

(3) the transition $f_1(1285)\to(K^*\bar K+\bar K^*K)\to(K^+K^-+
K^0\bar K^0)\pi^0\to f_0(980)\pi^0\to \pi^+\pi^-\pi^0$, where $K^*
=K^*(892)$;

(4) the transition $f_1(1285)\to(K^*_0\bar K+\bar
K^*_0K)\to(K^+K^-+K^0 \bar K^0)\pi^0\to f_0(980)\to\pi^+
\pi^-\pi^0$, where $K^*_0=K^*_0 (800)$ (or $\kappa$) and
$K^*_0(1430)$.

\begin{figure} 
\includegraphics[width=9cm]{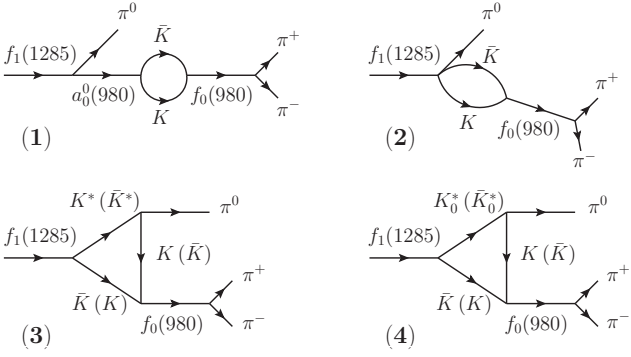}
\caption{\label{Fig7} Diagrams corresponding to the possible $K\bar
K$-loop mechanisms of the decay $f_1(1285)\to f_0(980)\pi^0\to
\pi^+\pi^-\pi^0$.}\end{figure}

Figure 7 illustrates the specified mechanisms in terms of respective
diagrams (1)--(4). The $f_1(1285$ meson first decays within each of
these mechanisms into $K\bar K\pi^0$ (according to the data in
\cite{PDG16}, the decay $f_1(1285)\to K\bar K\pi$ into all charge
modes occurs in about 9\% of cases). Next, owing to the final-state
interaction between $K$ and $\bar K$ mesons, i.e., the transitions
$K^+K^-\to f_0(980)\to\pi^+\pi^-$ and $K^0\bar K^0\to
f_0(980)\to\pi^+\pi^-$, the isospin-violating decay
$f_1(1285)\to(K^+K^-+K^0\bar
K^0)\pi^0$\,$\to$\,$f_0(980)\pi^0$\,$\to$\,$\pi^+\pi^-\pi^0$ is
initiated (see Fig. 7). This occurs because the contributions from
the production of $K^+K^-$ and $K^0\bar K^0$ pairs fail to fully
cancel each other due to the mass difference between $K^+$ and $K^0$
mesons. This mutual compensation is naturally minimal in the m
region between the $K^+K^-$ and $K^0\bar K^0$ thresholds. Regardless
of the specific mechanism of $K\bar K$-pair production, this results
in the emergence in the $\pi^+\pi^-$ mass spectrum of a narrow
resonance structure located in the vicinity of the $K\bar K$
thresholds, with the width $\approx 2m_{K^0}-2m_{K^+}\approx8$ MeV.
Observing this structure in an experiment would be a direct
indication of the general $K\bar K$-loop mechanism in isotopic
invariance violation.

(1) The main estimate that characterizes the contribution of
$a^0_0(980)-f_0(980)$ mixing to the decay $f_1(1285)\to
f_0(980)\pi^0\to\pi^+\pi^- \pi^0$ is already presented in Eqn (51)
(see also the discussion that follows the formula), and we only
elucidate some details of that calculation \cite{AKS16}.

First, we present the expression for the width of the $f_1(1285)\to
a^0_0(980)\pi^0\to\eta\pi^0\pi^0$ decay:
\begin{eqnarray}\label{EqIV-1}
\Gamma_{f_1\to
a^0_0\pi^0\to\eta\pi^0\pi^0}\qquad\qquad\qquad\nonumber \\
=\frac{g^2_{f_1 a^0_0\pi^0}g^2_{ a^0_0\eta
\pi^0}}{192\,\pi^3\,m^3_{f_1}}
\int\limits_{(m_\eta+m_\pi)^2}^{(m_{f_1}-m_\pi)^2}ds
\int\limits_{a_-(s)}^{a_+(s)}dt\ \mathcal{T}(s,t),
\end{eqnarray}
where $s$ ($t$) is the invariant mass squared of the $\eta\pi^0_1$
($\eta\pi^0_2$) pair in the decay $f_1(1285)\to\eta\pi^0_1 \pi^0_2$,
\begin{eqnarray}\label{EqIV-2}
\mathcal{T}(s,t)=\frac{p^2(s)}{|D_{a^0_0}(\sqrt{s})|^2}+
\mbox{Re}\,\frac{p(s)\,p(t)\,\cos\theta}{
D_{a^0_0}(\sqrt{s})D^*_{a^0_0}(\sqrt{t})}\,,\\
\label{EqIV-3}
a_\pm(s)=\frac{1}{2}(m^2_{f_1}+m^2_\eta+2m^2_\pi-s)\quad\ \nonumber\\
+\frac{(m^2_{f_1}-m^2_\pi)(m^2_\eta-m^2_\pi)}{2s}
\pm\frac{2m_{f_1}}{\sqrt{s}}\,p(s)q(s)\,,
\end{eqnarray}
\begin{eqnarray}
\label{EqIV-4}
p(s)=\sqrt{m^4_{f_1}-2m^2_{f_1}(s+m^2_\pi)+(s-m^2_\pi)^2}\Bigl/(2m_{f_1}),\ \\
\label{EqIV-5}
p(t)=\sqrt{m^4_{f_1}-2m^2_{f_1}(t+m^2_\pi)+(t-m^2_\pi)^2}\Bigl/(2m_{f_1}),\ \\
\label{EqIV-6}
q(s)=\sqrt{s^2-2s(m^2_\eta+m^2_\pi)+(m^2_\eta-m^2_\pi)^2}\Bigl/(2\sqrt{s}),\
\ \\ \label{EqIV-7}
p(s)\,p(t)\,\cos\theta=\frac{1}{2}(s+t-m^2_{f_1}-m^2_\eta)\quad\quad\nonumber \\
+\frac{(m^2_{f_1}+m^2_\pi-s)(m^2_{f_1}+m^2_\pi-t)}{4m^2_{f_1}}
\,.\qquad\qquad
\end{eqnarray}
As an effective vertex of the $f_1(1285)a^0_0(980)\pi^0$ coupling,
we here used the expression $V_{f_1a^0_0\pi^0 }=g_{f_1a^0_0\pi^0
}(\epsilon_{f_1 },p_{\pi^0}-p_{a^0_0})$, where $\epsilon_{f_1}$ is
the $f_1(1285)$ polarization 4-vector and $p_{\pi^0}$ and
$p_{a^0_0}$ are the 4-momenta of $\pi^0$ and $a^0_0(980)$. If
isotopic invariance holds, then $\Gamma_{f_1\to a^0_0 \pi^0\to
\eta\pi^0\pi^0}=\Gamma_{f_1\to a_0\pi\to\eta\pi\pi}/3$.

Similarly, we have the width of the $f_1(1285)\to a_0(980)\pi^0\to
K\bar K\pi$ decay
\begin{eqnarray}\label{EqIV-8}\Gamma_{f_1\to a_0\pi\to K\bar K\pi}
=6\,\Gamma_{f_1\to a^0_0\pi^0\to K^+K^-\pi^0}\ \ \  \nonumber \\
=\frac{g^2_{f_1 a^0_0\pi^0}}{\pi\,m^2_{f_1}}
\int\limits_{2m_{K^+}}^{ m_{f_1}-m_{\pi^0}}p^3(m^2)\,\frac{2
m^2\Gamma_{a^0_0\to K^+K^-}(m)}{ \pi|D_{a^0_0}(m)|^2}\,dm\,.
\end{eqnarray}
The width of the decay $f_1(1285)\to a^0_0(980)\pi^0\to
f_0(980)\pi^0\to \pi^+\pi^-\pi^0$ due to the $a^0_0(980)-f_0(980)$
mixing (see diagram (1) in Fig. 7) is
\begin{eqnarray}\label{EqIV-9}\Gamma_{f_1\to a^0_0\pi^0\to f_0
\pi^0\to\pi^+\pi^-\pi^0}=\frac{g^2_{f_1a^0_0\pi^0}}{6\pi\,
m^2_{f_1}}\qquad\qquad \nonumber \\
\hspace{-0.15cm}\times\int\limits_{0.9 \mbox{\scriptsize{
GeV}}}^{1.05\mbox{\scriptsize{ GeV}}} \left|G_{a^0_0f_0}(m)\right|^2
p^3(m^2)\,\frac{2m^2\Gamma_{f_0\to\pi^+ \pi^-}(m)}{\pi}\,dm.
\end{eqnarray}
The shape of the $\pi^+\pi^-$ mass spectrum that follows from the
integrand in (62) is virtually undistinguishable from the curves
shown in Fig. 3.

Numerical integration in (54) and (62) yields
\begin{eqnarray}\label{EqIV-10}
\frac{\Gamma_{f_1\to a^0_0\pi^0\to f_0 \pi^0\to\pi^+\pi^-
\pi^0}}{\Gamma_{f_1\to a^0_0\pi^0\to\eta\ \pi^0\pi^0}}
\qquad\qquad\quad  \nonumber\\ =\frac{BR(f_1\to a^0_0(980)\pi^0\to
f_0(980)\pi^0\to\pi^+\pi^-\pi^0)}{BR(f_1\to a^0_0(980)\pi^0\to
\eta\pi^0\pi^0)}\nonumber\\ \approx0.29\%.\ \ \ \ \qquad\qquad
\qquad\qquad \end{eqnarray} This result was discussed after Eqn
(51). From (54) and (61), we also obtain
\begin{eqnarray}\label{EqIV-11}\frac{\Gamma_{f_1\to a_0\pi\to K\bar
K\pi}}{\Gamma_{f_1\to a_0\pi\to\eta\pi\pi}}=\frac{BR(f_1\to
a_0(980)\pi\to K\bar K\pi)}{BR(f_1\to a_0(980)\pi
\to\eta\pi\pi)}\nonumber\\
\approx0.11. \qquad\qquad\qquad\qquad\qquad
\end{eqnarray}
It is noteworthy that the PDG data \cite{PDG16} for the ratio
\begin{eqnarray}\label{EqIV-12}\frac{BR(f_1\to K\bar K\pi)}{BR(f_1\to
a_0(980)\pi \to\eta\pi\pi)}\approx0.25\pm0.05 \end{eqnarray} are not
in disagreement with (64) and indicate that the transition
$f_1(1285)\to a_0(980)\pi\to K\bar K\pi$ may not be the single
source of the $f_1(1285)\to K\bar K\pi$ decay.

(2) A detailed calculation for the transition $f_1(1285)\to(K^+
K^-+K^0\bar K^0)\pi^0\to f_0(980)\pi^0\to\pi^+ \pi^-\pi^0$, which is
due to the pointlike decay $f_1 (1285)\to K\bar K\pi^0$ (see diagram
(2) in Fig. 7), yields \cite{AKS16}
\begin{eqnarray}\label{EqV-3}\frac{\Gamma_{f_1\to(K^+K^-+K^0\bar
K^0) \pi^0\to f_0\pi^0\to\pi^+\pi^-\pi^0}}{\Gamma_{f_1\to K\bar K
\pi}}\approx0.0022\,.\end{eqnarray}

This result is almost 15 times smaller than the corresponding
experimental value
\begin{eqnarray}\label{EqV-4}\frac{BR(f_1(1285)\to f_0(980)\pi^0
\to\pi^+\pi^-\pi^0)}{BR(f_1(1285)\to K\bar K \pi)}\nonumber
\\ =0.033\pm0.010\,,\qquad\qquad\ \ \end{eqnarray}
(which contains a significant error, however) following from the VES
\cite{Do11} [see (32)] and PDG \cite{PDG16} data.

The $\pi^+\pi^-$ mass spectrum in the decay $f_1(1285)\to(K^+K^-+K^0
\bar K^0)\pi^0\to f_0(980)\pi^0\to\pi^+\pi^-\pi^0$ is, for this
mechanism, similar to the curves in Fig. 3. It is clear, however,
that the pointlike mechanism of the decay $f_1(1285)\to K\bar K\pi$
cannot per se yield a significant probability of the decay
$f_1(1285)\to f_0(980)\pi^0\to\pi^+\pi^-\pi^0$.

(3) We now proceed to discussing the mechanism of the decay
$f_1(1285)\to(K^*\bar K+\bar K^*K)\to(K^+K^-+K^0\bar K^0)\pi^0\to
f_0(980)\pi^0\to\pi^+ \pi^-\pi^0$ [where $K^*=K^*(892)$] induced by
diagram (3) in Fig. 7.

The widths of the decays $\Gamma(f_1(1285)\to(K^*\bar K+ \bar
K^*K)\to K \bar K\pi)$ and $\Gamma(f_1(1285)\to(K^*\bar K+\bar
K^*K)\to(K^+K^-+K^0\bar K^0)\pi^0\to f_0(980)\pi^0\to\pi^+\pi^-
\pi^0)$ are described in this case by simple but rather cumbersome
formulas. We do not present them here (the interested reader can
find them in \cite{AKS16}), but only briefly discuss the assumptions
made in the calculations and report the calculation results.

Figure 8 shows a diagram that corresponds to the transition
$f_1(1285)\to(K^*\bar K+\bar K^*K)\to K\bar K\pi$ (the caption to
Fig. 8 specifies the notation for the particles involved in the
decay.) The $f_1(1285)\to K^*\bar K$ vertex is generically
determined by two independent effective coupling constants. Given
today's state of experimental data, the general structure of this
vertex is de facto unknown. Therefore, for certainty, we limit our
approach to a partial expression (in the spirit of effective chiral
Lagrangians \cite{EGLPR89,EGPR89,Bir96}) of the form
\begin{eqnarray}\label{EqVI-1} V_{f_1K^*\bar K}=g_{f_1K^*\bar K}
F^{(f_1)}_{\,\mu\nu}F^{(K^*)\mu\nu}\,, \end{eqnarray} where
$F^{(f_1)}_{\,\mu\nu}=p_{1\mu}\epsilon_{f_1\nu}-p_{1\nu}\epsilon_{f_1
\mu}$, $F^{(K^*)}_{\,\mu\nu}=k_{1\mu}\epsilon_{K^*\nu}-k_{1\nu}
\epsilon_{K^*\mu}$,\, $\epsilon_{f_1}$ and $\epsilon_{K^*}$ are the
respective polarization 4-vectors of $f_1(1285)$ and $K^*$ mesons.
\begin{figure} [!ht] 
\includegraphics[width=6.4cm]{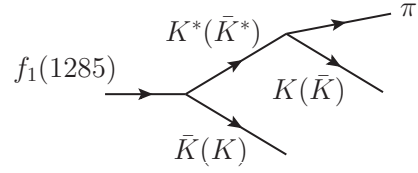}
\caption{\label{Fig8} Diagram of the decay $f_1(1285)\to(K^*\bar
K+\bar K^*K)\to K\bar K\pi$. The respective 4-momenta of
$f_1(1285)$, $K$ ($\bar K$), $K^*$ ($\bar K^*$), and $\pi$ are
$p_1$, $p_K$ ($p_{\bar K}$), $k_1$ ($k_2$), and $p_\pi$.
}\end{figure}
It is noteworthy that the $K^*$ resonance produced as a result of
such a transverse interaction carries spin 1 off the mass shell. The
amplitude of the $K^*\to K\pi$ decay is
\begin{eqnarray}\label{EqVI-2}
V_{K^*K\pi}=g_{K^*K\pi}(\epsilon_{K^*},p_\pi-p_K)\,, \end{eqnarray}
where $\,g_{K^{*+}K^+\pi^0}=-g_{\bar K^{*0}\bar K^0\pi^0}$ and
$g_{K^{*+}K^0\pi^+}=\sqrt{2}g_{{K^{*+}K^+\pi^0}}$. Similar formulas
hold for the decays $f_1(1285)\to\bar K^*K$ and $\bar K^*\to\bar
K\pi$.

As follows from Eqns (68) and (69), the product of the vertices in
the amplitude corresponding to diagram (3) in Fig. 7 is of the third
power in momenta. However, two momenta of the three are the momenta
of external particles, and therefore the diagram converges. We note
that this diagram in the physical region of the $f_1(1285)\to
f_0(980)\pi^0\to\pi^+\pi^-\pi^0$ decay does not contain the
logarithmic (triangular) singularity (all three particles in the
loop cannot be on the mass shell simultaneously)\,\footnote{Indeed,
the invariant mass of the $K\pi$ pair $(k^2_1)^{1/2}$ in the decay
$f_1(1285)\to K \bar K\pi$ ranges from 629 MeV to 788 MeV, while the
mass of the $K^*$ resonance is $m_{K^*}\approx895$ MeV and its decay
width is $\Gamma_{K^*}\approx 50$ MeV.} As a result, the nonzero
width of the $K^*(892)$ resonance affects the calculation of
$$\Gamma(f_1(1285)\to(K^*\bar K+ \bar K^*K)\to K \bar K\pi)$$ and
\begin{eqnarray}\Gamma(f_1(1285)\to(K^*\bar K+\bar K^*K)\to(K^+K^-+
K^0\bar K^0)\pi^0 \nonumber \\ \to f_0(980)\pi^0\to\pi^+\pi^-
\pi^0)\qquad\qquad\quad\nonumber\end{eqnarray} in a negligible way.

In obtaining numerical estimates in \cite{AKS16}, the contribution
to the propagator $1/D_{K^*}(k^2_{1(2)} )=1/(m^2_{K^*}-k^2_{1(2)}-
im_{K^*}\Gamma_{K^*})$ proportional to $\Gamma_{K^*}$ was simply
ignored.

\begin{figure} 
\includegraphics[width=6.5cm]{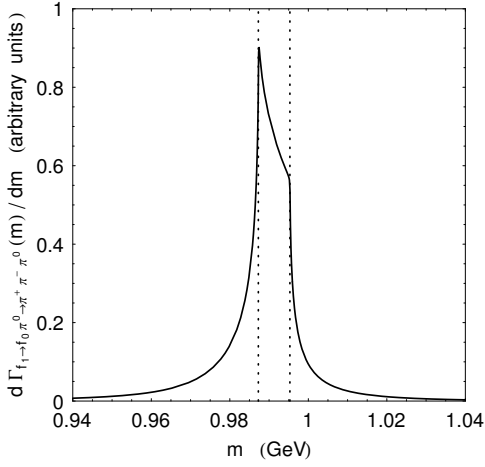}
\caption{\label{Fig9} The $\pi^+\pi^-$ mass spectrum in the decay
$f_1(1285)\to(K^*\bar K+\bar K^*K)\to(K^+K^-+K^0\bar K^0)\pi^0\to
f_0(980)\pi^0\to\pi^+\pi^-\pi^0$.}
\end{figure}

Figure 9 shows the $\pi^+\pi^-$ mass spectrum that corresponds to
the mechanism of the transition $f_1(1285)\to(K^*\bar K+\bar
K^*K)\to(K^+K^-+K^0 \bar K^0)\pi^0\to f_0(980)\pi^0\to\pi^+
\pi^-\pi^0$ \cite{AKS16}. Its shape is virtually the same as the
$\pi^+\pi^-$ mass spectrum for $a^0_0(980)-f_0(980)$ mixing
displayed in Fig. 3. Using the $f_1(1285)$ decay width corresponding
to the diagram in Fig. 8 and diagram (3) in Fig. 7, we integrate
numerically to obtain \cite{AKS16}
\begin{eqnarray}\label{EqVI-6}
\Gamma_{f_1\to(K^*\bar K+\bar K^*K)\to K\bar
K\pi}=\frac{g^2_{f_1K^{*+}K^-}g^2_{K^{*+}K^+\pi^0}}{4\,\pi^3}
\nonumber\\ \times\,0.976\,\times10^{-2}\,\mbox{GeV}^3\,,
\qquad\qquad\\ \label{EqVI-8} \Gamma_{f_1\to f_0\pi^0\to\pi^+\pi^-
\pi^0}=\frac{g^2_{f_1K^{*+}K^-}g^2_{K^{*+}K^+\pi^0}}{4\,\pi^3}
\nonumber\\ \times\,0.277\,\times10^{-4}\,\mbox{GeV}^3\,.
\qquad\qquad\end{eqnarray} The ratio of these values is
\begin{eqnarray}\label{EqVI-9}
\frac{\Gamma_{f_1\to f_0\pi^0\to\pi^+\pi^-\pi^0}}{\Gamma_{
f_1\to(K^*\bar K+\bar K^*K)\to K\bar K\pi}}\qquad\quad \nonumber\\
=\frac{BR(f_1\to f_0\pi^0\to\pi^+\pi^-\pi^0)} {BR(f_1\to(K^*\bar
K+\bar K^*K)\to K\bar K\pi)}\nonumber\\
=0.284\,\times10^{-2}\,.\qquad\quad\quad
\end{eqnarray}
Using this value and assuming that the entire value of $BR(f_1\to
K\bar K\pi)=(9.0\pm0.4)\%$ \cite{PDG16} is due to the
$f_1\to(K^*\bar K+\bar K^*K)\to K\bar K\pi$ decay mode, we obtain
the upper estimate
\begin{eqnarray}\label{EqVI-10}
BR(f_1\to f_0\pi^0\to\pi^+\pi^-\pi^0)\approx 0.255\,\times10^{-3}\,.
\end{eqnarray}
This result is approximately 12 times smaller than the central
experimental value (32) obtained by VES. Therefore, the single
transition mechanism $f_1(1285)\to(K^*\bar K+\bar K^*K)\to
(K^+K^-+K^0\bar K^0)\pi^0\to f_0(980)\pi^0\to\pi^+\pi^-\pi^0$ is
apparently insufficient for understanding the experimental results.

(4) Finally, we consider the $f_1(12 85)\to(K^*_0\bar K+\bar K^*_0
K)\to(K^+K^-+K^0\bar K^0)\pi^0\to f_0 (980)\pi^0\to\pi^+\pi^-\pi^0$
decay mechanism, which involves the scalar $K^*_0$ meson. This
mechanism is represented by diagram (4) in Fig. 7.

The version where the role of $K^*_0$ would be played by the
existing resonance $K^*_0(800)$ (or $\kappa$) \cite{PDG16} should be
disregarded. The point is that for the $\kappa$ resonance with a
mass $m_\kappa\lesssim$ 800 MeV and width $\Gamma_\kappa\approx
400-550$ MeV \cite{PDG16}, the shapes of the $K\pi$ and $K\bar K$
mass spectra in the decay $f_1(1285)\to(\kappa\bar K+\bar\kappa
K)\to K\bar K\pi$ are literally opposite to those observed in
experiment \cite{Bit84,Arm84,Arm87}. According to the data on the
$f_1(1285)\to K\bar K\pi$ decay \cite{PDG16,Bit84,Arm84,Arm87}, a
significant bump near the $K\bar K$ threshold is observed in the
$K\bar K$ spectrum, and a significant bump in the $K\pi$ spectrum is
observed near its upper edge, i.e., close to $m_{f_1}-m_K\approx788$
MeV. This picture agrees well with the $f_1(1285)\to a_0(980)\pi\to
K\bar K\pi$ decay mechanism and is consistent with the
$f_1(1285)\to(K^*\bar K+\bar K^*K)\to K\bar K\pi$ mechanism.
However, the $f_1(1285)\to(\kappa\bar K+\bar\kappa K)\to K\bar K\pi$
mechanism generates a bump near the upper end of the $K\bar K$
spectrum, i.e., close to $m_{f_1}-m_\pi\approx1147$ MeV and a bump
in the $K\pi$ spectrum near its threshold. This mechanism obviously
cannot be responsible for any significant contribution to the
$f_1(1285)\to K\bar K\pi$ decay. We are also unable to suggest any
special enhancement of the $f_1(1285)\to f_0(980)\pi^0\to\pi^+
\pi^-\pi^0$ decay due to that mechanism.

If the mass of the $K^*_0$ resonance is increased (moved away from
the $K\pi $ threshold, $m_K+m_\pi\approx0.629$ GeV), the
disagreement with the data on the $K\bar K$ and $K\pi$ spectra
gradually decreases. The resonance $K^*_0(1430)$ with the mass
$m_{K^*_0}\approx1425$ MeV and width $\Gamma_{K^*_0}\approx270$ MeV
\cite{PDG16} can be considered a candidate responsible for the
decays $f_1(1285)\to(K^*_0 \bar K+\bar K^*_0K)\to K\bar K\pi$ and
$f_1(1285)\to(K^*_0\bar K+\bar K^*_0K)\to(K^+K^-+K^0\bar
K^0)\pi^0\to f_0(980)\pi^0\to \pi^+\pi^- \pi^0$. It is then quite
natural that the $\pi^+\pi^-$ mass spectrum in the decay
$f_1(1285)\to(K^*_0\bar K+\bar K^*_0 K)\to(K^+K^-+K^0\bar K^0)\pi^0
\to f_0(980)\pi^0\to\pi^+\pi^-\pi^0$ is similar to the $\pi^+\pi^-$
mass spectra in Figs 3 and 9. The calculated ratio of the decay
widths corresponding to the mechanism involving the $K^*_0(1430)$
resonance is \cite{AKS16}
\begin{eqnarray}\label{EqVII-7}
\frac{\Gamma_{f_1\to f_0\pi^0\to\pi^+\pi^-\pi^0}}{\Gamma_{
f_1\to(K_0^*\bar K+\bar K_0^*K)\to K\bar K\pi}}\qquad\quad \nonumber\\
=\frac{BR(f_1\to f_0\pi^0\to\pi^+\pi^-\pi^0)}{BR(f_1\to(K_0^*\bar
K+\bar K_0^*K)\to K\bar K\pi)}\nonumber\\
=0.271\,\times10^{-2}\,.\qquad\quad\quad
\end{eqnarray}
Because estimate (74) virtually coincides with (72), the conclusions
regarding $BR(f_1\to f_0\pi^0\to\pi^+ \pi^- \pi^0)$ that follow Eqn
(72) hold in this case as well. Thus, the transition mechanism
\begin{eqnarray}f_1(1285)\to(K_0^*\bar K+ \bar K_0^*K)\to(K^+K^-+K^0\bar
K^0)\pi^0 \nonumber \\ \to f_0(980)\pi^0\to\pi^+\pi^-\pi^0 \qquad
\qquad\quad\nonumber\end{eqnarray} cannot per se explain the
experimental data.

Summarizing, we have considered four possible mechanisms of the
isospin-violating decay $f_1(1285)\to\pi^+\pi^-\pi^0$. The
conclusions we draw from the obtained estimates are as follows. The
experimental data can hardly be explained by any of the considered
mechanisms. On the other hand, the common feature of these
mechanisms is that in each of them the $\pi^+\pi^-$ mass spectrum in
the $f_1(1285)\to\pi^+\pi^-\pi^0$ decay is concentrated between the
$K^+K^-$ and $K^0\bar K^0$ thresholds due to the $K\bar K$-loop
character of isotopic symmetry violation. The considered mechanisms
are obviously a basis of the isospin violation in the
$f_1(1285)\to\pi^+\pi^-\pi^0$ decay; however, the situation requires
further elucidation. For example, it is still difficult to answer
the question whether the significant probability of the
$f_1(1285)\to\pi^+\pi^-\pi^0$ decay can be explained by the joint
action of the described mechanisms. Significant experimental efforts
are needed to remove uncertainties in the currently available data
(for example, it is desirable to measure different modes of the
$f_1(1285)$ decay concurrently at the same facility and identify the
mechanisms of the $f_1(1285)$ decay into $\eta\pi\pi$ and $K\bar
K\pi$ more accurately). Because data on the probability of the
$f_1(1285)\to\pi^+\pi^-\pi^0$ decay vary in a broad range [compare
(33), (52), and (53)], an improvement in the accuracy of these data
is a challenging and important task. \vspace*{0.3cm}

\noindent{\boldmath \bf 5.2. Consistency condition}

\noindent We use the decay $f_1(1285)\to f_0(980)\pi^0\to\pi^+
\pi^-\pi^0$ as an example to discuss a more general approach to
estimating the contribution from the $K\bar K$-loop mechanisms of
isotopic invariance violation. We formulate it as a consistency
requirement for data on the decays $f_1(1285)\to\pi^+\pi^-\pi^0$ and
$f_1(1285)\to K\bar K\pi$ based on the concept that isotopic
invariance is violated due to the mass difference between $K^+$ and
$K^0$ mesons.

We consider the diagrams shown in Figs 10 and 11. If $f_1(1285)$
decays into $K\bar K\pi$ (this occurs, as was noted above, in
approximately 9\% of cases \cite{PDG16}), then, due to final-state
interaction between $K$ and $\bar K$ mesons, i.e., transitions
$K^+K^-\to f_0(980)\to\pi^+\pi^-$ and $K^0\bar K^0\to f_0(980)\to
\pi^+\pi^-$, the isospin-violating decay $f_1(1285)\to(K^+K^-+
K^0\bar K^0)\pi^0$\,$\to$\,$ f_0(980)\pi^0$\,$\to$\,$\pi^+\pi^-
\pi^0$ is induced (see Fig. 11). Regardless of the specific
mechanism of $K\bar K$-pair production, this results, as has been
shown above, in the emergence in the $\pi^+\pi^-$ mass spectrum of a
narrow resonance structure with a width $\approx2m_{K^0}-2m_{K^+}
\approx8$ MeV, located in the region of $K\bar K$ thresholds.
Observing this structure experimentally would be a direct indication
of the general $K\bar K$-loop nature of the mechanisms responsible
for isotopic invariance violation.

\begin{figure} 
\includegraphics[width=5.9cm]{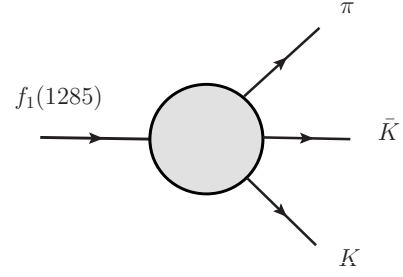}
\caption{\label{Fig10} Diagram of the decay $f_1(1285)\to K\bar
K\pi$.}\end{figure}
\begin{figure} 
\includegraphics[width=7.75cm]{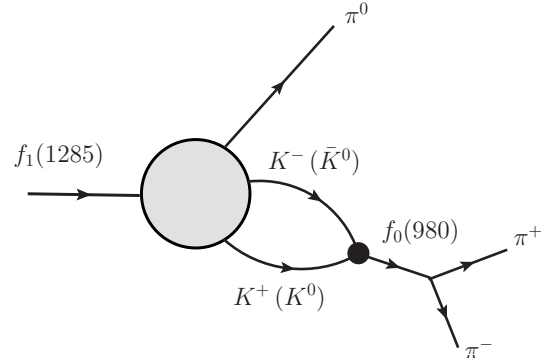}
\caption{\label{Fig11} Diagram corresponding to the combined
contribution of all possible $K\bar K$-loop mechanisms resulting in
the decay $f_1(1285)\to(K^+K^-+K^0\bar K^0)\pi^0\to f_0(980)\pi^0
\to\pi^+\pi^-\pi^0$.}\end{figure}

We expand the isospin-violating amplitude $\mathcal{M}_{f_1\to
f_0\pi^0}(s)$ near the $K\bar K$ thresholds in a power series in
$\rho_{K\bar K}(m)=\sqrt{1-4 m^2_K/m^2}$:
\begin{eqnarray}\label{EqVIII-2}
\hspace*{-25pt} & \mathcal{M}_{f_1(1285)\to
f_0(980)\pi^0}(m)=g_{f_0K^+K^-}\left\{A(m)\right.\ \qquad\nonumber\\
\hspace*{-25pt} & \times i[\rho_{K^+K^-}(m)-\rho_{K^0\bar K^0}(m)]
+B(m)[\rho^2_{K^+K^-}(m)\ \ \nonumber \\ \hspace*{-22.5pt} &
-\rho^2_{K^0 \bar K^0}(m)]+O[\rho^3_{K^+K^-}(m)-\rho^3_{K^0\bar
K^0}(m)]+\cdot\cdot\cdot \left.\right\}.\end{eqnarray} With good
accuracy,
\begin{eqnarray}
\mathcal{M}_{f_1(1285)\to f_0(980) \pi^0}(m)=g_{f_0K^+K^-}A(m)\nonumber\\
\times i[\rho_{K^+K^-}(m)-\rho_{K^0\bar K^0}(m)].\qquad\qquad
\end{eqnarray}
The amplitude $A(m)$ contains information about all possible
production mechanisms of the $K\bar K$ system with the isospin $I=1$
in the $S$-wave in the process $f_1(1285)\to K\bar K\pi$.

Data on the decay $f_1(1285)\to f_0(980)\pi^0\to\pi^+\pi^-\pi^0$ can
be used to extract information about $|A(m)|^2$ in the region of
$K^+K^-$ and $K^0\bar K^0$ thresholds:
\begin{eqnarray}\hspace*{10pt}\frac{d\Gamma_{f_1(1285)\to
f_0(980)\pi^0\to\pi^+ \pi^-\pi^0}(m)}{dm}\quad\nonumber \\
\hspace*{10pt}=\frac{1}{16\pi}|\mathcal{M}_{f_1(1285)\to
f_0(980)\pi^0}(m)|^2\quad\nonumber \\
\times p^3(m)\,\frac{2m^2\Gamma_{f_0\to\pi^+\pi^-}(m)}{
\pi|D_{f_0}(m)|^2}\,,\qquad\quad\end{eqnarray}  where
\begin{eqnarray}p(m)=\frac{[m^4_{f_1}-2m^2_{f_1}(m^2+m^2_\pi)+
(m^2-m^2_\pi)^2]^{1/2}}{2m_{f_1}}.\end{eqnarray} Information about
$|A(m)|^2$ at $m>2m_K$ can also be obtained from data on the $K\bar
K$ mass spectra measured in the $f_1(1285)\to K\bar K\pi$ decay.
Unfortunately, data on those spectra are still rather scant
\cite{Bit84,Arm84,Arm87}. But in the case of large statistics and
good resolution in the invariant mass of the $K\bar K$ system, a
straightforward scheme for obtaining data about $|A(m)|^2$ at $m$
above the $K^+K^-$, $K^0\bar K^0$, or $K^\pm K^0_S$ threshold could
be as follows.

The $K\bar K$ system in the $f_1(1285)\to K\bar K\pi$ decay (see
Fig. 10) is generated predominantly in the $S$-wave due to the very
restricted phase volume available for it ($2m_K<m<2m_K+150$ MeV).
Then, for example, the $K^+K^-$ mass spectrum in the decay
$f_1(1285)\to K^+K^-\pi^0$ can be represented in the form
\begin{eqnarray}\label{EqVIII-5a}
\frac{d\Gamma_{f_1(1285)\to K^+K^-\pi^0}(m)}{dm}\qquad\qquad \nonumber\\
\qquad =\frac{2\,m}{\pi}\,\rho_{K^+K^-}(m) \,p^3(m)\,|A(m)|^2\,.\ \
\end{eqnarray}
By approximating data on $d\Gamma_{f_1\to K^+K^-\pi^0}(m)/dm$, we
can determine the function $|A(m)|^2$ and use its value $|A(2m_{K^+}
)|^2$ at the $K^+K^-$ threshold (which certainly corresponds to the
contribution of the $S$-wave alone) to derive the following
approximate estimate for $\Gamma_{f_1\to f_0\pi^0\to\pi^+\pi^-
\pi^0}$ \cite{AKS16}:
\begin{eqnarray}\label{EqVIII-5} \Gamma_{f_1\to f_0\pi^0\to
\pi^+\pi^-\pi^0}=\int\frac{d\Gamma_{f_1\to
f_0\pi^0\to\pi^+\pi^-\pi^0} (m)}{dm}\,dm\nonumber\\ =|A
(2m_{K^+})|^2\int|\rho_{K^+K^-}(m)-\rho_{
K^0\bar K^0}(m)|^2\qquad \nonumber\\
\times p^3(m)\,\frac{g^2_{f_0K^+K^-}}{16\pi} \,
\frac{2m^2\Gamma_{f_0\to\pi^+ \pi^-}(m)}{\pi|D_{f_0}(m)|^2}\,dm \
\qquad\nonumber\\ \approx|A(2m_{K^+})|^2\,2.59 \times10^{-6}
\,\mbox{GeV}^5.\qquad\qquad
\end{eqnarray}
A comparison of this estimate with data on $f_1(1285)\to\pi^+\pi^-
\pi^0$ decay enables checking their consistency with the data on the
$f_1(1285)\to K\bar K\pi$ decay and the concept of isotopic
invariance violation due to the mass difference between $K^+$ and
$K^0$ mesons.

The considered approach can also be used to obtain estimates for
other decays of the same type.

If the isospin-violating amplitude contains logarithmic
singularities near the $K\bar K$ thresholds in the physical region
of kinematic variables (as is the case, for example, in the decay
$\eta(1405)\to(K^*\bar K+\bar K^*K)\to(K^+K^-+K^0\bar K^0)\pi^0\to
f_0(980)\pi^0\to\pi^+ \pi^-\pi^0$, which we discuss below), its
structure at $m\approx2m_K$ is more involved than expression (75).
The amplitude near the singularities cannot be straightforwardly
expanded in $\rho_{K\bar K}(m)$, and a simple consistency condition
such as that in (80) cannot be obtained. However, the $\pi^+\pi^-$
mass spectrum in the decay $\eta(1405)\to\pi^+\pi^-\pi^0$ is also
concentrated in the region between the $K^+K^-$ and $K^0\bar K^0$
thresholds, and therefore effectively corresponds to the shape
characteristic of the $K\bar K$-loop mechanism of isotopic
invariance violation \cite{AKS15}.\vspace*{0.3cm}

\noindent{\boldmath \bf 5.3. $K\bar K$-loop mechanism of isotopic
invariance violation in the decay $\eta(1405)$\,$\to$\,$f_0
(980)\pi^0$\,$\to$\,$3\pi$ and the role of anomalous Landau
thresholds}

\noindent In 2012, the BESIII collaboration measured the decays
$J/\psi\to\gamma\pi^+\pi^-\pi^0$ and $J/\psi\to\gamma\pi^0\pi^0
\pi^0$ and found a resonance peak about 50 MeV wide in the
three-pion mass spectra around 1.4 GeV \cite{Ab2}. The corresponding
$\pi^+\pi^-$ and $\pi^0\pi^0 $ mass spectra in the 990 MeV range
(i.e., in the vicinity of $K^+K^-$ and $K^0\bar K^0$ thresholds)
proved to contain a narrow structure whose width is about 10 MeV
\cite{Ab2}. Thus, the experiment was the first observation (with a
statistical confidence of 10s) of the isospin-violating decay
$J/\psi$\,$\to$\,$\gamma\eta(1405)$\,$ \to$\,$\gamma f_0(980)\pi^0$
followed by the decay $f_0(980)\to $\,$\pi^+\pi^-,\pi^0\pi^0$.
According to the data in \cite{Ab2},
\begin{eqnarray}\label{Eq5-3-1}
BR(J/\psi\to\gamma\eta (1405)\to\gamma
f_0(980)\pi^0\to\gamma\pi^+\pi^-\pi^0)\nonumber
\\ =(1.50 \pm0.11\pm0.11)\times10^{-5}\,.\qquad\quad\quad
\end{eqnarray}
Using the PDG data, the BESIII collaboration \cite{Ab2} also
obtained the ratio
\begin{eqnarray}\label{Eq5-3-2} \frac{BR(\eta
(1405)\to f_0(980)\pi^0\to\pi^+\pi^-\pi^0)}{BR(\eta
(1405)\to a^0_0(980)\pi^0\to\eta\pi^0\pi^0)} \nonumber \\
\ \ =(17.9\pm4.2 )\%\,,\qquad\qquad\qquad
\end{eqnarray}
whose value virtually precludes an explanation of the discovered
effect of isotopic invariance violation by the $a^0_0(980)-f_0(980)$
mixing. At the same time, the narrow resonance-like structure
discovered in $\pi^+\pi^-$ and $\pi^0 \pi^0$ mass spectra in the
decay $\eta(1405)$\,$\to$\,$\pi^+ \pi^-\pi^0$, $\,\pi^0\pi^0\pi^0$
in the vicinity of the $K^+K^-$ and $K^0\bar K^0 $ thresholds
suggests that the mechanism responsible for the production of the
$f_0 (980)$ resonance in the decay $\eta(1405)$\,$\to$\,$f_0(980)
\pi^0$\,$\to$\,$3\pi$ is similar to that of $a^0_0(980)-f_0(980)$
mixing, i.e., due to the $K\bar K$-loop transition $\eta(1405)$\,$
\to $\,$(K^+K^-+K^0\bar K^0)\pi^0$\,$\to $\,$f_0(980)\pi^0$\,$
\to$\,$3\pi$, whose amplitude does not vanish owing to the mass
difference between $K^+$ and $K^0$ mesons and has a significant
magnitude in the narrow region between the $K^+K^-$ and $K^0\bar K^0
$ thresholds.

Comparing the result in (81) obtained by BESIII with the PDG data
\cite{PDG16} for the dominant decay channel $J/\psi\to\gamma\eta
(1405/ 1475)\to\gamma K\bar K\pi$,
\begin{eqnarray}\label{Eq5-3-3}
BR(J/\psi\to\gamma\eta(1405/1475)\to\gamma K\bar K\pi)\nonumber
\\ =(2.8\pm0.6)\cdot10^{-3}\,,\qquad\qquad
\end{eqnarray} yields
\begin{eqnarray}\label{Eq5-3-4} \frac{BR(J/\psi\to\gamma\eta(1405)\to
\gamma f_0(980)\pi^0\to\gamma \pi^+\pi^-\pi^0)}{BR(J/\psi\to\gamma
\eta(1405/1475)\to\gamma K\bar K\pi)} && \nonumber \\ =(0.53\pm0.13
)\%\,.\qquad\qquad\qquad\ \ \ \ \,\end{eqnarray} The value of the
last ratio is also indicative of a very large violation of isospin
invariance in the decay $\eta(1405)$\,$\to$\,$f_0(980)\pi^0$.

\begin{figure} 
\includegraphics[width=8cm]{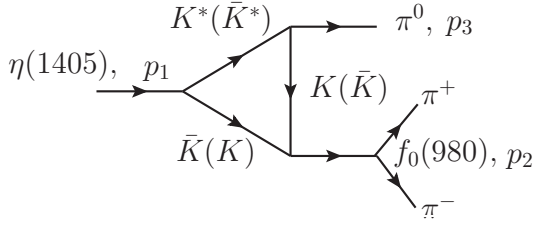} 
\caption{\label{Fig5-3-1} Diagram of the decay
$\eta(1405)\to(K^*\bar K+\bar K^*K)\to K\bar K\pi^0\to
f_0(980)\pi^0\to\pi^+\pi^-\pi^0$. All intermediate particles in the
triangle loop of this diagram in the region of $\eta(1405)$
resonance may be on the mass shell. As a result, a logarithmic
singularity emerges in the imaginary part of the triangle-diagram
amplitude in the case of a hypothetical stable $K^*$ meson
\cite{AKS15,AK1,AK2,AK3}. $p_1$, $p_2$, and $p_3$ denote the
4-momenta of the particle in the reaction; $p_1^2=s_1$ is the
invariant mass squared of $\eta(1405)$-resonance or the final
$\pi^+\pi^-\pi^0$ system; $p_2^2=s_2$ is the invariant mass squared
of $f_0(980)$ or the final $\pi^+\pi^-$ system; and
$p_3^2=m^2_{\pi^0}$.}\end{figure}

Below, we consider the theoretical possibility of explaining the
strong violation of isotopic invariance in the decay $\eta(1405)\to
f_0(980)\pi^0\to\pi^+\pi^-\pi^0$ by anomalous Landau thresholds (or
logarithmic triangular singularities) that are present in the
amplitude of the transition $\eta(1405)\to(K^*\bar K+\bar K^*K)\to
K\bar K\pi^0\to f_0(980)\pi^0\to\pi^+\pi^-\pi^0$ (Fig. 12) near the
$K\bar K$ thresholds. An attempt to explain the decay $\eta(1405)\to
f_0(980)\pi^0\to\pi^+\pi^-\pi^0$ invoking that mechanism was made in
\cite{WLZZ12,ALOWZ12,WWZZ13}. Shortly after that, we noted that in
the calculations, the vector $K^*$ meson $K^*(892)$ was considered a
stable particle in the intermediate state; we showed in \cite{AKS15}
that taking its finite width $\Gamma_{K^*}\approx \Gamma_{K^*\to
K\pi}\approx50$ MeV into account smoothes the logarithmic
singularities in the amplitude and reduces the calculated
probability of the decay $\eta(1405)\to f_0(980)\pi^0\to
\pi^+\pi^-\pi^0$ by a factor of 6 to 8 compared to that with
$\Gamma_{K^*} $\,=\,0. Also assuming the dominance of the decay
$\eta(1405)\to(K^*\bar K+\bar K^*K)\to K\bar K\pi^0$, we obtained
the estimate \cite{AKS15}
\begin{eqnarray}\label{Eq5-3-5} BR(J/\psi\to\gamma\eta(1405)\to\gamma
f_0(980)\pi^0\to\gamma\pi^+\pi^-\pi^0) \nonumber \\ \approx
1.12\cdot10^{-5}\,,\ \qquad\qquad\qquad\qquad \end{eqnarray} which
agrees with the BESIII data \cite{Ab2} quoted in (81) reasonably
well.

To show the effect of the $K^* $-meson width on the calculation of
the isospin-violating diagram displayed in Fig. 12 in the most
transparent way, we ignore the spin effects, which only make the
intermediate calculations significantly more complicated
\cite{AKS15}, having actually no effect whatsoever on the final
result \cite{AS18a} (i.e., we treat $K^* $ as a spinless particle in
what follows).\,\footnote{We note that convergence or divergence of
both the triangle diagram and the $K\bar K$ loops in the case of the
$a^0_0(980)\to(K^+K^-+ K^0\bar K^0)\to f_0(980)$ transition does not
have any relation to the effect of isospin violation under
consideration. The total of the subtraction constants for the
contributions from charged and neutral intermediate states to the
dispersion relation for the isospin-violating amplitude has a
natural order of smallness $\sim(m_{K^0}-m_{K^+})$ and cannot be
responsible for the enhanced symmetry violation in a narrow region
near the $K^+K^-$ and $K^0\bar K^0$ thresholds.}

For the amplitude of the triangle loop in Fig. 12, we introduce the
notation
\begin{eqnarray}\label{Eq5-3-6}
T=2\frac{g_1 g_2g_3}{16\pi}[F_+(s_1,s_2)-F_0(s_1,s_2)]\,,
\end{eqnarray} where $g_1$, $g_2$, and $g_3$ are the coupling constants
at the vertices (which are assumed to be the same for charged and
neutral channels), the amplitudes $F_+(s_1,s_2)$ and $F_0(s_1,s_2)$
describe the respective contributions of the charged and neutral
intermediate states, and the factor 2 appears because there are two
such contributions. We relate the isospin violation in the
considered diagram only to the difference in mass between charged
and neutral $K$ mesons and set $m_{K^{*+}}=m_{K^{*0}}=0.8955$ GeV.
The amplitude $F_+\equiv F_+(s_1, s_2)$ is given by
\begin{equation}\label{Eq5-3-7}
F_+=\frac{i}{\pi^3}\int\frac{d^4k}{D_1D_2D_3}\,,
\end{equation} where
$D_1=(k^2-m^2_{K^{*+}}+i\varepsilon)$, $D_2=((p_1-k)^2-m^2_{K^-}
+i\varepsilon)$, and $D_3=((k-p_3)^2-m^2_{K^+}+i\varepsilon)$ are
inverse propagators of loop particles. The imaginary part of $F_+$
in the region $s_1\geq(m_{K^{*+}}+m_{K^+} )^2$ and $s_2\geq4m^2_{
K^+}$ consists of the imaginary part due to the discontinuity on the
$K^{*+}K^-$ cut in the variable $s_1$ and the imaginary part due to
the discontinuity on the $K^+K^-$ cut in the variable $s_2$:
\begin{equation}\label{Eq5-3-8}
{\rm Im}F_+={\rm Im}F_+^{(K^{*+}K^-)}+{\rm Im}F_+^{(K^+K^-)}\,,
\end{equation}
\begin{equation}\label{Eq5-3-9}{\rm
Im}F_+^{(K^{*+}K^-)}=\frac{1}{\sqrt{\Delta}}\ln\left[\frac{\alpha_+
+\sqrt{\Delta\delta_+}}{\alpha_+-\sqrt{\Delta\delta_+}}\right]\,,
\end{equation}
\begin{equation}\label{Eq5-3-10}{\rm
Im}F_+^{(K^+K^-)}=\frac{1}{\sqrt{\Delta}}\ln\left[\frac{\alpha'_+
+\sqrt{\Delta\delta'_+}}{\alpha'_+-\sqrt{\Delta\delta'_+}}\right]\,,
\end{equation} where
\begin{eqnarray}\label{Eq5-3-11}
\Delta=s^2_1-2s_1(s_2+m^2_{\pi^0})+(s_2-m^2_{\pi^0})^2,\qquad\ \ \\
\label{Eq5-3-12} \alpha_+=s^2_1-s_1(s_2+m^2_{\pi^0}
+m^2_{K^{*+}}-m^2_{K^+})\qquad \nonumber
\\ +(s_2-m^2_{\pi^0})(m^2_{K^+}-m^2_{K^{*+}}),\qquad\qquad\  \\
\label{Eq5-3-13}
\delta_+=s^2_1-2s_1(m^2_{K^{*+}}+m^2_{K^+})+(m^2_{K^{*+}}-m^2_{K^+})^2,\\
\label{Eq5-3-14} \alpha'_+=s_2(s_2-s_1-m^2_{\pi^0}
-2m^2_{K^+}+2m^2_{K^{*+}}),\qquad \\
\label{Eq5-3-15} \delta'_+=s_2(s_2-4m^2_{K^+}).\qquad\qquad\qquad\
\end{eqnarray}
Replacing the index $+$ of the functions in (87)--(95) with the
index 0 and masses of charged intermediate particles with masses of
their neutral counterparts, we obtain everything necessary for
describing the amplitude $F_0\equiv F_0(s_1,s_2)$.


\begin{figure} 
\includegraphics[width=6.5cm]{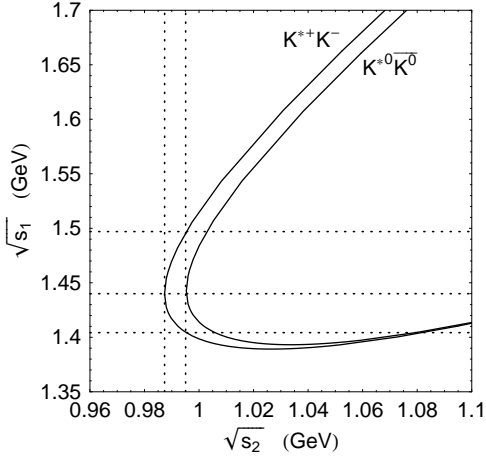} 
\caption{\label{Fig5-3-2} Solid curves in the
$(\sqrt{s_2}\,,\sqrt{s_1}\,)$ plane show the location of the
logarithmic singularity in the imaginary part of the triangle
diagram (see Fig. 12) that corresponds to the contributions of the
$K^{*+}K^-$ and $K^{*0}\bar K^0$ intermediate states. The dotted
vertical lines show the $K^+K^-$ and $K^0\bar K^0$ thresholds in the
variable $\sqrt{s_2}$ (i.e., its value $2m_{K^+}=0.987354$ GeV and
$2m_{K^0}= 0.99 5344$ GeV). The dotted horizontal lines show the
values of the variable $\sqrt{s_1}$ 1.404, 1.440, and 1.497 GeV. For
$1.404<\sqrt{s_1}<1.497$ GeV, the logarithmic singularity is located
in the case of the $K^{*+}K^-$ intermediate state at a value of
$\sqrt{s_2}$ between the $K^+K^-$ and $K^0\bar K^0$ thresholds,
while in the case of the $K^{*0}\bar K^0$ intermediate state, its
distance from the $K^0\bar K^0$ thresholds is no more than 6 MeV.
The singularities touch the $K\bar K$ thresholds at approximately
$\sqrt{s_1}=1.440$ GeV. }\end{figure}


A specific feature of the considered case is that all intermediate
particles in the triangle diagram in Fig. 12 can be on the mass
shell in the region of the $\eta(1405)$ resonance. This situation
occurs for the values of the kinematic variables $s_1$ and $s_2$
related to each other as
\begin{eqnarray}\label{Eq5-3-16}
\alpha_{+,0}=\pm\sqrt{\Delta\delta_{+,0}}
\end{eqnarray}
or equivalently \begin{eqnarray}\label{Eq5-3-17}
\alpha'_{+,0}=\pm\sqrt{\Delta\delta'_{+,0}}.
\end{eqnarray}
Therefore, in the hypothetical case of a stable $K^*$ meson, the
imaginary part of this triangle diagram contains a logarithmic
singularity \cite{AKS15,AK1,AK2,AK3}. Figure 13 shows the location
of the logarithmic singularities due to the contributions of the
$K^{*+}K^-$ and $K^{*0}\bar K^0$ intermediate states in the
$(\sqrt{s_2}\,,\sqrt{ s_1}\,)$ plane. It can be seen that they come
very close to the $K\bar K$ thresholds (whose location is shown in
this figure and in Figs 14--19 with vertical dotted lines) in the
region of the $\eta(1405)$ resonance. For example, if $\sqrt{s_1}
=1.420$ GeV, the singularities in the contributions of the
$K^{*+}K^-$ and $K^{*0}\bar K^0$ intermediate states manifest
themselves at the respective values $\sqrt{s_2}\approx0.989$ and
0.998 GeV of the $\pi^+\pi^-$ system invariant mass (see Fig. 13).
Figure 14 shows a typical behavior of the imaginary and real parts
of the amplitudes $F_+(s_1,s_2)$ and $F_0(s_1,s_2)$ as functions of
$\sqrt{s_2}$ in the vicinity of the $K\bar K$ thresholds and of
$\sqrt{s_1}$ in the vicinity of the $\eta(1405)$ resonance, namely,
at $\sqrt{s_1}=1.420$ GeV. This behavior is characterized by
singularities in $\mbox{Im}F_{+,0}(s_1,s_2)$ and discontinuities in
$\mbox{Re}F_{+,0}(s_1,s_2) $.

Because the singularities due to the charged and neutral
intermediate states, which are located in different places, do not
cancel each other, the mechanism under consideration can apparently
result in a catastrophic violation of isotopic symmetry in the decay
$\eta(1405)\to\pi^+\pi^-\pi^0$, as illustrated by Fig. 15.

\begin{figure} 
\includegraphics[width=7cm]{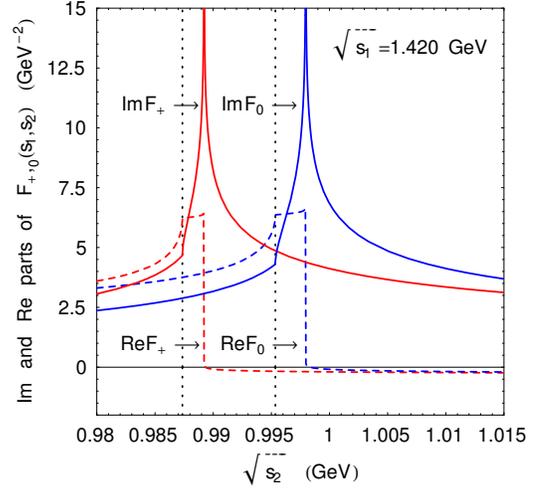}
\caption{\label{Fig5-3-3} Solid (dashed) curves show the imaginary
(real) part of the amplitude $F_+(s_1,s_2)$ for the charged
intermediate state and the amplitude $F_0(s_1,s_2)$ for the neutral
intermediate state in the triangle loop calculated for the
hypothetical case of a stable intermediate $K^*$ meson.}\end{figure}

\begin{figure} 
\includegraphics[width=7cm]{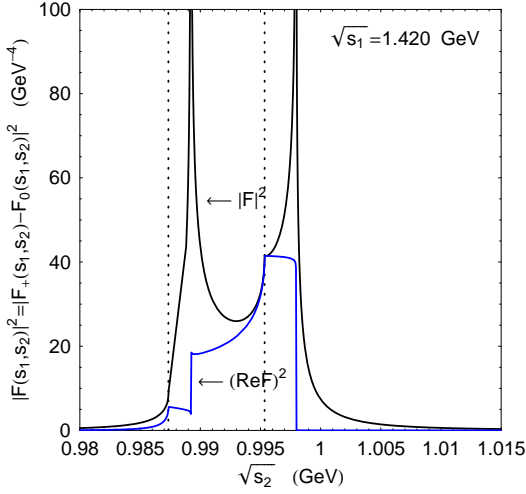}
\caption{\label{Fig5-3-4} Absolute value squared and the real part
squared of the triangle- loop isospin-violating amplitude
$F(s_1,s_2)\equiv F_+(s_1,s_2)-F_0(s_1,s_2)$ for a hypothetical
stable intermediate $K^*$ meson. The integral contributions from the
imaginary and real parts of the amplitude are approximately the same
here.}\end{figure}

\begin{figure} 
\includegraphics[width=6.7cm]{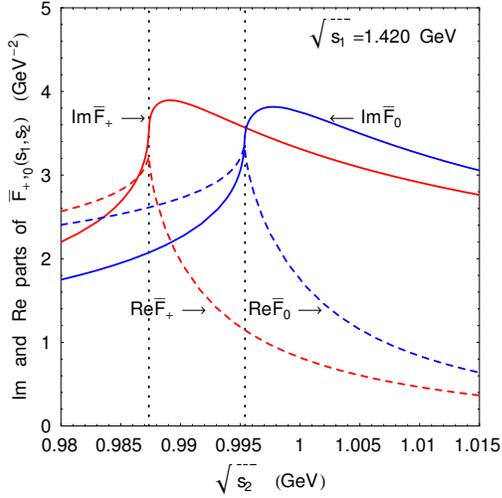}
\caption{\label{Fig5-3-5} Solid (dashed) curves show the imaginary
(real) parts of the amplitude $\bar{F}_+(s_1,s_2)$ for the charged
intermediate state and the amplitude $\bar{F}_0(s_1,s_2)$ for the
neutral intermediate state in the triangle loop calculated taking
the instability of the intermediate $K^*$ meson into
consideration.}\end{figure}

Such a `singular' scenario cannot occur in reality. Taking the
finite width of the $K^*$ resonance into account (i.e., averaging
the amplitude over a resonance Breit--Wigner distribution in
accordance with the K\"{a}ll\'{e}n--Lehmann spectral representation
for the propagator of an unstable $K^*$ meson \cite{AK1,AK2,AK3})
`smears out' the logarithmic singularities in the amplitude and in
this way enhances the mutual compensation of the contributions of
the $(K^{*+}K^-+ K^{*-}K^+)$ and $(K^{*0}\bar K^0+\bar K^{*0}K^0)$
intermediate states. This results in a significant decrease in the
calculated width of the $\eta(1405)\to\pi^+\pi^-\pi^0$ decay
compared to the case $\Gamma_{K^*\to K\pi}$\,=\,0 and in
localization of the main effect of isospin violation in the
$\pi^+\pi^-$ invariant mass region between the $K\bar K$ thresholds.

\begin{figure} 
\hspace*{-0.33cm}\includegraphics[width=9.6cm]{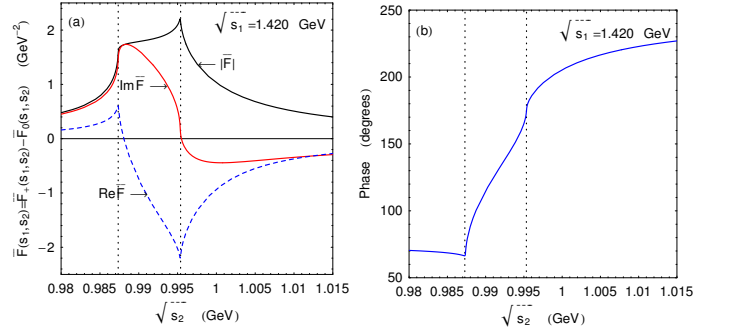}
\caption{\label{Fig5-3-6} (a) Absolute value, imaginary and real
parts of the triangle-loop amplitude
$\bar{F}(s_1,s_2)=\bar{F}_+(s_1,s_2)-\bar{F}_0(s_1,s_2)$ calculated
taking the instability of the intermediate $K^*$ meson into account.
(b) Phase of the amplitude $\bar{F}(s_1,s_2)$. }\end{figure}

As stated above, we write the propagator of an unstable $K^*$ meson
in the form of a spectral K\"{a}ll\'{e}n--Lehmann representation
\cite{AK1,AK2,AK3}
\begin{equation}\label{Eq5-3-18}
\frac{1}{m^2_{K^*}-k^2-im_{K^*}\Gamma_{K^*}}\to\int\limits^{\infty}_{
(m_{K}+m_\pi)^2}dm^2\frac{\rho(m^2)}{m^2-k^2-i\varepsilon}
\end{equation} and approximate $\rho(m^2)$ as
\begin{equation}\label{Eq5-3-19}
\rho(m^2)=\frac{1}{\pi}\frac{m_{K^*}\Gamma_{K^*}}
{(m^2-m^2_{K^*})^2+(m_{K^*}\Gamma_{K^*})^2}\,.
\end{equation}
Next, in the formulas for the amplitude $F_{+,0}(s_1,s_2)$, we
replace the $K^*$-meson mass squared $m^2_{K^*}$ with the variable
mass squared $m^2$ and define the amplitudes weighted with the
spectral density $\rho(m^2)$ \cite{AK1,AK2,AK3},
\begin{eqnarray}\label{Eq5-3-20}
\hspace*{-0.35cm}\bar{F}_{+,0}(s_1,s_2)=\int\limits^{\infty}_{
(m_{K}+m_\pi)^2}\rho(m^2)\,F_{+,0}(s_1,s_2;m^2)\,dm^2.
\end{eqnarray}
Figure 16 illustrates the behavior of the imaginary and real parts
of the weighted amplitudes $\bar{F}_+(s_1,s_2)$ and
$\bar{F}_0(s_1,s_2)$ as a function of $\sqrt{s_2}$ in the region of
$K\bar K$ thresholds at $\sqrt{s_1}=1.420$ GeV. A comparison of this
figure with Fig. 14 shows that the singular behavior of $F_+(s_1,s_2
)$ and $F_0(s_1,s_2)$ almost completely disappears if the $K^*$
meson instability is taken into account.

The absolute value, the imaginary and real parts, and the phase of
the amplitude of the triangle loop
$\bar{F}(s_1,s_2)\equiv\bar{F}_+(s_1,s_2)-\bar{F}_0(s_1,s_2)$ that
violates isotopic invariance calculated with the instability of the
intermediate $K^*$ meson taken into account are displayed in Fig.
17. We can see that all the characteristic features of the amplitude
$\bar{F}(s_1,s_2)$ are linked to the $K\bar K$ thresholds, and the
behavior of its absolute value and phase are essentially similar to
those of the $a^0_0(980)-f_0(980)$ mixing amplitude $\Pi_{a^0_0f_0}
(m)$ shown in Fig. 2.

Shown in Fig. 18 is the absolute value squared of the amplitude
$\bar{F}(s_1,s_2)=\bar{F}_+(s_1,s_2)-\bar{F}_0(s_1,s_2)$, which was
obtained taking the instability of the intermediate $K^*$ meson into
account. It should be compared with its analog for $\Gamma_{K^*}=0$
shown in Fig. 15. It is noteworthy that the areas below the
corresponding curves differ by about an order of magnitude. This is
the effect of the finite width $\Gamma_{K^*}=50$ MeV.

A similar situation occurs for all values of $\sqrt{s_1}$ in the
region of the $\eta(1405)$ resonance. Figure 19 shows the overall
view of the $\pi^+\pi^-$ mass spectrum in the decay
$\eta(1405)\to\pi^+\pi^-\pi^0$ calculated for the nominal mass of
$\eta(1405)$, i.e., at $\sqrt{s_1}=1.405$ GeV, using the formula
\begin{eqnarray}\label{Eq5-3-21}
\frac{dN}{d\sqrt{s_2}}=C\sqrt{\frac{\Delta}{s_1}}\left|\bar{F}_+
(s_1,s_2)-\bar{F}_0(s_1,s_2)\right|^2\nonumber \\
\times\frac{s_2\Gamma_{f_0\to\pi^+\pi^-}(\sqrt{s_2})}{\pi
|D_{f_0}(\sqrt{s_2})|^2},\qquad\qquad \end{eqnarray} where $C$ is a
normalization constant.

\begin{figure} 
\includegraphics[width=6.7cm]{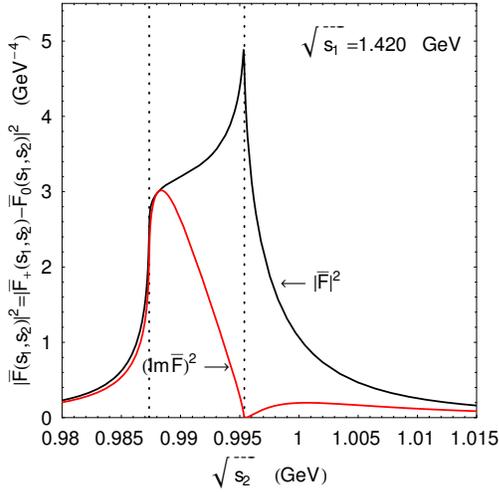}
\caption{\label{Fig5-3-7} Absolute value squared and the imaginary
part squared of the triangle-loop amplitude
$\bar{F}(s_1,s_2)=\bar{F}_+ (s_1,s_2)-\bar{F}_0(s_1,s_2)$ calculated
taking the instability of the intermediate $K^*$ meson into
consideration. The plots should be compared with Fig. 15.}
\end{figure}

\begin{figure} 
\vspace*{2mm}
\includegraphics[width=6.7cm]{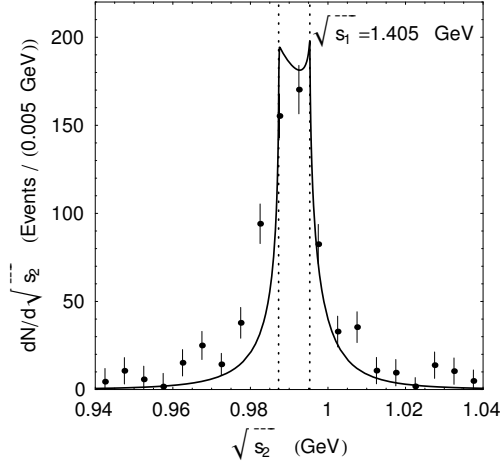}
\caption{\label{Fig5-3-8} Illustration of the $\pi^+\pi^-$ mass
spectrum shape in the decay $\eta(1405)\to\pi^+\pi^-\pi^0$ plotted
using Eqn (101), which corresponds to the contribution of the
diagram in Fig. 12. The dots with error bars represent the first
BESIII data for this decay \cite{Ab2}.}\end{figure}

In conclusion, we consider the high-level hierarchy of
isotopic-symmetry violations induced by the mass difference of $K^+$
and $K^0$ mesons discussed above.

The baseline isotopic symmetry violation in the process amplitude is
of the order of
\begin{equation}\label{Eq5-3-22}
\simeq\frac{m_{K^0}-m_{K^+}}{m_{K^0}}\approx\frac{1}{126}\,.
\end{equation}
Symmetry violation in the process amplitude in the region between
the $K^+K^-$ and $K^0\bar K^0$ thresholds owing to any mechanism of
production of $K\bar K$ pairs with a definite isospin in the
$S$-wave without anomalous Landau thresholds, in particular, as a
result of the $a^0_0(980)-f_0(980)$ mixing, is of the order of
\begin{equation}\label{Eq5-3-23}\simeq
\sqrt{\frac{2(m_{K^0}-m_{K^+})}{m_{K^0}}}\approx0.127\,.
\end{equation}
Symmetry violation in the amplitude of the decay $\eta(1405)\to
f_0(980)\pi^0\to\pi^+\pi^-\pi^0$ due to logarithmic triangular
singularities in the contributions of $(K^*\bar K+\bar K^*K)$
intermediate states in the $\sqrt{s_2}$ region between the $K^0\bar
K^0$ and $K^+ K^-$ thresholds \cite{AKS15} is of the order of
\begin{equation}
\label{Eq5-3-24}\simeq\left|\ln\left|\frac{\Gamma_{K^*}/2}
{\sqrt{m^2_{K^0}-m^2_{K^+}+\Gamma^2_{K^*}/4}}\right|\right|\approx1
\end{equation}
(this estimate of the uncompensated part of the contributions of
charged and neutral intermediate states in the triangle diagram,
which is in agreement with Fig. 17a, can be obtained, for example,
from Eqn (90) if $m^2_{K^*}$ at the singularity point is replaced
with $m^2_{K^*}-im_{K^*}\Gamma_{K^*}$).

In all of the cases of anomalous isotopic symmetry violation, the
phase of the symmetry-violating amplitude varies in the region
between the $K^+K^-$ and $K^0\bar K^0$ thresholds by approximately
$90^\circ$.


\vspace{0.3cm} \noindent{\large\bf\boldmath 6. Manifestation of
$a^0_0(980)-f_0(980)$ mixing in decays of charmed mesons}
\vspace{0.2cm}

\noindent Studies of the spectroscopy of light resonances (in
particular, $a_0(980)$ and $f_0(980)$) in weak hadronic decays of
$D$ and $D_s$ mesons is one of the main areas of experimental
activities in charmed particle physics at the CERN LHCb detector
\cite{Nogu15,Reis16}. We show below that observation of the effects
related to $a^0_0(980)-f_0(980)$ mixing in the decays
$D^+_s\to\eta\pi^0\pi^+$ \cite{AS17a}, $D^0\to K^0_S\pi^+\pi^-$ and
$D^0\to K^0_S\eta\pi^0$ \cite{AS17b} can shed additional light on
the mechanisms of $f_0(980)$ and $a_0(980)$ production in
three-particle hadronic decays of the $D^+_s$ and $D^0$ mesons (the
nature of these mechanisms is far from being well understood at the
current stage of research.) The huge statistics that can be
collected in LHCb experiments makes successful detection of the
$a^0_0(980)-f_0(980) $ mixing effects quite a feasible task.

\vspace*{0.3cm}\noindent{\boldmath\bf 6.1. Decay
$D^+_s\to\eta\pi^0\pi^+$}

\noindent Figure 20 displays data on the $S$-wave spectrum of the
$K^+K^-$ system mass in the decay $D^+_s\to K^+K^-\pi^+$ that were
obtained by the BaBar collaboration \cite{BaBar11}. The shape of
this spectrum and data on the shape of the $S$-wave $\pi^+\pi^-$
mass spectrum in the decay $D^+_s\to\pi^+\pi^-\pi^+$ \cite{BaBar09}
can be successfully approximated by the contribution of the
$f_0(980)$ resonance (see curves in Figs 20 and 21). Indeed, up to
overall normalization constants, the curves reproduce the absolute
value squared of the scalar resonance $f_0(980)$ propagator:
$|S_{K^+K^-} |^2\sim1/|D_{f_0}(m^2_{K^+ K^-})|^2$, where
$m_{K^+K^-}$ is the $K^+K^-$ invariant mass in the region above the
$K^+K^-$ threshold, $|S_{\pi^+\pi^-}|^2 \sim1/|D_{f_0} (m^2_{\pi^+
\pi^-})|^2$, and $m_{\pi^+\pi^-}$ is the $\pi^+\pi^-$ invariant
mass. The $f_0(980)$ propagator itself was taken from \cite{AKS16}
without any modifications.

\begin{figure}
\hspace*{-0.26cm}\includegraphics[width=17pc]{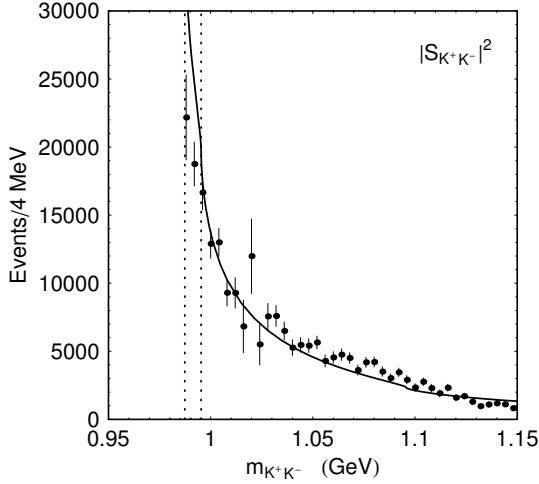}
\caption{\label{Figure1} $S$-wave $K^+K^-$-mass spectrum in the
decay $D^+_s\to K^+K^-\pi^+$ according to the BaBar data
\cite{BaBar11}; it corresponds to the absolute value squared of the
transition amplitude and does not include the phase space factor of
the $K^+K^-$ system in $D^+_s\to K^+K^-\pi^+$. The vertical dotted
lines show the location of the $K^+K^-$ and $K^0\bar K^0$
thresholds. The solid curve corresponds to the contribution of the
$f_0(980)$ resonance (see details in the text).}\end{figure}

\begin{figure}
\hspace*{-0.26cm}\includegraphics[width=17pc]{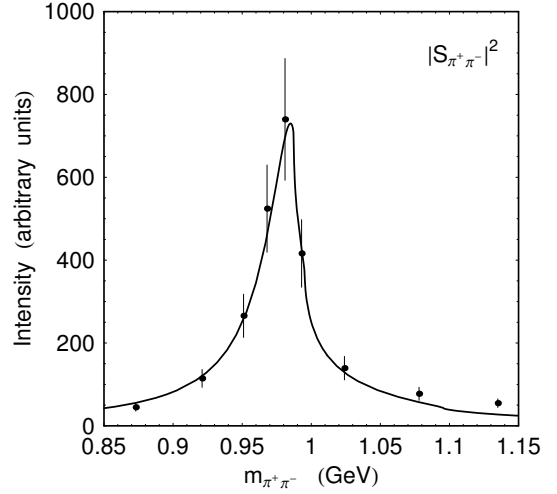}
\caption{\label{Figure2} $S$-wave $\pi^+\pi^-$ mass spectrum in the
decay $D^+_s\to\pi^+ \pi^-\pi^+$ according to the BaBar data
\cite{BaBar09}. The shape of the curve corresponds to the
contribution of the $f_0(980)$ resonance (see details in the
text).}\end{figure}

According to the PDG \cite{PDG16}, the decay probability is
\begin{eqnarray}\label{EqVI-1-1}
BR(D^+_s\to f_0(980)\pi^+\to K^+K^-\pi^+)\nonumber \\
=(1.15\pm0.32)\%\,.\qquad\qquad\end{eqnarray} The value itself and
its accuracy, which are based on the data of the BaBar
\cite{BaBar11} and CLEO \cite{CLEO09} experiments, need further
exploration. It was actually assumed in the initial analysis of
BaBar \cite{BaBar11} and CLEO \cite{CLEO09} that the possible
presence of the $a^0_0(980)$ resonance in the $K^+K^-$ system
produced in the $D^+_s\to K^+K^-\pi^+$ decay can be disregarded.
Therefore, the value in (105) effectively corresponds to the sum of
the $f_0(980)$ and $a^0_0(980)$ contributions to $D^+_s $-meson
decay (the discussion of the data in \cite{BaBar11,BaBar09,CLEO09}
is continued in footnote 8 and at the end of this section).

The reliability of result (105) can be assessed by exploring the
three-particle decays of the $D^+_s$ meson, which are akin to
$D^+_s\to K^+K^-\pi^+$. For example, we proposed in \cite{AS17a} to
explore manifestations of the $a^0_0(980)-f_0(980)$ mixing in the
$\eta\pi^0$ mass spectrum in the decay \footnote{It would be
reasonable to explore the contributions of various intermediate
states to the decay $D^+_s\to K^0_S K^0_S\pi^+$ in detail.}
$D^+_s\to\eta\pi^0\pi^+$ to obtain additional information about the
$f_0(980)$ and $a^0_0(980)$ resonance production mechanisms.

Guided by the value $BR(D^+_s\to f_0(980)\pi^+ \to K^+K^-\pi^+)$ in
(105) and the parameters of $f_0(980)$ and $a^0_0(980)$ resonances
discussed above, we obtained the following estimate for the
contribution to the decay $D^+_s\to\eta\pi^0\pi^+$ due to the $
a^0_0(980)-f_0(980)$ mixing:
\begin{eqnarray}\label{EqVI-1-2}
BR\left(D^+_s\to\left[f_0(980)\to(K^+K^-+K^0\bar K^0)\right.\right.
\nonumber \\ \left.\left.\to a^0_0(980)\right]\pi^+
\to\eta\pi^0\pi^+\right)=4.1\times10^{-4}\,.\ \
\end{eqnarray}
The corresponding amplitude of the transition
$D^+_s\to\left[f_0(980) \to(K^+K^-+K^0\bar K^0)\to
a^0_0(980)\right]\pi^+\to\eta\pi^0\pi^+$ is presented in Eqn (110).

Data on the decay $D^+_s\to\eta\pi^0\pi^+$ \cite{PDG16,
CLEO09a,CLEO13} indicate that it is almost completely dominated by
the $\eta\rho^+$ intermediate state:
\begin{eqnarray}\label{EqVI-1-3}
BR(D^+_s\to\eta\rho^+\to\eta\pi^0\pi^+)=(8.9\pm0.8)\%\,,\\
\label{EqVI-1-4} BR(D^+_s\to\eta\pi^0\pi^+)=(9.2\pm1.2)\%\,.\qquad
\end{eqnarray}

We let $A_{\eta\rho^+}$ and $A_{f_0 a^0_0}$ denote the amplitudes of
the respective transitions $D^+_s$\,$\to$\,$\eta\rho^+$\,$\to$\,$
\eta\pi^0\pi^+$ and $D^+_s$\,$\to$\,$\left[f_0(980)\right.$\,$\to
$\,$(K^+K^-+K^0\bar K^0)$\,$\to$\,$\left.a^0_0(980)\right]\pi^+
$\,$\to$\,$\eta\pi^0\pi^+$, and, to describe their dependence on
mass variables, use the expressions
\begin{eqnarray}\label{EqVI-1-5}
A_{\eta\rho^+}\equiv A_{\eta\rho^+}(m^2_{\eta\pi^0},m^2_{\eta\pi^+},
m^2_{\pi^0\pi^+})\equiv A_{\eta\rho^+}(s,t,u) \nonumber\\
=C_{D^+_s\eta\rho^+}\,\frac{s-t}{D_{\rho^+}(u)}F_\rho(u)
\sqrt{\frac{g^2_{\rho\pi\pi}}{ 16\pi}}\,,\qquad\qquad\end{eqnarray}
\begin{eqnarray}\label{EqVI-1-6}
A_{f_0a^0_0}\equiv A_{f_0a^0_0}(m^2_{\eta\pi^0})\equiv A_{f_0a^0_0}
(s)\qquad\quad\nonumber\\ =C_{D^+_s
f_0\pi^+}\,\frac{\Pi_{a^0_0f_0}(s)}{ D_{a^0_0}(s)
D_{f_0}(s)-\Pi^2_{a^0_0f_0}(s)}\sqrt{\frac{g^2_{a^0_0\eta\pi^0}}{16\pi
}}\,,
\end{eqnarray} where $s=m^2_{\eta\pi^0}$, $t=m^2_{\eta\pi^+}$, and
$u=m^2_{\pi^0\pi^+}$ are the invariant masses squared of the
specified meson pairs in the decay $D^+_s\to\eta\pi^0\pi^+$
($\Sigma=s+t+u=m^2_{D_s} +2m^2_\pi+m^2_\eta $); $D_{\rho^+}(u)$ and
$F_\rho(u)$ are the $\rho^+$-meson propagator and barrier factor in
the decay $\rho^+\to\pi^+\pi^0$ \cite{AS17a}; $C_{D^+_s\eta\rho^+}$
and $C_{D^+_s f_0\pi^+}$ are the invariant amplitudes of the decays
$D^+_s\to\eta\rho^+$ and $D^+_s\to f_0(980)\pi^+$. The effective
vertices $D^+_s\to\eta\rho^+$ and $\rho^+\to\pi^0 \pi^+$ were taken
in the form
\begin{eqnarray}\label{EqVI-1-7}
V_{D^+_s\eta\rho^+}=
C_{D^+_s\eta\rho^+}(\epsilon^*_{\rho^+},p_{D^+_s}+p_\eta)\,,\quad\\
\label{EqVI-1-8} V_{\rho^+\pi^0\pi^+}=g_{\rho\pi\pi}(\epsilon_{\rho^
+},p_{\pi^+}-p_{\pi^0})\,,\ \ \quad
\end{eqnarray}
where $\epsilon_{\rho^+}$ is the $\rho^+$-meson polarization
4-vector, and $p_{D^+_s}$, $p_\eta$, $p_{\pi^0}$, and $p_{\pi^+}$
are the 4-momenta of the mesons $D^+_s$, $\eta$, $\pi^0$, and
$\pi^+$ in the decay $D^+_s\to\eta\pi^0\pi^+$. Thus, the kinematic
factor $s-t$ in (109) is $(p_{D^+_s}+p_\eta,p_{\pi^0}-p_{\pi^+})$.
The amplitude $\,A_{f_0\pi}$ that is responsible for the decay
$D^+_s\to f_0(980)\pi^+\to K^+K^-\pi^+$ [see (105)] is given by
\begin{eqnarray}\label{EqVI-1-9}
A_{f_0\pi}\equiv A_{f_0\pi}(m^2_{K^+K^-})\equiv A_{f_0\pi}
(s)\qquad\nonumber\\ =C_{D^+_s f_0\pi^+}\,\frac{1}{D_{f_0}(s)
}\sqrt{\frac{g^2_{f_0K^+K^-}}{16\pi }}\,.\qquad
\end{eqnarray}
Each invariant amplitude $C_{D^+_s\eta\rho^+}$ and $C_{D^+_s
f_0\pi^+}$ is described by a pair of real numbers, the absolute
value and phase, which do not depend on mass variables:
$C_{D^+_s\eta\rho^+}=a_1e^{i\varphi_1}$ and $C_{D^+_s f_0\pi^+}=
a_2e^{i\varphi_2}$. A similar approximation for the amplitudes of
heavy-quarkonium decays containing light resonances in intermediate
states is widely used for fitting distributions of Dalitz plot
events (see, e.g., \cite{BaBar11,BaBar09,CLEO09}). It is
specifically this approximation that we use to obtain estimates.

As follows from (105) and (107), $$\frac{|C_{D^+_s f_0\pi^+}|}{
|C_{D^+_s \eta\rho^+}|}=\frac{a_2}{a_1}\approx4.5\mbox{ GeV}.$$ The
$\eta\pi^0$ and $\pi^0\pi^+$ mass spectra for the decay
$D^+_s\to\eta\pi^0\pi^+$, which correspond to a noncoherent sum of
contributions from two mechanisms, $D^+_s\to\eta\rho^+\to
\eta\pi^0\pi^+$ and $D^+_s\to \left[f_0(980)\to(K^+K^-+K^0\bar
K^0)\to a^0_0(980)\right] \pi^+\to\eta\pi^0\pi^+$, are plotted in
Figs 22a and b taking this relation into consideration. The
emergence of the sharp peak with a width $\approx2(m_{K^0}-
m_{K^+})\approx8$ MeV at $m_{\eta\pi^0}$ in the region of the
$K^+K^-$ and $K^0\bar K^0$ thresholds in Fig. 22a is due to the
$\eta\pi^0$-production mechanism owing to $a^0_0(980)-f_0(980)$
mixing. Figures 22c and d show, as an example, the distributions of
approximately $10^4$ Monte Carlo events on $s-u$ and $s-t$ Dalitz
diagrams for the decay $D^+_s\to\eta\pi^0\pi^+$, which also
correspond to the hypothetical case of noncoherent addition of the
two mechanisms. As follows from Eqn (109), the $s$-$u$ and $s$-$t $
distributions for the mechanism $D^+_s\to\eta\rho^+\to\eta\pi^0
\pi^+$ vanish on the dotted lines $u=m^2_{D_s}+2m^2_\pi+m^2_\eta
-2s$\, and \,$t=s$ respectively plotted in Figs 22c and d. Half of
the $D^+_s\to\eta\rho^+\to\eta\pi^0\pi^+ $ events are located to the
left and half to the right of these lines. The events that are due
to the $a^0_0(980)-f_0(980)$ mixing concentrate in $s-u$ and $s-t$
Dalitz diagrams at $s=m^2_{\eta\pi^0}\approx4 m^2_K$; they
constitute about 1\% of the half of the $D^+_s\to\eta\rho^+\to\eta
\pi^0\pi^+$ events [see (106) and (107)]. This value is too high for
the noncoherent isospin-violating contribution. As was noted above,
it would be natural to expect that contributions of that type to the
reaction amplitude are of the order of $(m_d-m_u)/\bar m$, where
$m_d$ and $m_u$ are the masses of constituent quarks, $\bar
m=(m_d+m_u)/2$ [or of the order of the electromagnetic constant
$\alpha=e^2/4\pi$], and hence of the order of $10^{-4}$ to the
absolute value of the amplitude squared.

\begin{figure}
\includegraphics[width=8.8cm]{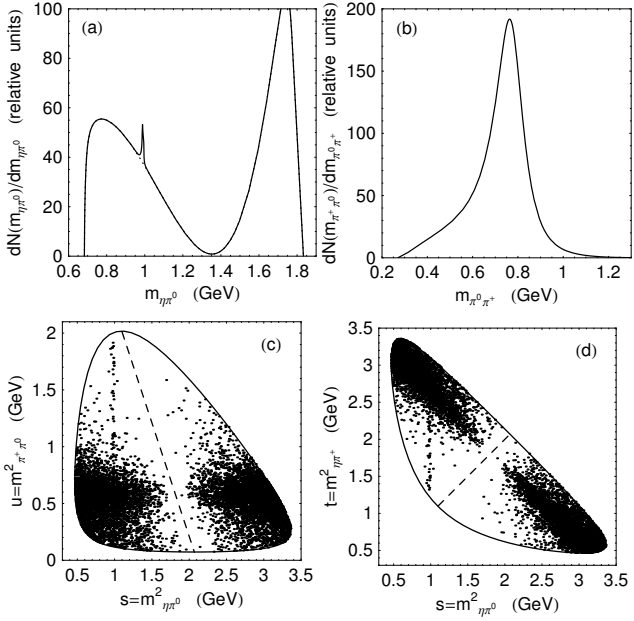} 
\caption{\label{FigDsDP1} Illustration of a manifestation of
$a^0_0(980)-f_0(98 0)$ mixing in the decay $D^+_s\to\eta\pi^0\pi^+$
on the background of its main mechanism
$D^+_s\to\eta\rho^+\to\eta\pi^0\pi^+$ in the case of noncoherent
addition of the contributions from $D^+_s\to\eta
\rho^+\to\eta\pi^0\pi^+$ and $D^+_s\to\left[f_0(980)\to(K^+
K^-+K^0\bar K^0)\to a^0_0(980)\right]\pi^+\to\eta\pi^0\pi^+$. (a, b)
Mass spectra of the $\eta\pi^0$ and $\pi^0\pi^+$ systems in the
decay $D^+_s\to\eta\pi^0 \pi^+ $. (c, d) Examples of Monte Carlo
distributions of the $D^+_s\to\eta\pi^0\pi^+$ decay events in $s-u$
and $s-t$ Dalitz diagrams, respectively.}\end{figure}

\begin{figure}
\includegraphics[width=9.0cm]{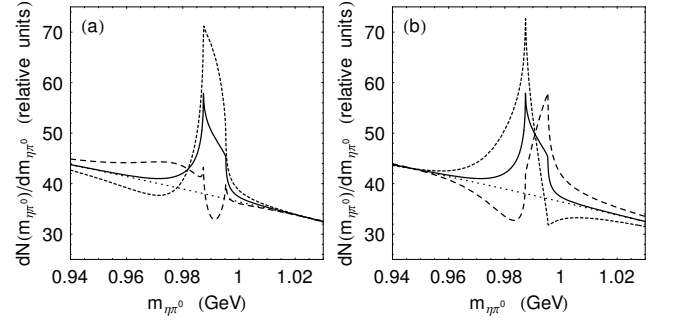}
\caption{\label{FigDs-3} $\eta\pi^0$ mass spectra in the region of
the $K^+K^-$ and $K^0\bar K^0$ thresholds for four versions of
interference between the amplitudes $D^+_s\to\eta \rho^+\to \eta
\pi^0\pi^+$ and $D^+_s\to \left[f_0(980)\to(K^+K^-+K^0\bar K^0)\to
a^0_0(980)\right]\pi^+ \to\eta\pi^0\pi^+$. These versions are
compared with the spectrum in the noncoherent case. All the curves
are described in the text.}\end{figure}

The mass spectrum of the $\eta\pi^0$ system in the decay
$D^+_s\to\eta\pi^0\pi^+$ in the case of coherent addition of the
amplitude $A_{\eta \rho^+}$ of the $D^+_s\to\eta \rho^+\to\eta\pi^0
\pi^+$ transition and the amplitude $A_{f_0a^0_0}$ due to the
$a^0_0(980)-f_0(980)$ mixing has the form
\begin{equation}\label{EqVI-1-10}
\frac{dN_{\eta\pi^0}}{dm_{\eta\pi^0}}=\int 
|A_{\eta\rho^+}+A_{f_0a^0_0}|^2\,2m_{\eta\pi^0}dm^2_{\pi^0\pi^+}\,,
\end{equation}
where integration is performed over the physical domain of the
variable $m^2_{\pi^0\pi^+}=u$. The relative phase of the amplitudes
$C_{D^+_s f_0\pi^+}$ and $C_{D^+_s \eta\rho^+}$ becomes of
importance here: $\varphi_{21}=\varphi_2-\varphi_1$ [or the
parameter $\xi=e^{i\varphi_{21}}$]. This phase is not known, and to
illustrate possible outcomes of interference between the amplitudes
$A_{\eta\rho^+}$ and $A_{f_0a^0_0}$, we set it equal to $0^\circ$,
$\pm90^\circ$, and $180^\circ$ (correspondingly, $\xi\,=\,1$, $\pm
i$ and $-1$). The short and long dashed lines in Fig. 23a show the
version of the $\eta\pi^0$ mass spectrum in the region of the
$K^+K^-$ and $K^0\bar K^0$ thresholds that correspond to the
respective values $\xi=1$ and $\xi=-1$; the dotted line shows the
contribution of the amplitude $A_{\eta\rho^+}$ alone, and the solid
curve represents the case of noncoherent addition of the two
mechanisms that was discussed above. The dotted and solid lines in
Fig. 23b show the same as in Fig. 23a, and the short and long dashed
curves display the versions that correspond to the respective values
$\xi=i$ and $\xi=-i$. It is clear that the interference of the
amplitude $A_{f_0a^0_0}$ with other contributions virtually always
manifests itself owing to rapid change of the phase of the
$a^0_0(980)-f_0(980)$ transition in the region between the $K^+K^-$
and $K^0\bar K^0$ thresholds (Fig. 2b).

As was noted in \cite{AS17a}, the decay $D^+_s\to\eta\pi^0\pi^+$ can
occur not only via the $\eta\rho^+$ intermediate state or owing to
the $a^0_0(980)-f_0(980)$ mixing but also via the intermediate state
$(a_0(980)\pi)^+$: $D^+_s\to[ a^+_0(980)\pi^0+ a^0_0(980)
\pi^+]\to\eta\pi^0\pi^+$. We can expect, however, that the
probability of that transition is not large. Using the data in (107)
and (108) as a very rough (upper) estimate, we find
$BR(D^+_s\to(a_0(980)\pi)^+\to\eta\pi^0\pi^+) \approx1\%
$.\,\footnote{The upper bound of $BR(D^+_s\to a^0_0(980)\pi^+\to
K^+K^-\pi^+)$ is, in this case, $\approx0.1\%$. It is noteworthy
that this estimate agrees with the initial assumption about the
dominance of the $D^+_s\to f_0(980)\pi^+\to K^+K^-\pi^+$ transition
involving the $f_0(980)$ resonance \cite{BaBar11, BaBar09,CLEO09}
[see Eqn (105)].} The case of three different mutually interfering
mechanisms of the decay $D^+_s\to\eta\pi^0 \pi^+$ seems to be more
realistic in principle. The corresponding combined amplitude of the
decay is
\begin{equation}\label{EqVI-1-11}
A_{D^+_s\to\eta\pi^0\pi^+}=A_{\eta\rho^+}+A_{ f_0a^0_0}+A_{a_0\pi
}\,, \end{equation} where $A_{a_0\pi}$ is the amplitude of the
transition $D^+_s\to(a_0(980)\pi)^+\to\eta\pi^0\pi^+$, which
[similarly to the amplitude $A_{\eta\rho^+}$; see (109)] is (in the
isotopic-invariance approximation) antisymmetric with respect to
permutations of $s$ and $t$ \cite{AS17a}. Given this feature, the
amplitude $A_{a_0\pi}$ can be naturally approximated using the
formula
\begin{eqnarray}\label{EqVI-1-12} A_{a_0\pi}\equiv A_{a_0\pi}(m^2_{\eta
\pi^0},m^2_{\eta\pi^+})\equiv A_{a_0\pi}(s,t)\qquad\nonumber\\
=C_{D^+_sa^0_0\pi^+}\,\left[\frac{1}{D_{a^0_0}(s)}-\frac{1}{D_{a^+_0}
(t)}\right]\sqrt{\frac{g^2_{a_0\eta\pi^0}}{16\pi}}\,,\quad
\end{eqnarray} where the amplitude $C_{D^+_sa^0_0\pi^+}=a_3e^{i
\varphi_3}$ is assumed to be a complex constant independent of $s$
and $t$. We note that on the $s-t$ Dalitz diagram, any coherent sum
of the amplitudes $A_{\eta\rho^+}$ and $A_{a_0\pi}$ yields a
distribution of $\eta\pi^0\pi^+$ events that is symmetric with
respect to the $t=s$ line (and vanishes on the line itself.)
Therefore, any asymmetry in the distribution of $\eta\pi^0\pi^+$
events in the $s-t$ Dalitz diagram (with respect to the $t=s$ line)
is only due to the $A_{f_0a^0_0}=A_{f_0 a^0_0}(s)$ amplitude, which
depends on $s$ alone, manifests itself in the region of the $K\bar K
$ thresholds, and is generated by the isospin-violating $a^0_0(980)-
f_0(980)$ mixing. An example of such an asymmetric $s-t$
distribution not involving the $A_{a_0\pi}$ amplitude is shown in
Fig. 22d. A variety of examples of $\eta\pi^0$ mass spectra and
asymmetric distributions of the $D^+_s\to\eta\pi^0\pi^+$ decay
events in the $s-t$ Dalitz diagrams, which illustrate possible
scenarios of interference among amplitudes $A_{\eta\rho^+}$,
$A_{f_0a^0_0}$, and $A_{a_0\pi}$ described by Eqns (109), (110), and
(116), are reported in \cite{AS17a}. Interested readers are referred
to that study for details.

Finding signatures of the mechanisms of the decay $D^+_s\to\left
(a^0_0(980) \pi^++a^+_0(980)\pi^0\right)\to\eta\pi^0\pi^+$ is a
challenging task for the physics of both weak hadronic decays of
$D^+_s$ mesons and light scalar $a_0(980)$ and $f_0(980)$ mesons.
Progress in these areas in the nearest future will naturally be
related to studies conducted by the LHCb, BaBar, CLEO, BESIII,
Belle, and Belle II facilities.

It is clear that the observation of $a^0_0(980)-f_0(980)$ mixing in
the decay $D^+_s\to\eta\pi^0\pi^+$ would provide independent
information about the mechanism of production of $f_0(980)$ mesons
(or the $K\bar K$ system with the isospin $I=0$ in an $S$-wave) and
would therefore facilitate elucidating its role in the decay channel
$D^+_s\to K^+K^-\pi^+$. This observation seems to be even more
important in relation to another phenomenon that has not been
explained yet. If $f_0(980)$ is produced in the $\pi^+\pi^-$ and
$K^+K^-$ channels as an isolated resonance without any background,
the CLEO result $BR(D_s^+\to\pi^+\pi^-\pi^+)=(1.11 \pm 0.04\pm
0.04)\%$ \cite{CLEO13} (see also \cite{BaBar09,PDG16}) disagrees
with the assumption made by BaBar \cite{BaBar11} and CLEO
\cite{CLEO09} that $f_0(980)$ dominates in the decay $D^+_s\to
f_0(980)\pi^+\to K^+K^- \pi^+$. We recall that the PDG's average
value is $BR(D^+_s\to f_0(980)\pi^+\to K^+K^-\pi^+)=(1.15\pm0.32)\%$
[see (105)]. Our estimates show that agreement with that value and
the shape of the line in Fig. 21 can be reproduced with $BR(D^+_s\to
f_0 (980)\pi^+\to \pi^+\pi^- \pi^+)\approx3.4\%$. The probability of
the decay of an isolated $f_0(980) $ resonance into $\pi^+\pi^-$ is
higher than that into $K^+K^-$ simply because of the $K^+K^-$ system
phase-space factor that suppresses the $f_0(980)\to K^+K^-$ decay
near the threshold. The situation undoubtedly requires further
exploration to elucidate the $f_0(980)$ resonance production
mechanisms in the decays $D_s^+\to f_0(980)\pi^+\to\pi^+\pi^-\pi^+ $
and $D^+_s\to f_0(980) \pi^+\to K^+K^-\pi^+ $.

\vspace*{0.3cm}\noindent{\boldmath\bf 6.2. Decays $D^0\to
K^0_S\pi^+\pi^-$ and $D^0\to K^0_S\eta\pi^0$}

\noindent Studies of the $a^0_0(980)-f_0(980)$ mixing in
three-particle decays of $D^0$ mesons, $D^0\to K^0_S\pi^+\pi^-$,
$D^0\to K^0_S\eta\pi^0$, $D^0\to\bar K^0K^-K^+$, $D^0\to K^-K^+\pi^0
$, and $D^0\to\pi^+\pi^-\pi^0$, are both promising and challenging.
We here analyze possible manifestations of that mixing in the decays
$D^0\to K^0_S\pi^+\pi^-$ and $D^0\to K^0_S\eta\pi^0$ \cite{AS17b}
and show that the $\pi^+\pi^-$ mass spectrum in the decay $D^0\to
K^0_S\pi^+ \pi^-$ is most strongly affected by the
$a^0_0(980)-f_0(980)$ mixing. Owing to that mixing, the shape of the
$f_0(980)$ peak in the decay $D^0\to K^0_Sf_0(980)\to
K^0_S\pi^+\pi^-$ can experience deformations comparable to its
magnitude. It is of importance that this effect significantly
depends on the relative phase of the $D^0\to K^0_Sf_0(980)$ and
$D^0\to K^0_Sa^0_0(980)$ decay amplitudes.

In the $D^0\to K^0_S \pi^+\pi^-$ and $D^0\to K^0_S\eta\pi^0$ decays,
the $\pi^+\pi^-$ and $\eta\pi^0$ mass spectra that are due to the
contribution of $f_0(980)$ and $a^0_0(980)$ resonances with the
$a^0_0(980)-f_0 (980)$ mixing taken into account are given by
\begin{eqnarray}\label{EqVI-2-1}
\frac{dN_{\pi^+\pi^-}}{dm}=2m^2P_{K^0_S}(m)\Gamma_{f_0\to\pi^+\pi^-}(m)
\ \ \nonumber\\
\times\left|\frac{C_1} {D_{f_0} (m)}+\frac{e^{i\varphi}\,C_2\,
\Pi_{a^0_0f_0}(m)} {D_{a^0_0} (m)D_{f_0}(m)-\Pi^2_{a^0_0f_0}(m)}
\right|^2\,,
\end{eqnarray}
\begin{eqnarray}\label{EqVI-2-2}
\frac{dN_{\eta\pi^0}}{dm}=2m^2P_{K^0_S}(m)\Gamma_{a^0_0\eta\pi^0}(m)
\ \ \nonumber\\
\times\left|\frac{e^{i\varphi} \,C_2}{D_{a^0_0}(m)}+\frac{C_1\,
\Pi_{a^0_0f_0}(m)}{D_{a^0_0} (m)D_{f_0}(m)-\Pi^2_{a^0_0f_0}(m)}
\right|^2\,,\end{eqnarray} where $m$ is the invariant mass of the
$\pi^+\pi^-$ or $\eta\pi^0$ system,
$$P_{K^0_S}(m)=\frac{[m^4_{D^0}-2m^2_{D^0}(m^2_{K^0}+m^2)
+(m^2_{K^0}-m^2)^2]^{1/2}}{2m_{D^0}},$$ and $\varphi$ is the
relative phase between the $D^0\to K^0_Sa^0_0(980)$ and $D^0\to
K^0_Sf_0(980)$ decay amplitudes. We have set the values of the
constants $C_1=0.047$ GeV$^{-1/2}$ and $C_2=0.095$ GeV$^{-1/2}$
given the CLEO Data \cite{CLEO02,CLEO04}, PDG information
\cite{PDG16}, and the relations
\begin{eqnarray}\label{EqVI-2-3}
BR(D^0\to K^0_Sf_0(980)\to K^0_S\pi^+\pi^-)\quad\quad\nonumber\\
=\left(1.23^{+0.40}_{-0.24}\right)\times10^{-3}\qquad\qquad\quad\nonumber\\
=\int\limits_{2m_\pi^+}^{m_{D^0}-m_{K^0}}
2P_{K^0_S}(m)\Gamma_{f_0\to\pi^+\pi^-}(m)
\left|\frac{m\,C_1}{D_{f_0}(m)}\right|^2dm,
\end{eqnarray}
\begin{eqnarray}\label{EqVI-2-4}
BR(D^0\to K^0_Sa^0_0(980)\to K^0_S\eta\pi^0)\quad\quad\nonumber\\
=\left(6.6\pm2.0\right)\times10^{-3}\qquad\qquad\quad\nonumber\\
=\int\limits_{m_\eta+m_{\pi^0}}^{m_{D^0}-m_{K^0}}
2P_{K^0_S}(m)\Gamma_{a^0_0\eta\pi^0}(m)
\left|\frac{m\,C_2}{D_{a^0_0}(m)}\right|^2dm,
\end{eqnarray}
in which the same parameters as before are used for the $f_0(980)$
and $a^0_0(980)$ resonances.

We discuss the situation with decays that are of interest to us in
more detail below. We only note here that the numerical values for
the decay probabilities used in (119) and (120) are based on limited
experimental statistics and an analysis of decay amplitudes in the
isobar model. More accurate information about these values is
definitely quite desirable. Unfortunately, in processing more recent
experimental data on the $D^0\to K^0_S\pi^+\pi^-$ decay, the BaBar
\cite{BaBar08} and Belle \cite{Belle14} collaborations (see also
\cite{BaBar10}) have not specifically separated the $f_0(980)$
resonance contribution from the entire array of $S$-wave
contributions. Nevertheless, the $f_0(980)$ peak is clearly seen in
the $\pi^+\pi^-$ mass spectrum in this decay \cite{BaBar08,
Belle14}.

We define the fractions of isospin-violating contributions to the
decays $D^0\to K^0_S\pi^+\pi^-$ and $D^0\to K^0_S\eta\pi^0$ due to
the $a^0_0(980)-f_0(980)$ mixing as
\begin{equation}\label{EqVI-2-5}
\Delta BR(\pi^+\pi^-)=\frac{1}{1.23\cdot10^{-3}}
\int\limits_{2m_\pi^+}^{m_{D^0}-m_{K^0}}
\frac{dN_{\pi^+\pi^-}}{dm}\,dm\ -1\,,
\end{equation}
\begin{equation}\label{EqVI-2-6}
\Delta BR(\eta\pi^0)=\frac{1}{6.6\cdot10^{-3}}
\int\limits_{m_\eta+m_{\pi^0}}^{m_{D^0}-m_{K^0}}
\frac{dN_{\eta\pi^0}}{dm}\,dm\ -1\,.
\end{equation}
The quantities $\Delta BR(\pi^+\pi^-)$ and $\Delta BR(\eta\pi^0)$ as
functions of the relative phase $\varphi$ between the $D^0\to
K^0_Sa^0_0(980)$ and $D^0\to K^0_Sf_0(980)$ decay amplitudes are
shown in Fig. 24 by respective solid and dashed lines. The
horizontal solid and dashed lines in the same figure show the
noncoherent contributions to $\Delta BR(\pi^+\pi^-)$ and $\Delta
BR(\eta\pi^0)$ from the $a^0_0(980)-f_0(980)$ mixing, i.e., the
contributions due to the absolute values squared of the second
summands in (117) and (118), which amount to $\approx1.7\%$ in the
$\pi^+\pi^-$ channel and $\approx0.17\%$ in the $\eta\pi^0$ channel.

As follows from Fig. 24, the maximum constructive (destructive)
interference of the contribution from $a^0_0(980)-f_0(980) $ mixing
in the $\pi^+\pi^-$ channel corresponds to the phase $\varphi\approx
245^\circ$ ($70^\circ$), and the maximum constructive (destructive)
interference of the contribution from $a^0_0(980)-f_0 (980) $ mixing
in the $\eta\pi^0$ channel corresponds to the phase $\varphi\approx
110^\circ$ ($290^\circ$). Figure 25 shows the $dN_{\pi^+\pi^-}/dm$
and $dN_{\eta\pi^0}/dm$ mass spectra for these limit interference
cases.

\begin{figure}
\includegraphics[width=6cm]{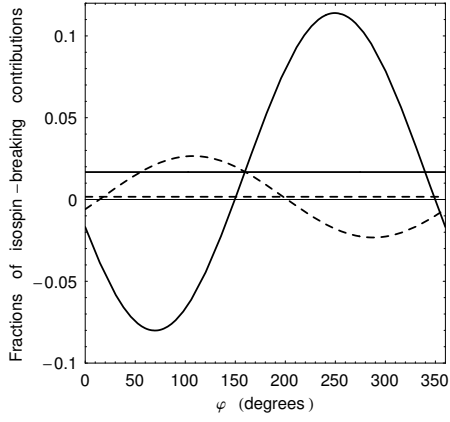}
\caption{\label{FigD0-1} Dependences of the isospin-violating
contributions $\Delta BR(\pi^+\pi^-)$ (solid curve) and $\Delta
BR(\eta\pi^0)$ (dashed curve) on the phase $\varphi$. The solid and
dashed horizontal lines show the respective noncoherent
contributions from the $a^0_0(980)-f_0(980)$ mixing in the
$\pi^+\pi^-$ and $\eta\pi^0$ channels.}\end{figure}

\begin{figure}[!ht]
\includegraphics[width=9cm]{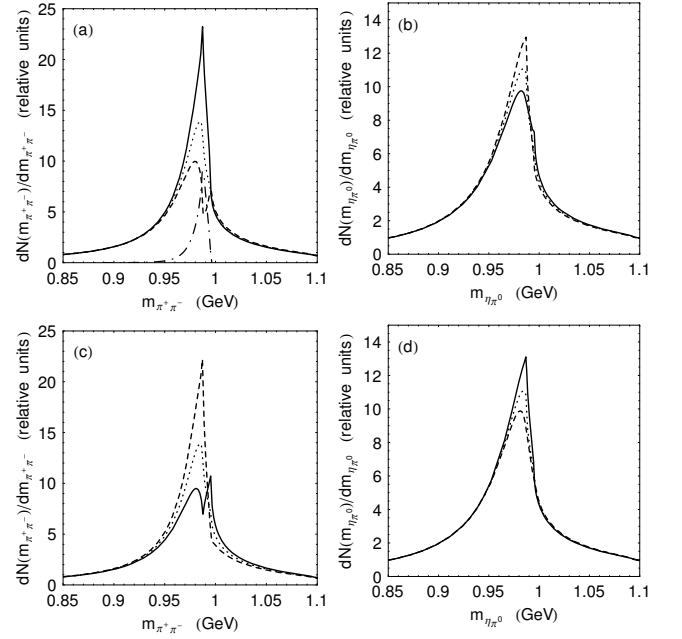}
\caption{\label{FigD0-2} Mass spectra of $\pi^+\pi^-$ (a) and
$\eta\pi^0$ (b) at $\varphi=245^\circ$ (solid curves) and
$\varphi=70^\circ$ (dashed curves); the dotted lines in Figs a and b
correspond to spectra without the $a^0_0(980)-f_0(980)$ mixing,
i.e., contributions from the respective $f_0(980)$ and $a^0_0(980)$
resonances. (c, d) The same as in Fig. a and b but for
$\varphi=110^\circ$ (solid curves) and $\varphi=290^\circ$ (dashed
curves). The dashed-dotted curve that corresponds to the difference
between the values of the solid and dashed curves is shown as an
example in Fig. a.}\end{figure}

The branching ratio $BR(D^0\to K^0_Sa^0_0(980)\to K^0_S\eta\pi^0)$
being approximately five times larger than $BR(D^0\to
K^0_Sf_0(980)\to K^0_S\pi^+\pi^-)$ [see (119) and (120)] results in
an enhancement of the $a^0_0(980)-f_0(980)$ mixing effect in the
$\pi^+\pi^-$ channel compared with this effect in the $\eta\pi^0$
channel. Owing to the interference, the integral effect of the
$a^0_0(980)-f_0(980)$ mixing in the $\pi^+\pi^-$ channel can be as
high as $\approx11\%$ (see Fig. 24), a value that is too large for
an isospin-violating effect. We can see from Fig. 25 that in the
$\pi^+\pi^-$ mass spectrum, the $a^0_0(980)-f_0(980)$ mixing can
result in a significant deformation of the shape of the $f_0(980)$
resonance peak, for example, narrowing by a factor of approximately
1.5, and increasing its height by up to 60\%, or even in the
emergence of two peaks. The effect significantly depends on the
relative phase of the $D^0\to K^0_Sf_0(980)$ and $D^0\to K^0_S
a^0_0(980)$ decay amplitudes. To detect these fine phenomena, high
resolution in the invariant mass m and large statistics of events
are certainly needed.

We now discuss the experimental situation \cite{PDG16,CLEO04,
BaBar05,CLEO02,BaBar08,Belle14,BaBar10}. The only experiment where
the $D^0\to K^0_S\eta\pi^0$ decay has been observed was performed by
the CLEO collaboration \cite{CLEO04} with statistics from 155
events. An analysis of the distribution of these events in the
Dalitz diagram showed in \cite{CLEO04} that the $D^0\to
K^0_S\eta\pi^0$ decay primarily occurs via the $K^0_S a^0_0(980)$
and$K^{*0}(892) \eta$ intermediate states, of which the former
dominates. It is noteworthy that $a^0_0(980)$ production was
observed not only in the decay $D^0\to K^0_Sa^0_0 (980)\to
K^0_S\eta\pi^0$ (for which $BR(D^0\to K^0_S a^0_0(980)\to
K^0_S\eta\pi^0)=(6.6\pm2.0) \times10^{-3}$ \cite{CLEO04,PDG16}) but
also in the channel $D^0\to K^0_S a^0_0(980)\to K^0_SK^+K^-$
\cite{BaBar05,BaBar10}. According to data in \cite{BaBar05}, $BR(D^0
\to K^0_Sa^0_0(980)\to K^0_SK^+ K^-)=(3.0\pm0.4)\times 10^{-3}$,
while it follows from the supplementary material in \cite{BaBar10}
(see Ref. [18] there) that the central value is $BR(D^0\to
K^0_Sa^0_0 (980)\to K^0_SK^+ K^-)\approx 2.3\times10^{-3}$. It is
noteworthy that the quoted branching ratios of the decays $D^0\to
K^0_Sa^0_0(980)\to K^0_S\eta\pi^0$ \cite{PDG16} and $D^0\to
K^0_Sa^0_0(980) \to K^0_SK^+ K^-$ agree with the $q^2\bar q^2$ model
of the $a^0_0(980)$ resonance \cite{ADS81,AI89,Ac98}.

The number of candidates for the events of the $D^0\to K^0_S\pi^+
\pi^-$ decay selected in the CLEO \cite{CLEO02}, BaBar
\cite{BaBar08}, and Belle \cite{Belle14} experiments was 5,299,
487,000, and 1,231,731, respectively. The Dalitz distributions of
the $D^0\to K^0_S\pi^+\pi^- $ events exhibit a rich structure. The
list of possible intermediate states includes $K^{*-}(982)\pi^+$,
$K^{*-}(1430)\pi^+$, $K^0_S\rho^0$, $K^0_Sf_0(980)$, $K^0_Sf_2(1270
)$, and $K^0_Sf_0(1370)$. An estimate in \cite{CLEO02} shows that
the contribution of the $K^0_S f_0(980)\to K^0_S\pi^+\pi^-$ channel
is $(4.3^{+1.4}_{-0.8})\%$. In total, production of $K^0_S$ together
with the $S$-wave $\pi^+\pi^-$ system, $K^0_S(\pi^+ \pi^-)_S$,
yields about 12\% of the $D^0\to K^0_S \pi^+\pi^-$ decay probability
\cite{BaBar08,Belle14}. The $\pi^+\pi^-$ invariant-mass step in the
mass region of 1 GeV was about 5 MeV in the BaBar \cite{BaBar08} and
Belle \cite{Belle14} experiments. It is noteworthy that the entire
visible $f_0(980)$ peak contains 6 to 7 points, implying that its
width is less than 25 MeV. The narrowness of the $f_0(980)$ peak can
be related to the effect of $a^0_0(980)-f_0(980)$ mixing. However,
interference with the background from other intermediate states
certainly cannot be ruled out a priori. Future studies will
hopefully resolve this issue. The most clear-cut information about
events corresponding to the $D^0\to K^0_S f_0(980)\to K^0_S\pi^+
\pi^-$ channel and the possible influence of background events on
that channel comes from the distribution of $D^0\to K^0_S\pi^+\pi^-$
events in the ($m^2_{\pi^+\pi^- },\,m^2_{K^0_S\pi^\pm}$) Dalitz
diagram.

\vspace{0.3cm} \noindent{\large\bf\boldmath 7. Bottomonium decay
$\Upsilon(10860)\to\Upsilon(1S)f_0(980)\to\Upsilon(1S)\eta\pi^0$}
\vspace{0.2cm}

\noindent The Belle collaboration has recently performed a complete
amplitude analysis of three-particle transitions $e^+e^-\to\Upsilon
(nS)\pi^+\pi^-$ ($n = 1, 2, 3$) for the energy $\sqrt{s}=10.865$ GeV
in the $e^+e^-$ c.m.s. and determined the relative weight of
different quasi-two-particle components of the three-particle
amplitudes and the spin and parity of two observed $Z_b$ states
\cite{Ga15}.\,\footnote{We are grateful to A E Bondar', who directed
our attention to the decay $\Upsilon(10860)\to\Upsilon(1S)f_0(980)
\to\Upsilon(1S)\eta\pi^0$ and data from the Belle collaboration
\cite{Ga15}.} The first data on the transition
$e^+e^-\to\Upsilon(1S)f_0 (980)$ have also been reported. According
to \cite{Ga15}, the fraction of the decay $\Upsilon(10860)\to
\Upsilon(1S)f_0(980)$ is
\begin{eqnarray}\label{EqVII-1}
&& \frac{BR(\Upsilon(10860)\to\Upsilon(1S)f_0
(980)\to\Upsilon(1S)\pi^+\pi^-)}{BR(\Upsilon(10860)\to\Upsilon(1S)
\pi^+\pi^-)}\nonumber\\ &&
\hspace*{2.2cm}=\left(6.9\pm1.6^{+0.8}_{-2.8}\right)\%
\,.\end{eqnarray}

Apart from decaying into $\pi^+\pi^-$, the resonance $f_0(980)$ can
also decay into $\eta\pi^0$ via the transition $$f_0(980)\to(K^+K^-+
K^0\bar K^0)\to a^0_0(980)\to\eta\pi^0,$$ i.e., owing to the
$a^0_0(980)-f_0(980)$ mixing. Guided by the central value of the
strength of the $f_0(980)\to a^0_0(980)$ transition measured in the
BES III experiment for the reaction $J/\psi\to\phi f_0(980)\to\phi
a^0_0(980)\to\phi\eta\pi^0$ [see (34)], i.e., assuming that
\begin{eqnarray}\label{EqVII-2}
\frac{BR(f_0(980)\to K\bar K\to a^0_0(980)\to\eta\pi^0)}
{BR(f_0(980)\to\pi^+\pi^-)}\approx0.009,
\end{eqnarray}
we arrive at the following estimate for the fraction of isospin-
violating decays of the $\Upsilon(10860)$ meson \cite{AS17c}:
\begin{eqnarray}\label{EqVII-3}
&& \frac{BR(\Upsilon(10860)\to\Upsilon(1S)f_0(980)\to\Upsilon
(1S)\eta\pi^0)} {BR(\Upsilon(10860)\to\Upsilon(1S)\pi^+\pi^-)}\ \
\nonumber\\ && \hspace*{2.5cm}\approx6.2\times10^{-4}\,.
\end{eqnarray}

We note that the $f_0(980)$ resonance in the $\pi^+\pi^-$ mass
spectrum in the decay $\Upsilon(10860)\to\Upsilon(1S)\pi^+\pi^-$ was
observed in \cite{Ga15} not as a peak but as a deep dip due to
destructive interference with a large and smooth background. The
dominance of a narrow resonance peak in the $\eta\pi^0$ mass
spectrum in the region of $K\bar K$ thresholds is supposed to be a
characteristic feature of the decay $\Upsilon(10860)\to\Upsilon
(1S)f_0(980)\to\Upsilon (1S)\eta\pi^0$ because there is no obvious
background in the $\eta\pi^0$ channel of the decay
$\Upsilon(10860)\to\Upsilon(1S) \eta\pi^0$. The $\eta\pi^0$ spectrum
is \cite{AS17c}
\begin{eqnarray}\label{EqVII-4}
&& \hspace*{-0.45cm}\frac{dN(\Upsilon(5S)\to\Upsilon(1S)\eta\pi^0)}
{dm}=Cp(m)\frac{ 2m^2\Gamma_{a^0_0\to\eta\pi^0}(m)}{\pi} \nonumber\\
&& \qquad\ \
\times\left|\frac{\Pi_{a^0_0f_0}(m)}{D_{a^0_0}(m)D_{f_0}(m)-
\Pi^2_{a^0_0f_0}(m)}\right|^2,
\end{eqnarray}
where $\Upsilon(5S)$ is a shorthand notation for $\Upsilon(10860)$,
$m$ is the invariant mass of the $\eta\pi^0$ system, $p(m)$ is the
$\eta\pi^0$ momentum in the $\Upsilon(5S)$ rest frame, and $C$ is a
normalization constant. The momentum $p(m)$ is weakly dependent on
$m$ in the mass range $m\sim1$ GeV in the decay $\Upsilon(5S)\to
\Upsilon(1S)\eta\pi^0$, and therefore the spectrum virtually
coincides with the $\eta\pi^0$ mass spectrum shown by the solid
curve in Fig. 3.

The number of events selected as candidates for the decay $\Upsilon
(10860)\to\Upsilon(1S)\pi^+\pi^-$ in the Belle experiment under
discussion in \cite{Ga15} was $2,090\pm115$. Thus, one or two events
corresponding to the decay $\Upsilon(10860)\to\Upsilon(1S)\eta
\pi^0$ could be produced owing to the $a_0(980)-f_0(980)$ mixing in
the course of that experiment. One may hope that the Belle II
detector at the $e^+e^-$ collider SuperKEK B to be launched shortly
will allow measuring this rare decay with an accuracy of no worse
than 10 percent. Given the instantaneous luminosity $8\times10^{35}$
cm$^{-2}$\,s$^{-1}$ of the SuperKEK B collider (40 times larger than
that of KEK), the Bell detector will be able to record approximately
100 $\Upsilon(10860)\to \Upsilon(1S)\eta\pi^0$ events (i.e., from 50
to 150 events) in a new experiment in a narrow region of the
$\eta\pi^0$ invariant mass near the $K\bar K$ thresholds within a
time-frame that was needed for the initial experiment of the Belle
collaboration \cite{Ga15}.\,\footnote{We estimate that about three
months was needed to collect the statistics in the experiment
\cite{Ga15}.} This implies that the anomalous violation of isotopic
invariance will enable studying the production mechanisms and nature
of light scalar mesons also in the bottomonium domain.

In conclusion, we make a comment that is primarily addressed to
experimentalists. In obtaining result (123) for the relative
fraction of the decay $\Upsilon(10860)\to\Upsilon(1S)f_0(980)
\to\Upsilon(1S)\pi^+\pi^-$, the Belle collaboration \cite{Ga15} used
the Flatt\'{e} propagator for the $f_0(980)$ resonance
\cite{Fl72,Fl76},
\begin{eqnarray}\label{EqVII-5}
\frac{1}{D_{f_0}(m)}=\frac{1}{M_{f_0}^2-m^2-iM_{f_0}(g_{\pi
\pi}q_\pi+g_{K\bar K}q_K)},
\end{eqnarray} where $q_\pi=\sqrt{m^2/4-m^2_\pi}$\, and
\[ q_K=\left\{\begin{array}{cl}
\sqrt{m^2/4-m^2_K}\,,\quad & \mbox{above the $K\bar K$ threshold},\\
i\,\sqrt{m^2_K-m^2/4}\,,\quad & \mbox{below the $K\bar K$
threshold}, \end{array}\right. \] with the mass $M_{f_0}=950$ MeV
and constants $g_{\pi\pi}=0.23$ and $g_{K\bar K}=0.75$ obtained in
\cite{Ga06} from the analysis of data on the $B^\pm\to K^\pm\pi^\pm
\pi^\pm$ decay. The mass $M_{f_0}=950$ MeV, which is 30--40 MeV
smaller than the value in tables \cite{PDG16}, may be perplexing.
However, we should take into account that at $M_{f_0}<2m_K$, the
value of $M^2_{f_0}$ in the Flatt\'{e} formula does not coincide
with the location of a zero of the real part of the resonance
propagator, in contrast to the case of the resonance mass squared if
the generally accepted definition is used [see, e.g., the definition
of $m_{f_0}$ in Eqn (5)]. The value of $m_{f_0}$ as the location of
a zero of Re$D_{f_0}(m)$ should be determined from the equation
\begin{eqnarray}\label{EqVII-5}
\mbox{Re}\left(M_{f_0}^2-m^2+M_{f_0}g_{K\bar K}\sqrt{
m^2_K-\frac{m^2}{4}}\right)=0. \end{eqnarray} We find
$m_{f_0}\approx 979$ MeV from this equation. Figure 26 clearly shows
that $m_{f_0}$ is shifted to values larger than $M_{f_0}$.
\begin{figure} 
\includegraphics[width=6.5cm]{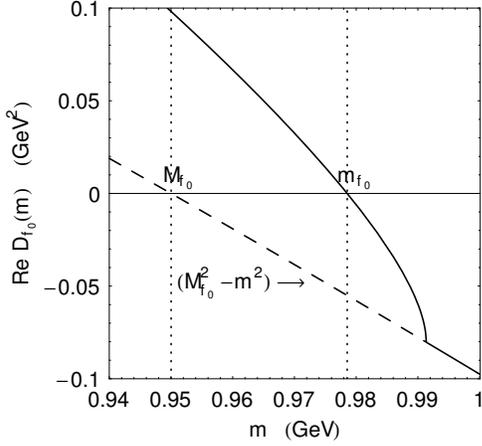}
\caption{\label{FigY-1} Illustration $m_{f_0}$ shift to values
larger than $M_{f_0}$. The solid curve shows $\mbox{Re}D_{f_0}
(m)=\mbox{Re}\left(M^2_{f_0} -m^2+M_{f_0}\,g_{K\bar K}(
m^2_K-m^2/4)^{1/2}\right)$ as a function of $m$.}\end{figure}

It has been noted many times that the Flatt\'{e} propagators are not
satisfactory for studying $f_0(980)$ and $a_0(980)$ resonances
\cite{AG95,AG97,AGShev97,AKi04}. If these propagators are employed
for the $f_0(980)$ and $a_0(980)$ states with masses below the
$K\bar K$ thresholds (see examples in \cite{Ga15,Ga06,BaBar11,
Ab05,Bon07,PDG16}), the resonance masses found by fitting must be
renormalized. Our propagators [see (5)--(9)], which were first
proposed in \cite{ADS79}, are not plagued with this problem.
Therefore, we once again recommend using them to determine the
parameters of $f_0(980)$ and $a_0(980)$ mesons.

\vspace{0.3cm} \noindent{\large\bf\boldmath 8. Reactions violating
isotopic invariance in the central region} \vspace{0.2cm}

\noindent The exclusive reactions of hadron production in the
central region\,\footnote{The value of the Feynman variable
$x_F\approx0$ can be considered a common feature of the particles
produced in the central region.} of high-energy $pp$ collisions,
$pp\to p(X^0)p$, were studied at the ISR (Intersecting Storage
Rings) and $Sp\bar pS$ (Super Proton--Antiproton Synchrotron) at
CERN and the Tevatron at the Fermi National Accelerator Laboratory
(Fermilab) using a fixed target, and are currently being explored at
the Large Hadron Collider (LHC) at CERN (see, e.g., reviews
\cite{KZ07,ACF10,Kir14,GuRe14,CMS17}). Mass spectra and production
cross sections have been measured in these experiments for a number
of hadronic systems $X^0$: $\pi\pi$ \cite{CMS17,Ake86,
AMP87,Arm91,Bar99b,Bar99c}, $K\bar K$ \cite{Arm91,Rey98,Bar99,
Bar99a}, $\eta\pi^+\pi^-$ \cite{Arm91a,Bar98}, $K\bar K\pi$
\cite{Bar97,Sos99}, $4\pi$ \cite{Bar00,Bar00a}, $\eta\pi^0$
\cite{Bar00b}, etc. Special attention was paid to studying resonance
contributions.
\begin{figure} [!ht]
\includegraphics[width=5.0cm]{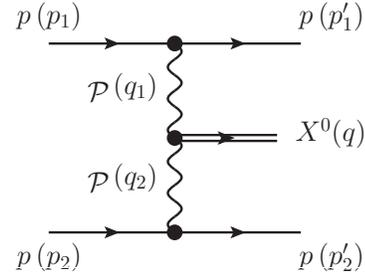}
\caption{\label{Fig20} Production of the $X^0$ hadron system in the
central region in the reaction $pp\to p(X^0)p$ due to the double
pomeron exchange, $\mathcal{P}\mathcal{P}$. Indicated in parentheses
are the 4-momenta of the initial and final protons,
$\mathcal{P}$-exchanges, and the $X^0$-system; the main kinematic
variables in this reaction are $s=(p_1+p_2)^2$, $s_1=(p'_1+q)^2$,
$s_2=(p'_2+q)^2$, $M^2=q^2=(q_1+q_2)^2$, and $t_1=q^2_1$,
$t_2=q^2_2$.}\end{figure}

Resonances are produced in the central region at high energies
primarily due to the double pomeron exchange (Fig. 27). Only
resonances with positive $C$-parity and the isotopic spin $I=0$ can
be produced in a two-pomeron collision. The cross sections of
processes driven by that mechanism do not decrease as a power of
energy \cite{KZ07,ACF10,AlGo81,GaRo80,Str86}. Therefore, the
observation of resonances in $X^0$ states with $I=1$  is a signal
that they are produced or decay with isotopic invariance violation.
Below, we present examples where such situations can occur
\cite{AS18b}.

\vspace*{0.3cm} \noindent{\boldmath\bf 8.1. Reactions $pp\to
p(f_1(1285)/f_1(1420))p\to p(\pi^+\pi^-\pi^0)p$}

\noindent Clear-cut signals from $f_1(1285)$ and $f_1(1420)$
resonances (with $I^G(J^{PC})=0^+(1^{++})$) produced in the central
region of the reaction $pp\to p(f_1(1285)/f_1(1420))p\to p(X^0)p$
have been observed in all of their main decay modes: $\eta\pi\pi$
\cite{Arm91a,Bar98}, $K\bar K\pi$ \cite{Bar97,Sos99,GuRe14}, and
$4\pi$ \cite{Bar00}. Experiments were performed at the momenta
$P^p_{lab}=85$, 300, 450, and 800 GeV$/c$ of the protons incident on
a fixed target, or, correspondingly, at the full energy
$\sqrt{s}=12.7$, 23.8, 29, and 40 GeV in the reaction center-of-mass
system. Data on the production cross sections of these resonances
are compatible with the $\mathcal{P}\mathcal{P}$-exchange mechanism
\cite{Arm91a,Bar98,Bar97,Sos99,Bar00,Kir14,GuRe14}. An important
complementary experimental observation is that the central-region
production of states with $I^G(J^{PC})=0^+(0^{-+})$ and masses
around 1.28 and 1.4 GeV was fully suppressed \cite{Arm91a,Bar98,
Bar97,Sos99,Bar00,Kir14,GuRe14}. In practical terms, this
circumstance may be helpful in determining the properties of the
$f_1(1285)$ and $f_1(1420)$ resonances more accurately than in other
experiments where states with both $J^{PC}=1^{++}$ and
$J^{PC}=0^{-+}$ are concurrently observed.

Thus, in studying the production of $f_1(1285)$ and $f_1(1420)$ in
the central region in $pp$ collisions, it is possible to determine
the probabilities of decays of the two resonances into all of their
main modes in a single experiment. We also note that due to the
isotopic neutrality of $\mathcal{P}\mathcal{P}$ exchange, the
reaction $pp\to p(f_1(1285))p\to p(\pi^+\pi^-\pi^0)p$ provides a
unique possibility of studying the isospin-violating decay
$f_1(1285)\to f_0(980)\pi^0\to\pi^+\pi^-\pi^0$ in a setup that is
virtually free of any background \cite{AS18b}. Owing to a
substantially different arrangement of the experiment, such a study
would be a very efficient test of the first results from VES
\cite{Do11} [see (32) and (33)] and BESIII \cite{Ab3} [see (52)],
which indicate a very strong violation of isospin invariance in that
decay. A search for isospin-violating events $pp\to
p(\pi^+\pi^-\pi^0)p$ in the vicinity of the $f_1(1420)$ resonance is
also of interest.

Regarding options to measure the reaction $pp\to p(f_1 (1285))p\to
p(\pi^+\pi^-\pi^0)p$, we note the following. In principle, one could
use the data recorded by the Omega spectrometer at CERN and the CDF
detector (Collider Detector at Fermilab) to extract information
about the events $pp\to p(\pi^+ \pi^-\pi^0)p$ in the central region.
However, enthusiasts are needed for this task, because the
facilities themselves were decommissioned long ago. The reaction
$pp\to p(f_1 (1285))p\to p(\pi^+\pi^-\pi^0)p$ could currently be
measured at the LHC using the CMS (Compact Muon Solenoid) detector.
The CMS collaboration has recently reported data on $\pi^+\pi^-$
production in the central region in $pp$ collisions at $\sqrt{s}=7$
TeV \cite{CMS17}. At an energy this immense, the energies
$\sqrt{s_1}$ and $\sqrt{s_2}$ of the subprocesses
$p(p_1)\mathcal{P}(q_2) \to p(p'_1)X^0(q)$ and
$p(p_2)\mathcal{P}(q_1)\to p(p'_2)X^0(q)$ (see Fig. 27) are also
very large (and correspond to the energy range where exchange by
secondary Regge trajectories, $\mathcal{R}$, can de facto be
disregarded compared to the $\mathcal{P}$ exchange). Setting
$s_1\approx s_2$, $M\approx1$ GeV, and $\sqrt{s}=7$ TeV and using
the relation $s_1s_2\approx M^2s$ (which holds for processes in the
central region \cite{AlGo81,GaRo80,Str86}), we obtain
$\sqrt{s_1}\approx \sqrt{s_2}\approx84$ GeV. Thus, the dominance of
the $\mathcal{P} \mathcal{P}$ exchange in LHC experiments is a very
good approximation. We note for comparison that in the fixed-target
experiments performed at CERN and the Tevatron, the corresponding
values of $\sqrt{s_1}\approx\sqrt{s_2}$ were $\approx3.6$, 4.9, 5.4,
and 6.3 GeV. Therefore, in interpreting the experimental results, it
was necessary in some cases to take possible contributions from
$\mathcal{R}\mathcal{P}$ and $\mathcal{R}\mathcal{R}$ exchanges into
account, in addition to the $\mathcal{P}\mathcal{P}$ exchange.

Production of the $f_1(1285))$ resonance and its subsequent decay
into $\pi^+\pi^-\pi^0$ could also be studied in the central region
of $pp$, $pA$, $\pi^-p$, and $\pi^-A$ collisions in the accelerator
of the Institute of High Energy Physics in Protvino.

\vspace*{0.3cm} \noindent{\boldmath\bf 8.2. Reaction $pp\to
p(a^0_0(980))p\to p(\eta\pi^0)p$}

\noindent The $a^0_0(980)$ resonance production at LHC energies in
the central region in the reaction $pp\to p(a^0_0(980))p\to
p(\eta\pi^0)p$ is supposed to primarily occur owing to the mechanism
shown in Fig. 28 \cite{AS18b}. Due to an incomplete cancellation of
the $K^+K^-$ and $K^0\bar K^0$ intermediate states produced in
$\mathcal{P}\mathcal{P}$ collisions, the isospin-violating amplitude
of $a^0_0(980)$ production does not decrease as energy grows. The
mass spectrum of the final $\eta\pi^0$ system should be a narrow
resonance peak concentrated in the region of the $K\bar K$
thresholds, similarly to that depicted with a solid curve in Fig. 3.
Judging by available data, the amplitude of the $\mathcal{P}
\mathcal{P}\to K\bar K$ transition, which drives the process shown
in Fig. 28, is dominated by the contribution from $f_0(980)$
resonance production \cite{Kir14,GuRe14,Ake86,AMP87,Arm91,Bar99b,
Bar99c,Rey98,Bar99,Bar99a}, $\mathcal{P} \mathcal{P}\to f_0(980)\to
K\bar K$ (Fig. 29).
\begin{figure} 
\includegraphics[width=6.1cm]{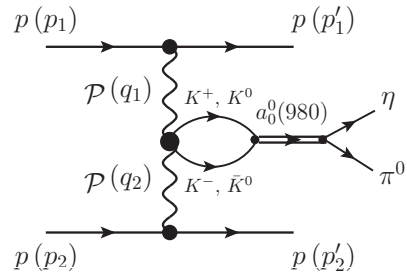}
\caption{\label{Fig21} $K\bar K$-loop mechanism of $a^0_0(980)$
production in the central region in $\mathcal{P}\mathcal{P} $
collisions.}\end{figure}
\begin{figure} 
\includegraphics[width=5.3cm]{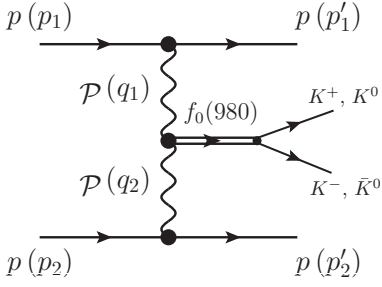}
\caption{\label{Fig22} Production of the $f_0(980)$ resonance in
$K\bar K$ decay channels in the central region in
$\mathcal{P}\mathcal{P} $ collisions.}\end{figure}
Indeed, the mass spectra of $K^+K^-$ and $K^0\bar K^0$ pairs exhibit
a powerful enhancement in the vicinity of their thresholds
\cite{Rey98, Bar99,Bar99a}. The resonance $f_0(980)$ also clearly
manifests itself in the $\pi^+\pi^-$ and $\pi^0\pi^0$ mass spectra,
albeit in the form of a narrow dip that occurs as a result of
destructive interference with a large and smooth coherent background
\cite{Kir14,Ake86,AMP87,Arm91,Bar99b, Bar99c,CMS17}. Thus,
$a^0_0(980)$ is quite probably produced due to the
$a^0_0(980)-f_0(980)$ mixing, $\mathcal{P}\mathcal{P}\to
f_0(980)\to(K^+K^-+K^0\bar K^0)\to a^0_0(980)\to\eta\pi^0$. The
corresponding cross section as a function of the invariant mass
$M\equiv m$ of the $\eta\pi^0$ system is given by
\begin{eqnarray}\label{Eq7-2-1a}
\sigma(\mathcal{P}\mathcal{P}\to f_0(980)\to(K^+K^-+K^0\bar K^0)
\qquad\nonumber\\ \to
a^0_0(980)\to\eta\pi^0;m)=|C_{\mathcal{P}\mathcal{P}\to
f_0}|^2\,m\Gamma_{a^0_0\to \eta\pi^0}(m)\nonumber\\
\times\left|\frac{\Pi_{a^0_0f_0}(m)}{D_{a^0_0}(m)D_{f_0}(m)-
\Pi^2_{a^0_0f_0}(m)}\right|^2\,,\qquad\qquad\
\end{eqnarray} where $C_{\mathcal{P}\mathcal{P}\to
f_0}$ is the $\mathcal{P}\mathcal{P}\to f_0(980)$ transition
amplitude.

The $a^0_0(980)$ production cross section in $\mathcal{P}
\mathcal{P}$ collisions followed by the decay into $\eta\pi^0$ can
be estimated without going into details of the transition
$\mathcal{P}\mathcal{P}\to K\bar K$ (see Fig. 29). The latter can,
in principle, be due to not only the contribution of the $f_0(980)$
resonance (see Fig. 29) but also a nonresonance mechanism of $K\bar
K$ production.

For this, we use the relation
\begin{eqnarray}\label{Eq7-2-1}
\sigma(\mathcal{P}\mathcal{P}\to a^0_0(980)\to\eta\pi^0;m)
\qquad\qquad \nonumber\\
\ \ \ \approx|\widetilde{A}(2m_{K^+})|^2|\rho_{K^+K^-}(m)-
\rho_{K^0\bar K^0}(m)|^2\quad \nonumber\\
\times\,\frac{g^2_{a^0_0K^+K^-}}{16\pi}\,\frac{m\Gamma_{a^0_0\to
\eta\pi^0}(m)}{|D_{a^0_0}(m)|^2}\,,\qquad\qquad\quad
\end{eqnarray}
in which $|\widetilde{A}(2m_{K^+})|^2$ should be determined using
data on the $K^+K^-$ production cross section in the near-threshold
region:
\begin{eqnarray}\label{Eq7-2-2}
\sigma(\mathcal{P}\mathcal{P}\to K^+K^-;m)=\rho_{K^+K^-}(m)\,|
\widetilde{A}(m)|^2\,.\ \
\end{eqnarray}
An order-of-magnitude estimate in the region of $m$ between the
$K^+K^-$ and $K^0\bar K^0$ thresholds yields
\begin{eqnarray}\label{Eq7-2-3} \sigma(\mathcal{P}\mathcal{P}\to
a^0_0(980)\to \eta\pi^0;m) \approx0.05|\widetilde{A}(2m_{K^+})|^2.
\end{eqnarray}
A comparison of this estimate with data on $\sigma(\mathcal{P}
\mathcal{P}\to a^0_0(980)\to\eta\pi^0;m)$ allows verifying their
consistency with data on $\sigma(\mathcal{P}\mathcal{P}\to
K^+K^-;m)$ and the hypothesized violation of isotopic invariance due
to the mass difference of $K^+$ and $K^0$ mesons. It should be kept
in mind that a similar way to check the consistency of the results
of measurements for the decays $f_1(1285)\to\pi^+\pi^-\pi^0$ and
$f_1(1285)\to K\bar K\pi$ was discussed in Section 5.2. Formulas
relating $\sigma(\mathcal{P}\mathcal{P}\to X^0;m)$ to the
experimentally measured cross section of the reaction $pp\to
p(X^0)p$ can be found, e.g., in \cite{AlGo81,GaRo80,Str86}.

Data about the central production of $a^0_0(980)$ in the reaction
$pp\to p(\eta \pi^0)p$ are only available from measurements made by
the Omega spectrometer at CERN at $\sqrt{s}=29$ GeV
\cite{Bar00b,Sob01} (see the discussion on the interpretation of
those data in \cite{CK1,AK02,AS04a,WZZ07}). We note the following.
Clear-cut peaks at $\sqrt{s}=29$ GeV have been observed in the
$\eta\pi^0$ mass spectrum that correspond to $a^0_0(980)$ and
$a^0_2(1320)$ resonances with widths typical of those mesons,
$\Gamma(a_0(980))=72\pm16$ MeV and $\Gamma(a_2(1320))=115\pm20$ MeV
(see PDG data \cite{PDG16}). This picture at the energies
$\sqrt{s_1}\approx\sqrt{s_2}\approx\left(
m^2_{a^0_0}s\right)^{1/4}\approx\left(29^2\
\mbox{GeV}^4\right)^{1/4}\approx5.4$ GeV indicates a significant
role of secondary Regge trajectories, for which the $\eta\pi^0$
production in the central region is not forbidden by $G$-parity. For
example, the $a^0_0(980)$ resonance can be produced in merging the
$\eta\pi^0$, $a^0_2f_2$, and $a^0_2\mathcal{P}$ Regge exchanges.
Contributions from the secondary Regge trajectories significantly
taper off at LHC energies, and it is natural to expect that the
$a^0_0(980)$ resonance will only be produced owing to the mechanism
described above (see Fig. 28), which does not `fade' and violates
isotopic invariance. The change in the mechanism of central
production of $a^0_0(980)$ will be signalled by a narrowing of the
$a^0_0(980)$ peak in the $\eta\pi^0$ channel with increasing energy.

\vspace{0.3cm} \noindent{\large\bf 9. Conclusion} \vspace{0.2cm}

\noindent The phenomenon of $a^0_0(980)-f_0(980)$ mixing
\cite{ADS79} gave a boost to the experimental search for its effects
in reactions (24)--(28) as performed by the VES \cite{Do08,Do11} and
BES III \cite{Ab1,Ab2,Ab3} collaborations. These activities clearly
showed that studies of the $a^0_0(980)-f_0(980)$ mixing in various
reactions and a more general phenomenon, the $K\bar K$-loop
mechanism of isotopic invariance violation, can be very helpful in
identifying both the production mechanism of light scalar mesons and
the nature of those mesons \cite{A2016,AS16}.

We emphasize once again that the mass spectrum of light scalar
mesons $\sigma(600)$, $\kappa(800)$, $a_0(980)$, and $f_0(980)$
indicates their four-quark, $q^2\bar q^2$, structure. The rates and
mechanisms of the production of $a_0(980)$ and $f_0(980) $
resonances in radiative decays of the $\phi(1020)$ meson, four-quark
transitions $\phi(1020)\to K^+K^-\to\gamma[a_0(980)/f_0(980)]$,
indicate their $q^2\bar q^2$ nature. The rates and mechanisms of
two-photon production of light scalars, four-quark transitions
$\gamma\gamma\to\pi^+\pi^-\to\sigma(600)$, $\gamma\gamma\to
\pi^0\eta\to a_0(980)$, and $\gamma\gamma\to K^+K^-\to f_0(980)/a_0
(980)$, are also evidence in favor of their $q^2\bar q^2$ nature.

We also note that these states cannot be loosely bound molecules
\cite{A2016}. The decay $\phi(1020)\to K^+K^-\to\gamma
a^0_0(980)/f_0(980)$ was shown in \cite{AGShev97,AK07b,AK08} to
involve virtual momenta of $K(\bar K)$ mesons over 2 GeV$/c$, while
in the case of loosely bound molecules with a binding energy of
about 20 MeV the corresponding momenta would be approximately 100
MeV$/c$. It is noteworthy that the production of scalar mesons in
$\pi N $ collisions at large transferred momenta also indicates a
compact structure of these particles \cite{AS98}. Promising
prospects are presented in \cite{A2016} for studying the nature of
$a_0(980)$ and $f_0(980) $ states in $\gamma\gamma$ and
$\gamma\gamma^*$ collisions (where $\gamma^*$ is the virtual
gamma-quantum), in $J/\psi$ decays, and in $\pi N$ collisions, and
when comparing the production of light scalar and pseudoscalar
mesons in semileptonic decays of $D_s$ and $D$ mesons in c$-\tau$
and super-c$-\tau$ factories and semileptonic decays of $B$ mesons
in super-b factories.

The high statistical accuracy of modern-day experiments raises hopes
that data on $a^0_0(980)-f_0(980)$ mixing will be elucidated (as
confirmed by the recent study of the BESIII collaboration
\cite{Ab18c}), and new precise data on the decays $f_1(1285)\to
f_0(980) \pi^0\to\pi^+\pi^-\pi^0$ and $\eta(1405)\to
f_0(980)\pi^0\to\pi^+ \pi^-\pi^0$ will be obtained. The isotopic
invariance violation discussed in this review will hopefully be
discovered in polarization experiments, weak hadronic decays of
charmed $D^+_s$ and $D^0 $ mesons, the bottomonium decay
$\Upsilon(10860)\to\Upsilon(1S) f_0(980)\to\Upsilon(1S)\eta\pi^0$,
and the production of hadrons in the central region. It is quite
probable that other interesting cases of strong isospin violation in
the production of $a^0_0(980)$ and $f_0(980)$ resonances related to
the difference in masses between $K^+$ and $K^0$ mesons will be
observed, in particular, in $B$ and $B_s$ decays.

\vspace*{0.2cm}
This study was partially supported by the grant no. 16-02-00065 from
the Russian Foundation for Basis Research and the grant no.
0314-2019-0021 from Program no. II.15.1 of fundamental scientific
research of the Siberian Branch of the Russian Academy of Sciences.


\end{document}